\documentclass[preprint,12pt,3p]{elsarticle}

\usepackage{amssymb}
\usepackage{lineno,hyperref}
\usepackage{amsmath}
\usepackage{bm}
\usepackage{subcaption}
\usepackage{graphicx}
\usepackage{svg}
\usepackage{rotating}
\usepackage{lscape}
\usepackage{array,multirow}
\usepackage[title]{appendix}
\usepackage{float}
\usepackage{stmaryrd}
\usepackage{textcomp}
\usepackage{algorithm}
\usepackage[noend]{algpseudocode}
\usepackage{amsthm}
\usepackage{listings}
\usepackage{xcolor}

\newcommand{\rev}[1]{{{\textcolor{black}{#1}}}}

\theoremstyle{definition}

\newtheorem{remark}{Remark}

\usepackage{lineno}

\makeatletter
\def\BState{\State\hskip-\ALG@thistlm}
\makeatother

\journal{Elsevier}

\begin{document}

\begin{frontmatter}

\title{Enriched Galerkin Discretization for Modeling Poroelasticity and Permeability Alteration in Heterogeneous Porous Media}

\author[label1,label2]{T. Kadeethum\corref{cor1}}
\address[label1]{Technical University of Denmark, Denmark}
\address[label2]{Cornell University, New York, USA}

\cortext[cor1]{corresponding author}

\ead{teekad@dtu.dk}

\author[label1]{Nick H. M.}
\ead{hamid@dtu.dk}

\author[label3]{Lee S.}
\address[label3]{Florida State University, Florida, USA}
\ead{lee@math.fsu.edu}

\author[label4]{Ballarin F.}
\address[label4]{mathLab, Mathematics Area, SISSA, Italy}
\ead{francesco.ballarin@sissa.it}

\begin{abstract}
Accurate simulation of the coupled fluid flow and solid deformation in porous media is challenging, especially when the media permeability and storativity are heterogeneous. \rev{We apply the enriched Galerkin (EG) finite element method for the Biot\textquotesingle s system}. Block structure used to compose the enriched space and linearization and iterative schemes employed to solve the coupled media permeability alteration are illustrated. \rev{The open-source platform used to build the block structure is presented and illustrate that it helps the enriched Galerkin method easily adaptable to any existing discontinuous Galerkin codes.} Subsequently, we compare the EG method with the classic continuous Galerkin (CG) and discontinuous Galerkin (DG) finite element methods. While these methods provide similar approximations for the pressure solution of Terzaghi\textquotesingle s one-dimensional consolidation, the CG method produces spurious oscillations in fluid pressure and volumetric strain solutions at material interfaces that have permeability contrast and does not conserve mass locally. As a result, the flux approximation of the CG method is significantly different from the one of EG and DG methods, especially for the soft materials. The difference of flux approximation between EG and DG methods is insignificant; still, the EG method demands approximately two and three times fewer degrees of freedom than the DG method for two- and three-dimensional geometries, respectively. \rev{Lastly, we illustrate that the EG method produces accurate results even for much coarser meshes.} 

\end{abstract}


\end{frontmatter}


\section{Introduction}

The volumetric displacement of a porous medium caused by the changing in pore pressure is relevant for many applications, including groundwater flow, underground heat mining, fossil fuel production, earthquake mechanics, and biomedical engineering  \cite{nick2013reactive,bisdom2016impact,lee2016pressure,juanes2016were,vinje2018fluid,kadeethum2020well,kadeethum2020flow}. The volumetric deformation may impact the hydraulic storability and permeability of porous material, which in turn, influences the fluid flow field. This coupling between fluid flow and solid deformation can be captured through the application of Biot\textquotesingle s equation of poroelasticity \cite{biot1941general, biot1957elastic}.

The Biot's equation is often discretized using the two-field formulation and classical continuous Galerkin (CG) finite element methods because the implementation is straightforward and results in a small number of degrees of freedom (DOF) \cite{Haga2012,SlatlemVik2018,salimzadeh2019effect,kadeethum2019investigation}. The CG method, in its classical form, is unable to provide solutions with local mass conservation or fluid pressure oscillation free at material interfaces with high permeability contrast \cite{kadeethum2020finite,vermeer1981accuracy, murad1994stability,scovazzi2017analytical}. \par

To mitigate fluid pressure oscillations at material interfaces and assuring local mass conservation, the following approaches have been proposed:
$\mathrm{\bm{(1)}}$ an element based post-processing technique applied to the classic CG method \cite{deng2017locally}, $\mathrm{\bm{(2)}}$ mixed finite element (MFE), i.e. three- and four-field formulations, developed by adding fluid velocity and solid pressure to primary variables \cite{phillips2007coupling2,wheeler2014coupling, Haga2012}, $\mathrm{\bm{(3)}}$ hybrid-finite-element-finite-volume (FEFV) established by combing the CG finite element and cell-centered finite volume discretizations \cite{flemisch2011dumux,nick2011comparison},  $\mathrm{\bm{(4)}}$ a discontinuous Galerkin (DG) method utilizing an interior penalty approach to stabilize the solution and preserve the local mass \cite{phillips2008coupling, liu2009modeling}, and  $\mathrm{\bm{(5)}}$ a weak Galerkin (WG), which employs the first order Bernardi-Raugel elements on quadrilaterals mesh \cite{harper2018lowest, liu2018lowest}. In recent years, physics-informed neural networks (PINN) have also been applied to solve this problem as a meshfree method \cite{Kadeethum2020ARMA,kadeethum2020pinn}.
\par

Although the above six proposed methods have many benefits, there are disadvantages to each technique. The FEFV method creates a dual-mesh, which may result in smearing of mass in heterogeneous reservoirs \cite{nick2011hybrid, salinas2018discontinuous}. The MFE method, on the other hand, increases the number of primary variables. Therefore, it may require more computational resources compared to the two-field formulation, especially in a three-dimensional domain. Moreover, it involves the inversion of the permeability tensor, which may lead to an ill-posed problem \cite{choo2018enriched, Haga2012}. Even though the DG method is developed based on the two-field formulation, it is a rather computationally expensive method \cite{sun2009locally}.

\begin{figure}[H]
   \centering
        \includegraphics[width=9.5cm, height=4.8cm]{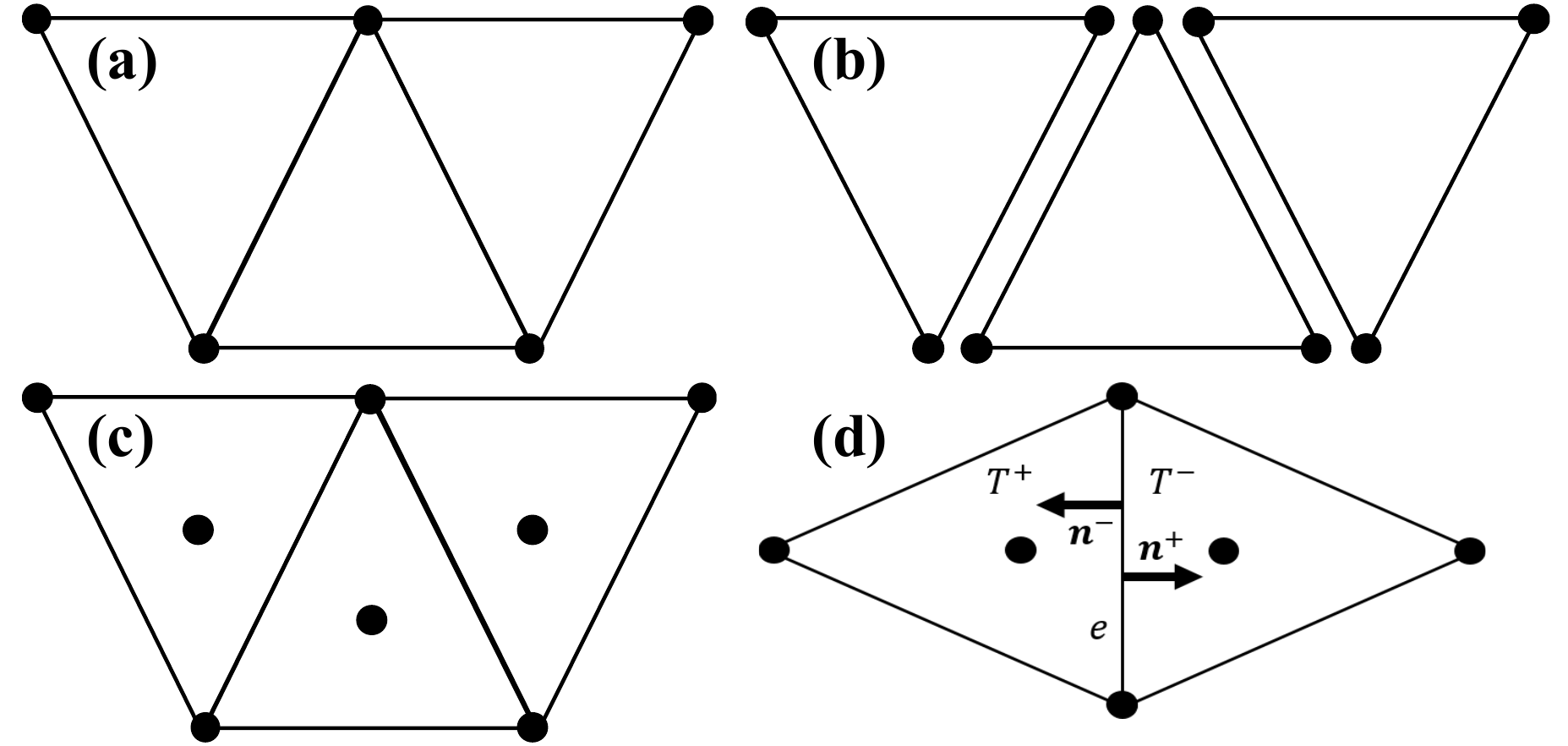}
   \caption{Comparison of degrees of freedom for linear polynomial case among (\textbf{a}) continuous Galerkin (CG), (\textbf{b}) discontinuous Galerkin (DG), (\textbf{c}) enriched Galerkin (EG) function spaces, and (\textbf{d}) EG elements, where $T$ is a triangular element, $e$ is a face of $T$, and $\bm{n}$ is a normal unit vector to $e$.}
   \label{fig:function_spaces}
\end{figure}

\rev{To overcome some of the disadvantages discussed above, we apply the enriched Galerkin (EG) method, which is composed of the CG function space augmented by a piecewise-constant function at the center of each element.} This method has the same interior penalty type bi-linear form as the DG method \cite{sun2009locally,lee2016locally}. As the EG method only requires to have discontinuous constants, it has fewer DOF than the DG method (see Figures \ref{fig:function_spaces}a-c for the comparison of DOF among CG, EG, and DG methods). \rev{The EG method has been developed to solve general elliptic and parabolic problems with dynamic mesh adaptivity  \cite{lee2017adaptive,lee2018enriched,lee2018phase} and recently extended to address the poroelastic problem \cite{choo2018enriched,choo2018large, choo2019stabilized}. Previous studies illustrate that the EG method provides local mass conservation without any post-processing operations, and delivers smooth solutions across interfaces where material permeability are highly different \cite{choo2018enriched,choo2018large, choo2019stabilized}.} \par

\rev{In this study, we illustrate the block structure built to compose the EG function space and the iterative scheme utilized to incorporate the media permeability alteration resulted from the volumetric displacement into the system. Besides, we present the open-source platforms \cite{AlnaesBlechta2015a,Ballarin2019} used to build this system, and emphasize their capabilities to simplify the process of creating the EG solver. Subsequently, we illustrate the performance of the EG method in both structural and fully-random heterogeneous media and gauge the sensitivity of its solutions to mesh refinement.} As we aim to provide comprehensive comparisons among the CG (in its classical form), EG, and DG methods, the following six quantities of interest are carefully discussed: local mass conservation, flux approximation, permeability and volumetric strain alteration, number of iterations, and DOF. 

\rev{In summary, the presented work based on \cite{choo2018enriched} has the following novelties:} 
\begin{enumerate}
    \item \rev{This paper proposes a numerical model that incorporates fluid flow in heterogeneous poroelastic media and matrix permeability alteration. The permeability alteration model used in this paper is applicable to underground energy harvesting, especially oil and gas engineering \cite{abou2013petroleum,Du2007}. The presented model couples the permeability and the displacement, and the solution algorithm to avoid numerical instability is established. Besides, the effect of permeability alteration from the solid deformation is illustrated and compared its intensity with different material properties.}
    \item \rev{Several detailed numerical results comparing CG, DG, and EG methods, in a single computational platform is presented. Moreover, we also show the effects of material properties on the accuracy with several numerical examples, including an actual North Sea field for different material properties in two- and three-dimensional spaces. 
    }
    \item \rev{A new computational framework of the EG method by employing the multiphenics framework \cite{Ballarin2019}, which is built upon the FEniCS platform \cite{AlnaesBlechta2015a} is developed. This framework is based on the block structures, which simplifies the implementation of the EG method. The scripts for creating the EG solver is partially illustrated in this manuscript and placed in the multiphenics repository.}
    
    \item \rev{We illustrate that the EG method allows using a relatively coarse mesh to maintain accuracy. This is an essential feature of the EG method that makes it more suitable for large scale simulations.}
    
\end{enumerate}

The rest of the paper is organized as follows. The methodology section includes model description, mathematical equations, and their discretizations for CG, EG, and DG methods. Subsequently, the solution strategy, including block formulations and iteration scheme, is illustrated. The results and discussion section begins with an analysis of the error convergence rate between the EG and DG methods to verify the developed block structure. Next, the established numerical schemes, CG, EG, and DG methods, for solving the coupled solid deformation and fluid flow system are validated using Terzaghi\textquotesingle s 1D consolidation model \cite{terzaghi1951theoretical}. \rev{Subsequently, we highlight the positive and negatives aspects of applying CG, EG, and DG methods for four examples: 1D consolidation of a two-layered material, 2D flow in a deformable media with structured heterogeneity, 2D flow in a deformable media with random porosity and permeability, 3D flow in a deformable media with random porosity and permeability, and a North Sea hydrocarbon field.} Summary and conclusion are finally provided.\par

\section{Methodology}\label{sec:methodology}

\subsection{Governing equations}

We are interested in solving Biot\textquotesingle s equation on $\Omega$, which amounts to time-dependent multiphysics problem coupling the solid deformation with fluid flow problem. Let $\Omega \subset \mathbb{R}^d$ be the domain of interest in $d$-dimensional space where $d = 1,$ $2,$ or $3$ bounded by boundary, $\partial \Omega$. $\partial \Omega$ can be decomposed to displacement and traction boundaries, $\partial \Omega_u$ and $\partial \Omega_t$, respectively, for the solid deformation problem. For the fluid flow problem, $\partial \Omega$ is decomposed to pressure and flux boundaries, $\partial \Omega_p$ and $\partial \Omega_q$, respectively. The time domain is denoted by $\mathbb{T} = \left(0,\tau\right]$ with $\tau>0$.  \par


The coupling between the fluid flow and solid deformation can be captured through the application of Biot\textquotesingle s equation of poroelasticity, which is composed of linear momentum and mass balance equations \citep{biot1941general}. The linear momentum balance equation can be written as follows:  \par

\begin{equation}
\nabla \cdot \bm{\sigma}(\bm{u},p) =\bm{f},
\end{equation}

\noindent
where $\bm{u}$ is displacement, $p$ is fluid pressure, $\bm{f}$ is body force, which is neglected in this study. The bold-face letters or symbols denote tensors and vectors, and the normal letters or symbols denote scalar values. Here, $\bm{\sigma}$ is total stress, which is defined as:

\begin{equation}
\bm{\sigma}:=\bm{\sigma^{\prime}} - \alpha p \bm{I},
\end{equation}

\noindent
where $\bm{I}$ is the identity tensor and $\alpha$ is Biot\textquotesingle s coefficient defined as \cite{jaeger2009fundamentals}:

\begin{equation} \label{eq:biot_coeff}
\alpha:=1-\frac{K}{K_{{s}}},
\end{equation}

\noindent
with the bulk modulus of a rock matrix $K$ and the solid grains modulus $K_s$. In addition, $\bm{\sigma^{\prime}}$ is an effective stress written as:

\begin{equation}
\bm{\sigma^{\prime}}:=2 \mu_{l} \bm{\varepsilon}+\lambda_{l} \nabla \cdot \boldsymbol{u} \bm{I},
\end{equation}

\noindent
where $\lambda_{l}$ and $\mu_{l}$ are Lam\'e constants, $\bm{\varepsilon}$ is strain,  which is  defined as:

\begin{equation}
\bm{\varepsilon} :=\frac{1}{2}\left(\nabla \boldsymbol{u}+\nabla^{T} \boldsymbol{u}\right),
\end{equation}
by assuming infinitesimal displacements.

\noindent
To this end, we can write the linear momentum balance and its boundary conditions as:

\begin{equation}\label{eq:linear_balance}
\nabla \cdot \bm{\sigma^{\prime}}(\bm{u}) -\alpha \nabla \cdot (p \bm{I}) = \bm{f} \text { in } \Omega \times \mathbb{T},
\end{equation}
\begin{equation}
\boldsymbol{u}=\boldsymbol{u}_{D} \text { on } \partial \Omega_{u} \times \mathbb{T},
\end{equation}
\begin{equation}
\boldsymbol{\sigma} \cdot \boldsymbol{n}=\bm{\sigma_{D}} \text { on } \partial \Omega_{t} \times \mathbb{T},
\end{equation}
\begin{equation}
\bm{u}=\bm{u}_{0} \text { in } \Omega \text { at } t = 0,
\end{equation}
\noindent



\noindent
where $\bm{u_D}$ and ${\bm{\sigma_D}}$ are prescribed displacement and traction at boundaries, respectively, and $t$ is time.

The second equation is the mass balance equation, which is written as \cite{coussy2004poromechanics}:

\begin{equation} \label{eq:mass_balance}
\rho\left(\phi c_{f}+\frac{\alpha-\phi}{K_{s}}\right) \frac{\partial p}{\partial t}+\rho \alpha \frac{\partial \left(\nabla \cdot \boldsymbol{u}\right)}{\partial t}-\nabla \cdot \bm{\kappa}(\nabla p-\rho \mathbf{g})=g \text { in } \Omega \times \mathbb{T},
\end{equation}



\begin{equation}
p=p_{D} \text { on } \partial \Omega_{p} \times \mathbb{T},
\end{equation}

\begin{equation}
-\nabla \cdot \bm{\kappa}(\nabla p-\rho \mathbf{g}) \cdot \boldsymbol{n}=q_{D} \text { on } \partial \Omega_{q} \times \mathbb{T},
\end{equation}

\begin{equation}
p=p_{0} \text { in } \Omega \text { at } t = 0,
\end{equation}

\noindent
where $\rho$ is a fluid density, $\phi$ is an initial porosity and remains constant throughout the simulation (the volumetric deformation is represented by $\frac{\partial \left(\nabla \cdot \boldsymbol{u}\right)}{\partial t}$), $c_f$ is fluid compressibility, $\mathbf{g}$ is a gravitational vector, $g$ is sink/source, $p_D$ and $q_D$ are specified pressure and flux, respectively, and $\bm{\kappa}$ is defined as:

\begin{equation}
\bm{\kappa}:=\frac{\rho {\bm{k_{m}}}}{\mu},
\label{eq:kappa}
\end{equation}

\noindent
$\bm{k_m}$ is matrix permeability tensor and $\mu$ is fluid viscosity. By assuming that the rock volumetric displacement can cause the matrix permeability alteration, $\bm{k_m}$ is defined as \citep{Du2007, abou2013petroleum}:

\begin{equation}\label{eq:wong_perm}
\bm{k_{m}}=\bm{k_{m0}}\frac{\left(1+\frac{\varepsilon_{v}}{\phi}\right)^{3}}{1+\varepsilon_{v}},
\end{equation}

\noindent
where $\bm{k_{m0}}$ represent initial rock matrix permeability. Here, $\varepsilon_{v}$ is the total volumetric strain, which is defined as:

\begin{equation} \label{eq:volumetric_str}
\varepsilon_{v} :=\operatorname{tr}({\bm{\varepsilon}})=\sum_{i=1}^{d} \varepsilon_{i i}.
\end{equation}

\noindent
Throughout this paper, we assume $\rho$ and $\mu$ are constants; thus, $\bm{\kappa}$ varies only as a function of $\bm{k_m}$ throughout the simulation. \rev{Note that the inverse of Biot's modulus ($1/M$) is equivalent to $\phi c_{f}+\frac{\alpha-\phi}{K_{s}}$ term of \eqref{eq:mass_balance} \cite{yoon2018spatial}.}



\subsection{Numerical discretization}

The domain $\Omega$ is partitioned into $n_{e}$ elements, $\mathcal{T}_{h}$, which is the family of elements $T$ (triangles in 2D, tetrahedrons in 3D). We will further denote by $e$ a face of $T$, and $\bm{n}$ is a normal unit vector to $e$ as illustrated in Figure \ref{fig:function_spaces}d. We denote $h_T$ as the diameter of $T$ and define $h :=\max \left(h_{T}\right)$, which we may refer as mesh size hereafter. Let $\mathcal{E}_{h}$ denotes the set of all facets, $\mathcal{E}_{h}^{0}$ the internal facets, $\mathcal{E}_{h}^{D}$ the Dirichlet boundary faces, and $\mathcal{E}_{h}^{N}$ denotes the Neumann boundary faces. $\mathcal{E}_{h}=\mathcal{E}_{h}^{0}$ $\cup$ $\mathcal{E}_{h}^{D}$ $\cup$ $\mathcal{E}_{h}^{N}$, $\mathcal{E}_{h}^{0}$ $\cap$ $\mathcal{E}_{h}^{D}=\emptyset$, $\mathcal{E}_{h}^{0}$ $\cap$ $\mathcal{E}_{h}^{N}=\emptyset$, and $\mathcal{E}_{h}^{D} \cap \mathcal{E}_{h}^{N}=\emptyset$.  \par

This study focuses on three function spaces, arising from CG, EG, and DG discretizations, respectively. We begin with defining the CG function space for displacement, $\bm{u}$ as

\begin{equation}
\mathcal{U}_{h}^{\mathrm{CG}_{k}}\left(\mathcal{T}_{h}\right) :=\left\{\bm{\psi_u} \in \mathbb{C}^{0}(\Omega{; \mathbb{R}^d}) :\left.\bm{\psi_u}\right|_{T} \in \mathbb{P}_{k}(T{; \mathbb{R}^d}), \forall T \in \mathcal{T}_{h}, \bm{u} = \bm{u}_D \text{ on } \mathcal{E}_{h}^{D} \right\},
\label{eq:CG_U}
\end{equation}

\noindent
where $\mathcal{U}_{h}^{\mathrm{CG}_{k}}\left(\mathcal{T}_{h}\right)$ is the space for the $\mathrm{CG}$ approximation with $k^{th}$ degree polynomials for the $\bm{u}$ unknown, $\mathbb{C}^0(\Omega{; \mathbb{R}^d})$ denotes the space of vector-valued piecewise continuous polynomials, $\mathbb{P}_{k}(T{; \mathbb{R}^d})$ is the space of polynomials of degree at most $k$ over each element $T$, and $\bm{\psi_u}$ denotes a generic function of $\mathcal{U}_{h}^{\mathrm{CG}_{k}}\left(\mathcal{T}_{h}\right)$. Subsequently, the CG function space for pressure, $p$, is written as follows:

\begin{equation}\label{eq:CG_P}
\mathcal{P}_{h}^{\mathrm{CG}_{k}}\left(\mathcal{T}_{h}\right) :=\left\{\psi_p \in \mathbb{C}^{0}(\Omega) :\left.\psi_p\right|_{T} \in \mathbb{P}_{k}(T), \forall T \in \mathcal{T}_{h}\right\},
\end{equation}

\noindent
using a similar notation as in \eqref{eq:CG_U}, the only exception being the scalar-valued unknown. Furthermore, we define the following DG function space for the pressure $p$:

\begin{equation}\label{eq:DG_P}
\mathcal{P}_{h}^{\mathrm{DG}_{k}}\left(\mathcal{T}_{h}\right) :=\left\{\psi_p \in L^{2}(\Omega) :\left.\psi_p\right|_{T} \in \mathbb{P}_{k}(T), \forall T \in \mathcal{T}_{h}\right\},
\end{equation}

\noindent
where $\mathcal{P}_{h}^{\mathrm{DG}_{k}}\left(\mathcal{T}_{h}\right)$ is the space for the $\mathrm{DG}$ approximation with $k^{th}$ degree polynomials for the $p$ space and $L^{2}(\Omega)$ is the space of square integrable functions. Finally, we define the EG function space for $p$ as:

\begin{equation}\label{eq:EG_P}
\mathcal{P}_{h}^{\mathrm{EG}_{k}}\left(\mathcal{T}_{h}\right) :=\mathcal{P}_{h}^{\mathrm{CG}_{k}}\left(\mathcal{T}_{h}\right)+\mathcal{P}_{h}^{\mathrm{DG}_{0}}\left(\mathcal{T}_{h}\right),
\end{equation}

\noindent
where $\mathcal{P}_{h}^{\mathrm{EG}_{k}}\left(\mathcal{T}_{h}\right)$ is the space for the $\mathrm{EG}$ approximation with $k^{th}$ degree polynomials for the $p$ space and $\mathcal{P}_{h}^{\mathrm{DG}_{0}}\left(\mathcal{T}_{h}\right)$ is the space for the $\mathrm{DG}$ approximation with $0^{th}$ degree polynomials, in other words, a piecewise constant approximation.
Note that EG discretization is expected to be beneficial for an accurate approximation of the pressure $p$ near interfaces where high permeability contrast is observed, for which a CG discretization may result in unphysical oscillations. In contrast, the mechanical properties that dictate the displacement field, $\bm{u}$, are assumed to be homogeneous, and the solid is compressible. Therefore, the CG discretization of the displacement unknown will suffice in the following.

We now seek approximate solutions $p_h \in \mathcal{P}_{h}^{\mathrm{EG}_{k}}\left(\mathcal{T}_{h}\right)$ and $\bm{u_h} \in \mathcal{U}_{h}^{\mathrm{CG}_{k}}\left(\mathcal{T}_{h}\right)$ of $p$ and $\bm{u}$, respectively.
The linear momentum balance equation, \eqref{eq:linear_balance}, is discretized using the CG method as follows \cite{choo2018enriched, lee2020choice,SlatlemVik2018}:

\begin{equation} \label{eq:linear_balance_dis}
\begin{split}
\sum_{T \in \mathcal{T}_{h}} \int_{T} \boldsymbol{\sigma}^{\prime}\left(\bm{u}_{h}\right) : \nabla^{s} \bm{\psi_u} \: d V & +  \sum_{T \in \mathcal{T}_{h}} \int_{T} \alpha  \nabla p_{h}  . \nabla \bm{\psi_u} \: d V \\ & =\sum_{T \in \mathcal{T}_{h}} \int_{T} \bm{f} \bm{\psi_u} \: d V+\sum_{e \in \mathcal{E}_{h}^{N}} \int_{e} \bm{\sigma_{D}} \bm{\psi_u} \: d S, \quad \forall \bm{\psi_u} \in \mathcal{U}_{h}^{\mathrm{CG}_{k}}\left(\mathcal{T}_{h}\right),
\end{split}
\end{equation}

\noindent
where $\int_{T} .  d V$ and $\int_{e} .  d S$ refer to volume and surface integrals, respectively. On the contrary, the mass balance equation, \eqref{eq:mass_balance}, is discretized using the EG method with symmetric internal penalty approach \cite{choo2018enriched,Kadeethum2019ARMA,kadeethum2020finite}, and the space discretization is presented as follows::

\begin{equation} \label{eq:mass_discretization}
\begin{split}
\sum_{T \in \mathcal{T}_{h}} &\int_{T} \rho\left(\phi c_{f}+\frac{\alpha-\phi}{K_{s}}\right) \frac{\partial p_{h}}{\partial t} \psi_{p} \: d V+\sum_{T \in \mathcal{T}_{h}} \int_{T} \rho \alpha \frac{\partial \nabla \cdot \boldsymbol{u}_{h}}{\partial t} \psi_{p} \: d V \\
& +\sum_{T \in \mathcal{T}_{h}} \int_{T} \bm{\kappa}\left(\nabla p_{h}-\rho \mathbf{g}\right) \cdot \nabla \psi_{p} \: d V - \sum_{e \in \mathcal{E}_{h}^{0} \cup \mathcal{E}_{h}^{D}} \int_{e}\left\{\bm{\kappa}\left(\nabla p_{h}-\rho \mathbf{g}\right)\right\}_{\delta_{e}} \cdot \llbracket \psi_p \rrbracket \: d S \\
& -\sum_{e \in \mathcal{E}_{h}^{0} \cup \mathcal{E}_{h}^{D}} \int_{e}\left\{\bm{\kappa}  \nabla \psi_{p}\right\}_{\delta_{e}} \cdot \llbracket p_h \rrbracket \: d S + \sum_{e \in \mathcal{E}_{h}^{0} \cup \mathcal{E}_{h}^{D}} \int_{e} \frac{\beta}{h_{e}} {\kappa}_{{e}}  \llbracket p_h \rrbracket \cdot \llbracket \psi_p \rrbracket \: d S \\
& =\sum_{T \in \mathcal{T}_{h}} \int_{T} g \psi_{p} \: d V+\sum_{e \in \mathcal{E}_{h}^{N}} \int_{e} q_{D} \psi_{p} \: d S -\sum_{e \in \mathcal{E}_{h}^{D}} \int_{e} \bm{\kappa} \nabla \psi_{p} \cdot p_{D} \boldsymbol{n} \: d S \\
&+\sum_{e \in \mathcal{E}_{h}^{D}} \int_{e} \frac{\beta}{h_{e}} {\kappa}_{{e}} \llbracket \psi_p \rrbracket \cdot p_D \bm{n} \: d S, \quad \forall \psi_p \in \mathcal{P}_{h}^{\mathrm{EG}_{k}}\left(\mathcal{T}_{h}\right).
\end{split}
\end{equation}

\noindent
Here, $\llbracket X \rrbracket$ is a jump operator defined as:

\begin{equation}
\begin{split}
&\llbracket X \rrbracket =X^{+} \boldsymbol{n}^{+}+X^{-} \boldsymbol{n}^{-}, \\
& \llbracket  \bm{X}  \rrbracket  =\boldsymbol{X}^{+} \cdot \boldsymbol{n}^{+}+\boldsymbol{X}^{-} \cdot \boldsymbol{n}^{-},
\end{split}
\end{equation}

\noindent
where $\boldsymbol{n}^{+}$ and $\boldsymbol{n}^{-}$ are the unit normal vector of $T^+$  and $T^-$, respectively (see Figure \ref{fig:function_spaces}d). In addition, $\{X\}_{\delta_{e}}$ is a weighted average operator defined as follows:

\begin{equation}
\{X\}_{\delta e}=\delta_{e} X^{+}+\left(1-\delta_{e}\right) X^{-},
\end{equation}

\noindent
where $\delta_{e}$, in this study, is calculated as \cite{lee2016locally}:

\begin{equation}
\delta_{e} :=\frac{{\kappa}^{-}_e}{{\kappa}^{+}_e+{\kappa}^{-}_e},
\end{equation}

\begin{equation}
\begin{split}
&{\kappa}^{+}_e :=\left(\boldsymbol{n}^{+}\right)^{T} \cdot \bm{\kappa}^{+} \cdot \boldsymbol{n}^{+}, \\
&{\kappa}^{-}_e :=\left(\boldsymbol{n}^{-}\right)^{T} \cdot \bm{\kappa}^{-} \cdot \boldsymbol{n}^{-},
\end{split}
\end{equation}

\noindent
${\kappa_e}$ is a harmonic average of $\kappa^{+}_e$ and ${\kappa}^{-}_e$ read as:

\begin{equation}
{\kappa_{e}}:= \frac{2{\kappa}^{+}_e {\kappa}^{-}_e}{\left({\kappa}^{+}_e+{\kappa}^{-}_e\right)}.
\end{equation}

\noindent
The interior penalty parameter, $\beta$, is a function of polynomial degree approximation, $k$. Here, $h_e$ is a characteristic length calculated as:

\begin{equation}
h_{e} :=\frac{\operatorname{meas}\left(T^{+}\right)+\operatorname{meas}\left(T^{-}\right)}{2 \operatorname{meas}(e)},
\end{equation}

\noindent
where meas($.$) represents a measurement operator, measuring length, area, or volume.

\begin{remark}
Note that the surface integral terms of interior facets, $e \in \mathcal{E}_{h}^{0}$, naturally become zero if $p_h$ were sought in $\mathcal{P}_{h}^{\mathrm{CG}_{k}}\left(\mathcal{T}_{h}\right)$. In other words, if the CG method is used instead of EG discretization of the pressure, $\llbracket p_h \rrbracket$ and $\llbracket \psi_p \rrbracket$ are equal to zero. Thus, eventually, the CG, EG, and DG methods have the same (bi-)linear forms for both \eqref{eq:linear_balance_dis} and \eqref{eq:mass_discretization}, but a different function space for the pressure is employed.
\end{remark}

\begin{remark}
\rev{While the Neumann boundary condition is naturally applied on the boundary faces that belong to the Neumann boundary domain, $e \in \mathcal{E}_{h}^{N}$, the Dirichlet boundary condition is strongly enforced in \eqref{eq:CG_U} but weakly enforced for \eqref{eq:CG_P}, \eqref{eq:EG_P} and \eqref{eq:DG_P} on the Dirichlet boundary faces, $e \in \mathcal{E}_{h}^{D}$.}
\end{remark}

The time domain, $\mathbb{T} = \left(0,\tau\right]$, is partitioned into $N$ open subintervals such that, $0=: t^{0}<t^{1}<\cdots<t^{N} := \tau$. The length of the subinterval, $\Delta t^n$, is defined as $\Delta t^n=t^{n}-t^{n-1}$ where $n$ represents the current time step. In this study, implicit first-order time discretization is utilized for a time domain of \eqref{eq:mass_discretization} as shown below for both $p_h^n$ and $\bm{u_h}^n$:

\begin{equation} \label{eq:time_discre}
\frac{\partial p_{h}}{\partial t} \approx\frac{p_{h}^{n}-p_{h}^{n-1}}{\Delta t^n},
\text{ and }
\frac{\partial \boldsymbol{u}_{h}}{\partial t} \approx\frac{\boldsymbol{u}_{h}^{n}-\boldsymbol{u}_{h}^{n-1}}{\Delta t^n}.
\end{equation}

\subsection{Fully discrete form and block formulation}

Combining space and time discretizations from the previous section, we propose the fully discrete form and block formulation using displacement, $\bm{u}$, and pressure, $p$, as the primary variables. From the linear momentum balance equation \eqref{eq:linear_balance_dis}, we can define the following forms:

\begin{equation}
{\mathcal{A}_u\left((\bm{u}_{h}^{n}, p_{h}^{n}), \bm{\psi_u} \right)} := a\left(\bm{u}_{h}^{n}, \bm{\psi_u} \right)+b\left(p_{h}^{n}, \bm{\psi_u} \right),
\end{equation}

\noindent
with

\begin{equation}
a\left(\bm{u}_{h}^{n}, \bm{\psi_u} \right) := \sum_{T \in \mathcal{T}_{h}} \int_{T} \boldsymbol{\sigma}^{\prime}\left(\bm{u}_{h}^n\right) : \nabla^{s} \bm{\psi_u} \: d V,
\end{equation}

\begin{equation}
b\left(p_{h}^{n}, \bm{\psi_u} \right) := \sum_{T \in \mathcal{T}_{h}} \int_{T} \alpha  \nabla p_{h}^n  . \nabla \bm{\psi_u} \: d V,
\end{equation}

\noindent
and

\begin{equation}
\mathcal{L}_u\left(\bm{\psi_u} \right) := \sum_{T \in \mathcal{T}_{h}} \int_{T} \bm{f} \bm{\psi_u} \: d V+\sum_{e \in \mathcal{E}_{h}^{N}} \int_{e} \bm{\sigma_{D}} \bm{\psi_u} \: d S.
\end{equation}

\noindent
and, equivalently, solve

\begin{equation}
\mathcal{A}_u\left((\bm{u}_{h}^{n}, p_{h}^{n}), \bm{\psi_u} \right) = \mathcal{L}_u\left(\bm{\psi_u} \right), \quad\forall \bm{\psi_u} \in \mathcal{U}_{h}^{\mathrm{CG}_{k}}\left(\mathcal{T}_{h}\right)
\label{eq:A_u_L_u}
\end{equation}
at each time step $t^n$.

\noindent
Subsequently, from Equations \ref{eq:mass_discretization} and \ref{eq:time_discre}, we define the following forms:

\begin{equation}
{\mathcal{A}_p\left((\bm{u}_{h}^{n}, p_{h}^{n}), \psi_p; \bm{u}_{h}^{n} \right)} := c\left(\bm{u}_{h}^{n}, \psi_p \right)+d\left(p_{h}^{n}, \psi_p{; \bm{u}_{h}^{n}} \right),
\label{eq:A_p}
\end{equation}

\noindent
with

\begin{equation}
c\left(\bm{u}_{h}^{n}, \psi_p \right) := \sum_{T \in \mathcal{T}_{h}} \int_{T} \rho \alpha \nabla \cdot \boldsymbol{u}_{h}^n\psi_{p} \: d V,
\end{equation}

\begin{equation}
\begin{split}
d\left(p_{h}^{n}, \psi_p{; \bm{u}_{h}^{n}} \right)  :=  &  \sum_{T \in \mathcal{T}_{h}} \int_{T} \rho\left(\phi c_{f}+\frac{\alpha-\phi}{K_{s}}\right)  p_{h}^n \psi_{p} \: d V + \Delta t^n \sum_{T \in \mathcal{T}_{h}} \int_{T} \bm{\kappa}^n\left(\nabla p_{h}^n-\rho \mathbf{g}\right) \cdot \nabla \psi_{p} \: d V \\ & - \Delta t^n \sum_{e \in \mathcal{E}_{h}^{0} \cup \mathcal{E}_{h}^{D}} \int_{e}\left\{\bm{\kappa}^n\left(\nabla p_{h}^n-\rho \mathbf{g}\right)\right\}_{\delta_{e}} \cdot \llbracket \psi_p \rrbracket \: d S \\ & - \Delta t^n \sum_{e \in \mathcal{E}_{h}^{0} \cup \mathcal{E}_{h}^{D}} \int_{e}\left\{\bm{\kappa}^n  \nabla \psi_{p}\right\}_{\delta_{e}} \cdot \llbracket p_h^n \rrbracket \: d S \\ & +  \Delta t^n \sum_{e \in \mathcal{E}_{h}^{0} \cup \mathcal{E}_{h}^{D}} \int_{e} \frac{\beta}{h_{e}} {\kappa}^n_{{e}}  \llbracket p_h^n \rrbracket \cdot \llbracket \psi_p \rrbracket \: d S,
\end{split}
\label{eq:d}
\end{equation}

\noindent
and

\begin{equation}
\begin{split}
\mathcal{L}_p\left(\psi_p {; \bm{u}_{h}^{n}} \right) := & \sum_{T \in \mathcal{T}_{h}} \int_{T} \rho \alpha \nabla \cdot \boldsymbol{u}_{h}^{n-1}\psi_{p} \: d V+ \sum_{T \in \mathcal{T}_{h}} \int_{T} \rho\left(\phi c_{f}+\frac{\alpha-\phi}{K_{s}}\right)  p_{h}^{n-1} \psi_{p} \: d V \\ & + \Delta t^n \sum_{T \in \mathcal{T}_{h}} \int_{T} g \psi_{p} \: d V+ \Delta t^n \sum_{e \in \mathcal{E}_{h}^{N}} \int_{e} q_{D} \psi_{p} \: d S \\ &  -\Delta t^n \sum_{e \in \mathcal{E}_{h}^{D}} \int_{e} \bm{\kappa}^{n} \nabla \psi_{p} \cdot p_{D} \boldsymbol{n} \: d S + \Delta t^n \sum_{e \in \mathcal{E}_{h}^{D}} \int_{e} \frac{\beta}{h_{e}} {\kappa}^{n}_{{e}} \llbracket \psi_p \rrbracket \cdot p_D \bm{n} \: d S.
\end{split}
\label{eq:L_p}
\end{equation}

\noindent
and, equivalently, solve

\begin{equation}
\mathcal{A}_p\left((\bm{u}_{h}^{n}, p_{h}^{n}), \psi_p; \bm{u}_{h}^{n} \right) = \mathcal{L}_p\left(\psi_p; \bm{u}_{h}^{n} \right), \quad\forall \psi_p \in \mathcal{P}_{h}^{\mathrm{EG}_{k}}\left(\mathcal{T}_{h}\right)
\label{eq:A_p_L_p}
\end{equation}

at each time step $t^n$. We remark that \eqref{eq:A_p}, \eqref{eq:d} and \eqref{eq:L_p} are nonlinear forms in the displacement $\bm{u}_{h}^{n}$ (the nonlinear variable being reported after a semicolon for the sake of clarity), since $\bm{\kappa}^{n}:=\bm{\kappa}(\bm{u}_{h}^{n})$ depends on the strain $\bm{\varepsilon} = \bm{\varepsilon}(\bm{u}_{h}^{n})$ (see \eqref{eq:kappa}--\eqref{eq:volumetric_str}). Thus, two linearization strategies are sought in this work: \par

\begin{enumerate}
    \item \emph{Pressure-independent $\bm{k_m}$ model}, which amounts to replacing $\bm{\kappa}^{n}$ with $\bm{\kappa}^{0}$ in \eqref{eq:d} and \eqref{eq:L_p}, that is, employing a semi-implicit time-stepping scheme. It preserves the compaction effect, i.e. changing in $\varepsilon_{v}$, but it neglects the effect of the solid deformation on $\bm{k_m}$ alteration as illustrated in \eqref{eq:wong_perm}. Therefore, we replace \eqref{eq:A_p_L_p} with

    \begin{equation}
    \mathcal{A}_p\left((\bm{u}_{h}^{n}, p_{h}^{n}), \psi_p; \bm{u}_{h}^{0} \right) = \mathcal{L}_p\left(\psi_p; \bm{u}_{h}^{0} \right), \quad\forall \psi_p \in \mathcal{P}_{h}^{\mathrm{EG}_{k}}\left(\mathcal{T}_{h}\right),
    \label{eq:A_p_L_p_minus_1}
    \end{equation}

    and solve for \eqref{eq:A_u_L_u} and \eqref{eq:A_p_L_p_minus_1}.
    The resulting system of equations is linear, and its algebraic formulation is summarized hereafter in Remark \ref{remark:algebraic}. The algorithm used to solve this model is presented in Algorithm \ref{al:perm_noniteration}. A direct solver is used for every time step \cite{petsc-user-ref}. For the sake of brevity, in Algorithm \ref{al:perm_noniteration} (and in the following one) we omit the subscript $h$ as well as superscripts $\text{CG}_k, \text{DG}_k$ or $\text{EG}_k$, $\mathcal{W}$ represents a mixed function space between $\mathcal{U}$ $\times$ $\mathcal{P}$, and $w$ is a solution vector composed of $\bm{u}$ and $p$.
    \item \emph{Pressure-dependent $\bm{k_m}$ model}, which includes the nonlinear effect of the current solid deformation on $\bm{k_m}$ alteration by performing Picard iterations. $\bm{k_m}$ is affected by the change in $\varepsilon_{v}$ as reflected in \eqref{eq:wong_perm}. Note that $\bm{\kappa}$ is calculated using $\bm{k_m}$, $\rho$, and $\mu$. Given that $\rho$ and $\mu$ are constants, $\bm{\kappa}$ is only a function of $\bm{k_m}$.
    Thus, we replace \eqref{eq:A_u_L_u} and \eqref{eq:A_p_L_p} by

    \begin{equation}
    \begin{cases}
    \mathcal{A}_u\left((\bm{u}_{h}^{n,\iota}, p_{h}^{n,\iota}), \bm{\psi_u} \right) = \mathcal{L}_u\left(\bm{\psi_u} \right), \quad\forall \bm{\psi_u} \in \mathcal{U}_{h}^{\mathrm{CG}_{k}}\left(\mathcal{T}_{h}\right)\\
    \mathcal{A}_p\left((\bm{u}_{h}^{n,\iota}, p_{h}^{n,\iota}), \psi_p; \bm{u}_{h}^{n,\iota-1} \right) = \mathcal{L}_p\left(\psi_p; \bm{u}_{h}^{n,\iota-1} \right), \quad\forall \psi_p \in \mathcal{P}_{h}^{\mathrm{EG}_{k}}\left(\mathcal{T}_{h}\right),
    \end{cases}
    \end{equation}

    where $\iota$ denotes the iteration counter and $(\bm{u}_{h}^{n,\iota}, p_{h}^{n,\iota})$ represents the solution of time step $n$ at Picard iteration $\iota$.
    Nonlinear iterations are initialized with $(\bm{u}_{h}^{n,0}, p_{h}^{n,0}) = (\bm{u}_{h}^{n-1}, p_{h}^{n-1})$.
    The algorithm utilized to solve this model is presented in Algorithm \ref{al:perm_iteration}. The iteration tolerance, $\xi$, is set to be $1 \times 10^{-6}$, and a direct solver is used inside each iteration for solving the resulting system of linear equations \cite{petsc-user-ref}.
    The $L^2$ norm of the difference, $\zeta$, between the current iteration, $w^{\iota}:= (\bm{u}_{h}^{n,\iota}, p_{h}^{n,\iota})$, and the previous iteration, $w^{\iota-1}:= (\bm{u}_{h}^{n,\iota-1}, p_{h}^{n,\iota-1})$ is calculated as follows:

    \begin{equation}\label{eq:l2_norm_diff}
    \zeta=||w^{\iota} - w^{\iota-1}||,
    \end{equation}

    where $||\cdot||$ denotes the $L^2$ norm on the product space $\mathcal{W}$.
\end{enumerate}

\begin{algorithm}[H]
\caption{Solving algorithm for pressure-independent $\bm{k_m}$ model}\label{al:perm_noniteration}
$\mathcal{U}$ = VectorFunctionSpace(`CG', k = 2) \newline
$\mathcal{P}$ = FunctionSpace(`CG/EG/DG', k = 1) \newline
$\mathcal{W}$ = MixedFunctionSpace({$\mathcal{U}$ $\times$ $\mathcal{P}$}) \newline
$\bm{u}$ = Function($\mathcal{U}$) \Comment{the vector solution of $\mathcal{U}$} \newline
$p$ = Function($\mathcal{P}$) \Comment{ the vector solution of $\mathcal{P}$} \newline
$w$ = Function($\mathcal{W}$) \Comment{the vector solution composed of $\bm{u}$ and $p$}
\begin{algorithmic}[1]
\State Initialize all input parameters \Comment{initial reservoir pressure ($p_0$) must be provided.}
\State Solve the equilibrium state for $\bm{u_0}$
\While {$t$ $<$ $\tau$}

\State $t$ $\mathrel{{+}{=}}$ $\Delta t^n$
\State Solve time step  $n$

\State Post-processing, i.e. $\bm{\sigma}$, residual, etc.
\State Assign $w^{n} \rightarrow w^{n-1}$ \Comment{update time step $n-1$}
\State Output
\EndWhile

\end{algorithmic}
\end{algorithm}

\begin{algorithm}[H]
\caption{Solving algorithm for pressure-dependent $\bm{k_m}$ model}\label{al:perm_iteration}
$\mathcal{U}$ = VectorFunctionSpace(`CG', k = 2) \newline
$\mathcal{P}$ = FunctionSpace(`CG/EG/DG', k = 1) \newline
$\mathcal{W}$ = MixedFunctionSpace({$\mathcal{U}$ $\times$ $\mathcal{P}$}) \newline
$\bm{u}$ = Function($\mathcal{U}$) \Comment{the vector solution of $\mathcal{U}$} \newline
$p$ = Function($\mathcal{P}$) \Comment{the vector solution of $\mathcal{P}$} \newline
$w$ = Function($\mathcal{W}$) \Comment{the vector solution composed of $\bm{u}$ and $p$}
\begin{algorithmic}[1]
\State Initialize all input parameters \Comment{initial reservoir pressure ($p_0$) must be provided.}
\State Solve the equilibrium state for $\bm{u_0}$
\State Update $\varepsilon_{v0}$ and $\bm{\kappa}_0$
\Comment{\eqref{eq:volumetric_str} and \eqref{eq:wong_perm}}
\While {$t$ $<$ $\tau$}

\State $t$ $\mathrel{{+}{=}}$ $\Delta t^n$
\State Solve time step  $n$
\State Update $\varepsilon_{v}^0$ and $\bm{\kappa}^0${, assign $\iota \rightarrow 0$}  \Comment{iteration zero}

\While {$\zeta$ $>$ $\xi$}

\State $\iota$ $\mathrel{{+}{=}}$ $1$
\State Solve time step  $n$ and iteration $\iota$
\State Update $\varepsilon_{v}^\iota$ and $\bm{\kappa}^\iota$  \Comment{ iteration $\iota$}
\If {$\bm{\kappa}^\iota$ $<$ $\bm{\kappa}_{r}$}
$\bm{\kappa}^\iota$ = $\bm{\kappa}_{r}$ \Comment{$\bm{\kappa}_{r}$ is a minimum value of $\bm{\kappa}$}
\EndIf
\State Update $\zeta$ \Comment{\eqref{eq:l2_norm_diff}}
\State Assign $w^{\iota} \rightarrow w^{\iota-1}$

\EndWhile

\State Assign $\varepsilon_{v}^\iota$ $\rightarrow$ $\varepsilon_{v}^n$ \Comment{time step $n$}
\State Assign $\bm{\kappa}^\iota$ $\rightarrow$ $\bm{\kappa}^n$ \Comment{ time step $n$}
\State Post-processing, i.e. $\bm{\sigma}$, residual, etc.
\State Assign $w^{n} \rightarrow w^{n-1}$ \Comment{update time step $n-1$}
\State Output
\EndWhile

\end{algorithmic}
\end{algorithm}

\begin{remark}[Algebraic formulation]
\label{remark:algebraic}
We discuss the algebraic formulation of the pressure-independent $\bm{k_m}$ model for CG, EG, and DG pressure function spaces. A similar formulation can be derived for the pressure-dependent case.

\noindent
Let $\{\bm{\psi_u}^{(i)}\}_{i=1}^{\mathcal{N}_{\bm{u}}}$
denote the set of basis functions of $\mathcal{U}_{h}^{\mathrm{CG}_{k}}\left(\mathcal{T}_{h}\right)$,
i.e. $\mathcal{U}_{h}^{\mathrm{CG}_{k}}\left(\mathcal{T}_{h}\right) = \text{span}\{\bm{\psi_u}^{(i)}\}_{i=1}^{\mathcal{N}_{\bm{u}}}$,
having denoted by $\mathcal{N}_{\bm{u}}$ the number of DOF for the CG displacement space. In a similar way, let
$\{\psi_{p,\pi}^{(m)}\}_{m=1}^{\mathcal{N}_{p, \pi}}$ be the set of basis functions for the space $\mathcal{P}_{h}^{\pi_{k}}\left(\mathcal{T}_{h}\right)$, where $\pi$ can mean either CG, EG, or DG.
Hence, three mixed function spaces, $\bm{(1)}$ $\mathrm{CG}_k \times \mathrm{CG}_k$, $\bm{(2)}$ $\mathrm{CG}_k \times \mathrm{EG}_k$, and $\bm{(3)}$ $\mathrm{CG}_k \times \mathrm{DG}_k$ where $k$ represents the degree of polynomial approximation, are possible, and will be compared in the numerical examples in the next section. The Jacobian matrix corresponding to the left-hand sides of \eqref{eq:A_u_L_u} and \eqref{eq:A_p_L_p_minus_1} is assembled composing the following blocks:

{\begin{equation}
\begin{aligned}
&\left[\mathcal{J}_{u u}^{\mathrm{CG}_k \times \mathrm{CG}_k}\right]_{ij} := a\left(\bm{\psi_u}^{(j)}, \bm{\psi_u}^{(i)} \right), \; i = 1, \hdots, \mathcal{N}_{\bm{u}}, j = 1, \hdots, \mathcal{N}_{\bm{u}},\\
&\left[\mathcal{J}_{u p}^{\mathrm{CG}_k \times \pi_k}\right]_{im} := b\left(\psi_{p,\pi}^{(m)}, \bm{\psi_u}^{(i)} \right), \; i = 1, \hdots, \mathcal{N}_{\bm{u}}, m = 1, \hdots, \mathcal{N}_{p,\pi},\\
&\left[\mathcal{J}_{p u}^{\pi_k \times \mathrm{CG}_k}\right]_{lj} := c\left(\bm{\psi_u}^{(j)}, \psi_{p,\pi}^{(l)} \right), \; l = 1, \hdots, \mathcal{N}_{p,\pi}, j = 1, \hdots, \mathcal{N}_{\bm{u}},\\
&\left[\mathcal{J}_{p p}^{\pi_k \times \pi_k}\right]_{lm} := d\left(\psi_{p,\pi}^{(m)}, \psi_{p,\pi}^{(l)}; \bm{u}_{h}^n \right)\; l = 1, \hdots, \mathcal{N}_{p,\pi}, m = 1, \hdots, \mathcal{N}_{p,\pi}.
\end{aligned}
\end{equation}}

\noindent
In a similar way, the right-hand side of \eqref{eq:A_u_L_u} and \eqref{eq:A_p_L_p_minus_1} gives rise to a block vector of components

\begin{equation}
\begin{aligned}
&\left[\mathcal{L}_u^{\mathrm{CG}_k}\right]_{i} := \mathcal{L}_u\left(\bm{\psi_u}^{(i)}\right), \; i = 1, \hdots, \mathcal{N}_{\bm{u}},\\
&\left[\mathcal{L}_p^{\pi_k}\right]_{l} := \mathcal{L}_p\left(\psi_{p,\pi}^{(l)}; \bm{u}_{h}^n \right)\; l = 1, \hdots, \mathcal{N}_{p,\pi}.
\end{aligned}
\end{equation}

\noindent
The resulting block structure is thus

\begin{equation} \label{eq:generic_block}
\left[ \begin{array}{ll}{\mathcal{J}_{u u}^{\mathrm{CG}_k \times \mathrm{CG}_k}} & {\mathcal{J}_{up}^{\mathrm{CG}_k \times \pi_k}} \\ {\mathcal{J}_{pu}^{\pi_k \times \mathrm{CG}_k}} & {\mathcal{J}_{p p}^{\pi_k \times \pi_k}}\end{array}\right]
\left\{\begin{array}{l}{\left(\bm{u}_{h}^{n}\right)^{\mathrm{CG}_k}} \\ {\left({p_{h}}^{n}\right)^{\pi_k}}\end{array}\right\}
= \left\{\begin{array}{l}{\mathcal{L}_u^{\mathrm{CG}_k}} \\ {\mathcal{L}_p^{\pi_k}}\end{array}\right\}.
\end{equation}
\noindent
where $\left(\bm{u}_{h}^{n}\right)^{\mathrm{CG}_k}$ and $\left({p_{h}}^{n}\right)^{\pi_k}$ collect the degrees of freedom for displacement and pressure.

\noindent
Finally, we remark that (owing to \eqref{eq:EG_P}) the case $\pi = \mathrm{EG}$ can be equivalently decomposed into a $\mathrm{CG}_k$ $\times$ $(\mathrm{CG}_k$ $\times$ $\mathrm{DG}_0)$ mixed function space, resulting in:

\begin{equation} \label{eq:eg_block}
\left[ \begin{array}{lll}
{\mathcal{J}_{u u}^{\mathrm{CG}_k \times \mathrm{CG}_k}}
& {\mathcal{J}_{u p}^{\mathrm{CG}_k \times \mathrm{CG}_k}}
& {\mathcal{J}_{u p}^{\mathrm{CG}_k \times \mathrm{DG}_0}}
\\ {\mathcal{J}_{p u}^{\mathrm{CG}_k \times \mathrm{CG}_k}}
& {\mathcal{J}_{p p}^{\mathrm{CG}_k \times \mathrm{CG}_k}}
& {\mathcal{J}_{p p}^{\mathrm{CG}_k \times \mathrm{DG}_0}}
\\ {\mathcal{J}_{p u}^{\mathrm{DG}_0 \times \mathrm{CG}_k}}
& {\mathcal{J}_{p p}^{\mathrm{DG}_0 \times \mathrm{CG}_k}}
& {\mathcal{J}_{p p}^{\mathrm{DG}_0 \times \mathrm{DG}_0}}
\end{array} \right]
\left\{\begin{array}{l}{\left(\bm{u}_{h}^{n}\right)^{\mathrm{CG}_k}} \\ {\left({p_{h}}^{n}\right)^{\mathrm{CG}_k}}
\\ {\left({p_{h}}^{n}\right)^{\mathrm{DG}_0}}
\end{array}\right\}
= \left\{\begin{array}{l}
{\mathcal{L}_u^{\mathrm{CG}_k}}
\\ {\mathcal{L}_p^{\mathrm{CG}_k}}
\\ {\mathcal{L}_p^{\mathrm{DG}_0}}
\end{array}\right\}.
\end{equation}

\noindent
\rev{This implementation makes the EG methodology easily implementable in any existing DG codes}.
Matrices and vectors are built using FEniCS form compiler \cite{AlnaesBlechta2015a}. The block structure is setup using multiphenics toolbox \cite{Ballarin2019}\rev{, in which users are only required to provide weak forms associated to each block, as it will be illustrated in the results and discussion section.} Random fields, porosity ($\phi$) and permeability ($\bm{k_{m0}}$), are populated using SciPy package \cite{Jones2001}. $\beta$, penalty parameter, is set at 0.95 and 0.9 for DG and EG methods, respectively.
\end{remark}

\begin{remark}
As discussed above, Picard iteration is employed to solve the pressure-dependent $\bm{k_m}$ model (Algorithm \ref{al:perm_iteration}). \rev{The Picard iteration is employed in this study because the permeability calculation resulted from \eqref{eq:wong_perm} can be negative. Therefore, we want to enforce the minimum value of $\bm{k_m}$ through Picard iteration, as illustrated in Algorithm \ref{al:perm_iteration}. Throughout this paper, we set $\bm{\kappa}_{r}$, a minimum value of $\bm{\kappa}$, as $10^{-16}$. Note that $\bm{\kappa}$ is a function of $\bm{k_m}$, $\rho$, and $\mu$ as presented in \eqref{eq:kappa}.} As a future investigation to reduce the number of iterations, the Newton method could be applied to solve this system of nonlinear equations while enforcing the $\bm{\kappa}_{r}$ value.
\end{remark}

\section{Results and discussion}\label{sec:results}

\subsection{Error convergence analysis}

To verify and illustrate the implementation of the developed block structure utilized to compose the EG function space using multiphenics framework \cite{Ballarin2019}, we start by analyzing and comparing the error convergence of the EG method with the DG method for the Poisson equation with a source term as follows:

\begin{equation} \label{eq:poisson}
-\nabla^2 p=g,
\end{equation}

\noindent
where $p$ is primary variable and $g$ is sink/source term. Note that \eqref{eq:poisson} is a specific case of \eqref{eq:mass_balance} at the steady-state solution, and $\alpha = \phi = 0$ (uncoupled from \eqref{eq:linear_balance}). We take $\Omega = \left[ 0,1\right]^d$, and $d= 2,3$ and choose the exact solution in $\Omega$ as:

\begin{equation}\label{eq:exact_solution}
\begin{cases}
{p(x,y) =} -\cos(x+y), & \text{if} \ d =  2 \\
{p(x,y, z) =} -\cos(x+y+z), & \text{if} \ d =  3
\end{cases}
\end{equation}

\noindent
where $x$, $y$, and $z$ represent points in x-, y-, z-direction, respectively. Subsequently, $g$ is chosen as:

\begin{equation}\label{eq:exact_source}
\begin{cases}
{g(x,y) = }-2\cos(x+y), & \text{if} \ d =  2 \\
{g(x,y,z) = }-3\cos(x+y+z), & \text{if} \ d =  3
\end{cases}
\end{equation}

\noindent
to satisfy the exact solution. The block structure of the EG method is formed as:

{\begin{equation} \label{eq:poisson_block}
\left[ \begin{array}{ll}{\mathcal{J}_{pp}^{\mathrm{CG}_k \times \mathrm{CG}_k}} & {\mathcal{J}_{pp}^{\mathrm{CG}_k \times \mathrm{DG}_0}} \\ {\mathcal{J}_{pp}^{\mathrm{DG}_0 \times \mathrm{CG}_k}} & {\mathcal{J}_{pp}^{\mathrm{DG}_0 \times \mathrm{DG}_0}}\end{array}\right]
\left\{\begin{array}{l}{\left({p_{h}}\right)^{\mathrm{CG}_k}} \\ {\left({p_{h}}\right)^{\mathrm{DG}_0}}\end{array}\right\}
= \left\{\begin{array}{l}{\mathcal{L}_p^{\mathrm{CG}_k}} \\ {\mathcal{L}_p^{\mathrm{DG}_0}}\end{array}\right\}.
\end{equation}}

\begin{figure}[H]
   \centering
        \includegraphics[width=8.0cm, height=7.0cm]{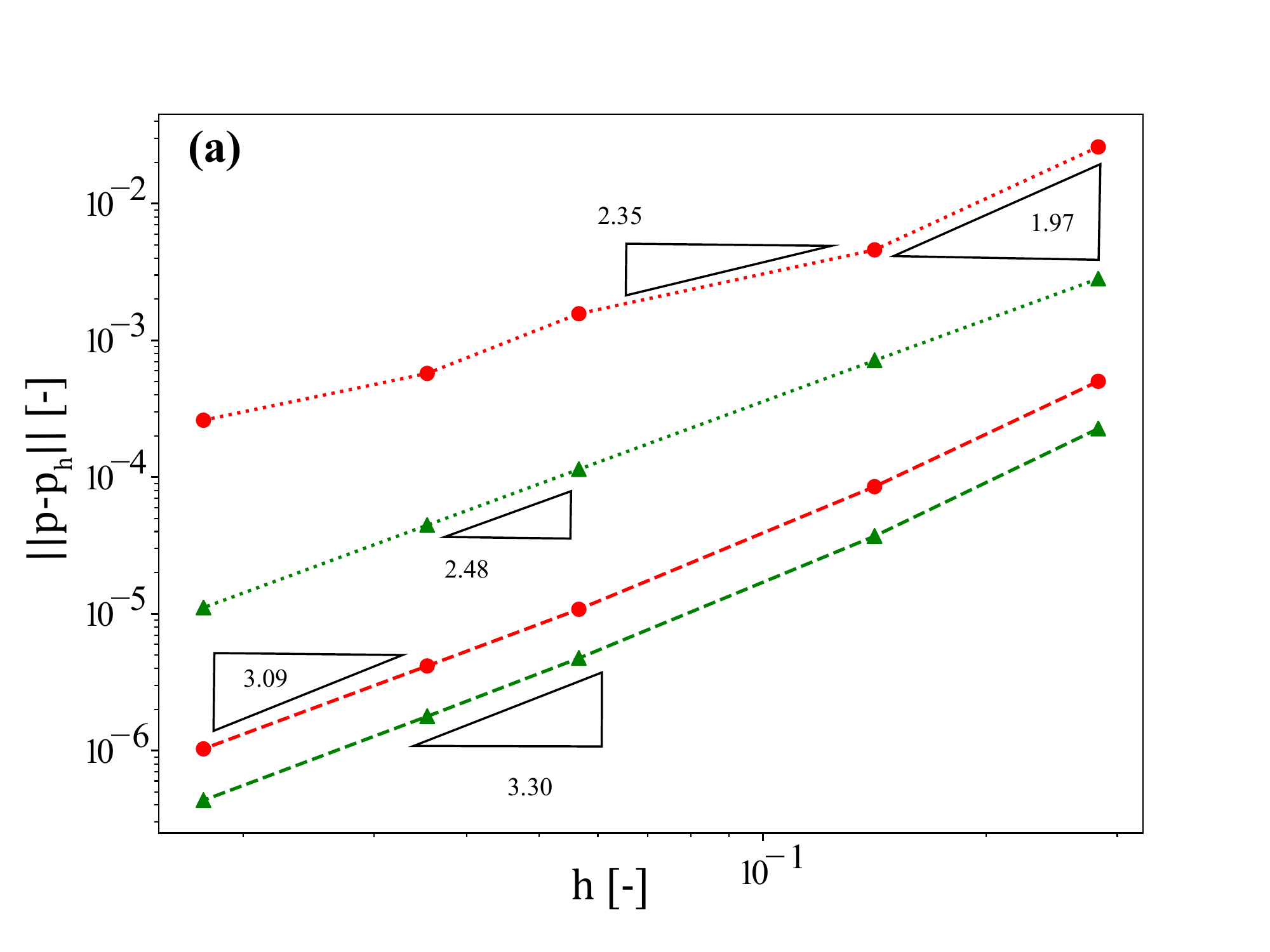}
        \includegraphics[width=8.0cm, height=7.0cm]{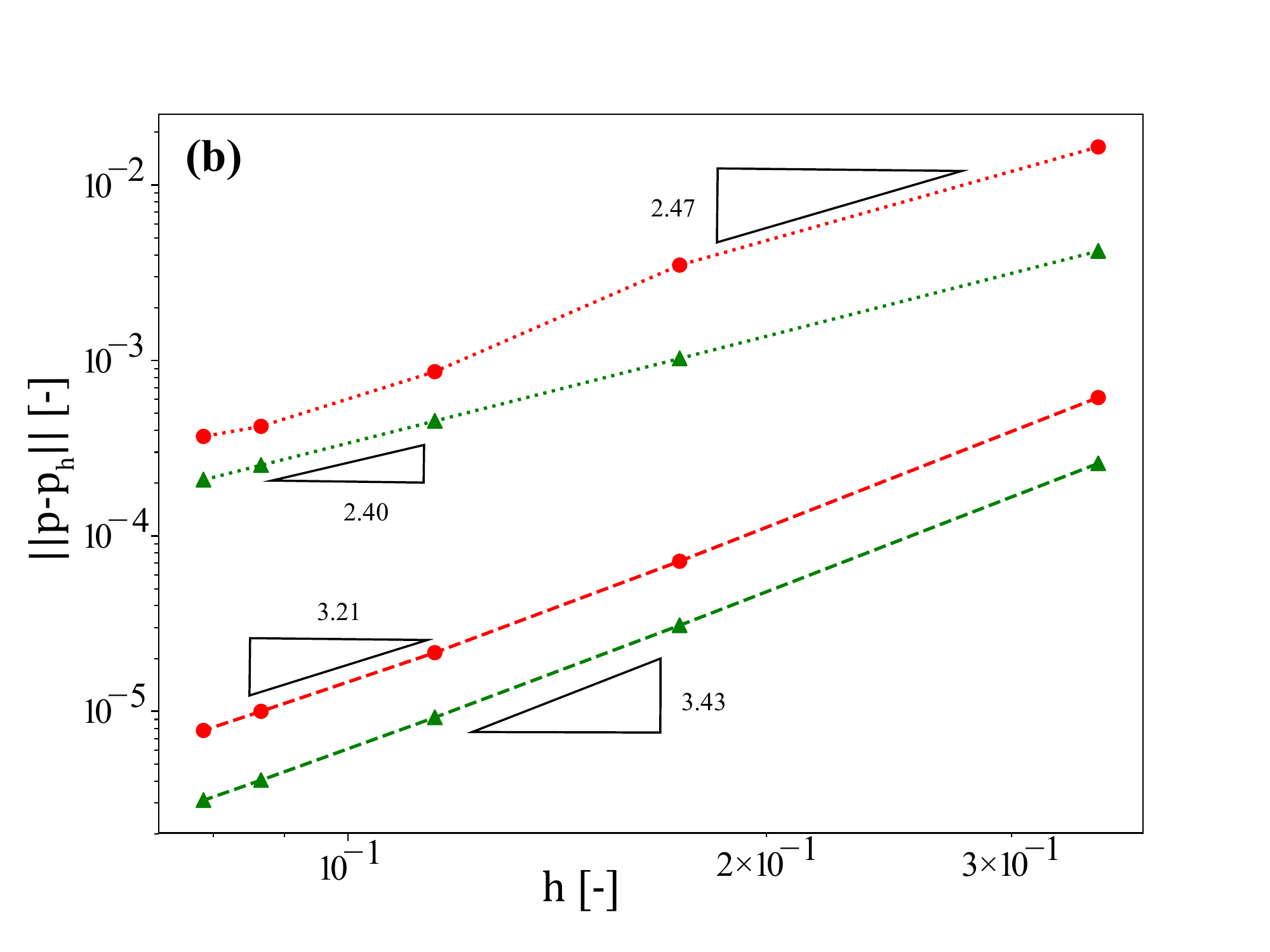}
        \includegraphics[width=8.0cm, height=0.9cm]{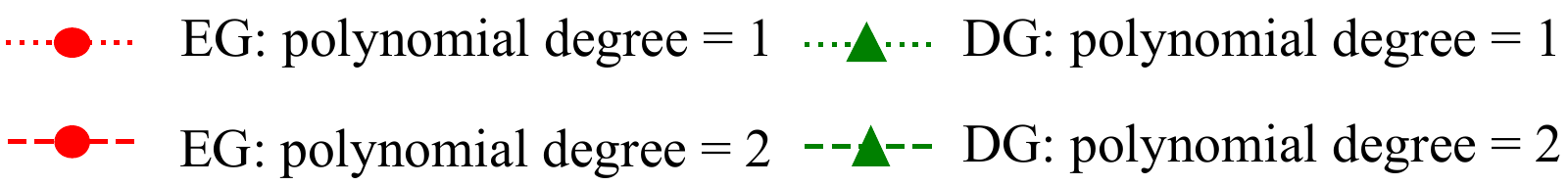}
   \caption{Error convergence plot between EG and DG methods with polynomial approximation degree 1 and 2 of (\textbf{a}) 2-Dimension and (\textbf{b}) 3-Dimension cases for the Poisson equation.}
   \label{fig:err_analysis}
\end{figure}

We calculate $L^2$ norm of the difference between the exact solution, $p$, and approximated solution, $p_h$ (\ref{fig:err_analysis}a-b for $d=2,3$). These plot show $L^2$ norm of the difference, $|| p-p_h|| $, versus the mesh size, $h$. For both $d=2,3$ cases, the EG and DG methods provide the expected convergence rate of two and three for polynomial degree approximation, $k$, of one and two, respectively \cite{babuvska1973finite}.

\rev{
We also want to emphasize the simplicity of the EG implementation using multiphenics framework \cite{Ballarin2019}. This open-source platform is built on top of FEniCS \cite{AlnaesBlechta2015a}, and it takes care of assembling the mass matrix and the right-hand side of a block structure. As illustrated in the code snippet below, users are only required to provide weak formulations associated to each block. This simplicity will enable a broader range of users to employ the EG implementation in their research. }

\definecolor{codegreen}{rgb}{0,0.6,0}
\definecolor{codegray}{rgb}{0.5,0.5,0.5}
\definecolor{codepurple}{rgb}{0.58,0,0.82}
\definecolor{backcolour}{rgb}{0.95,0.95,0.92}

\lstdefinestyle{mystyle}{
  backgroundcolor=\color{backcolour},   commentstyle=\color{codegreen},
  keywordstyle=\color{magenta},
  numberstyle=\tiny\color{codegray},
  stringstyle=\color{codepurple},
  basicstyle=\ttfamily\footnotesize,
  breakatwhitespace=false,
  breaklines=true,
  captionpos=b,
  keepspaces=true,
  numbersep=5pt,
  showspaces=false,
  showstringspaces=false,
  showtabs=false,
  tabsize=2
}

\lstset{style=mystyle}

\begin{lstlisting}[language=Python, caption= Illustration of the EG implementation using multiphenics]
from dolfin import *
import multiphenics as mp

#start initialization
#....
#end initialization

CGk = FunctionSpace(mesh, 'CG', k)
DG0 = FunctionSpace(mesh, 'DG', 0)

W = mp.BlockFunctionSpace([CGk, DG0])

CGkDG0_trial = mp.BlockTrialFunction(W)
(p_CGk, p_DG0) = mp.block_split(CGkDG0_trial)
CGkDG0_test = mp.BlockTestFunction(W)
(psi_CGk, psi_DG0) = mp.block_split(CGkDG0_test)

J11 = inner(grad(p_CGk),grad(psi_CGk))*dx \
     - inner(grad(p_CGk),n)*psi_CGk*ds \
     + beta/h*p_CGk*psi_CGk*ds

J12 = beta/h*p_DG0*psi_CGk*ds

J21 = - inner(avg(grad(p_CGk)),jump(psi_DG0,n))*dS \
    - inner(grad(p_CGk),n)*psi_DG0*ds \
    + beta/h*p_CGk*psi_DG0*ds

J22 = beta/h_e*inner(jump(p_DG0,n),jump(psi_DG0,n))*dS \
    + beta/h*p_DG0*psi_DG0*ds

lhs = [[J11, J12],
       [J21, J22]]

f_CGk = Expression('3.0*cos(x[0]+x[1]+x[2])', element = CGk.ufl_element())
f_DG0 = Expression('3.0*cos(x[0]+x[1]+x[2])', element = DG0.ufl_element())
p_CGk = Expression('cos(x[0]+x[1]+x[2])', element = CGk.ufl_element())
p_DG0 = Expression('cos(x[0]+x[1]+x[2])', element = DG0.ufl_element())

l1 = f_CGk*psi_CGk*dx + beta/h*p_CGk*psi_CGk*ds
l2 = f_DG0*psi_DG0*dx + beta/h*p_DG0*psi_DG0*ds

rhs = [l1, l2]

A = mp.block_assemble(rhs)
F = mp.block_assemble(lhs)

sol = BlockFunction(W)
block_solve(A, sol.block_vector(), F)

#start post-processing
#....
#end post-processing
\end{lstlisting} \label{list:eg_3d}


\subsection{Terzaghi\textquotesingle s 1D consolidation problem}

\begin{figure}[H]
   \centering
    \includegraphics[width=7.0cm, height=7.0cm]{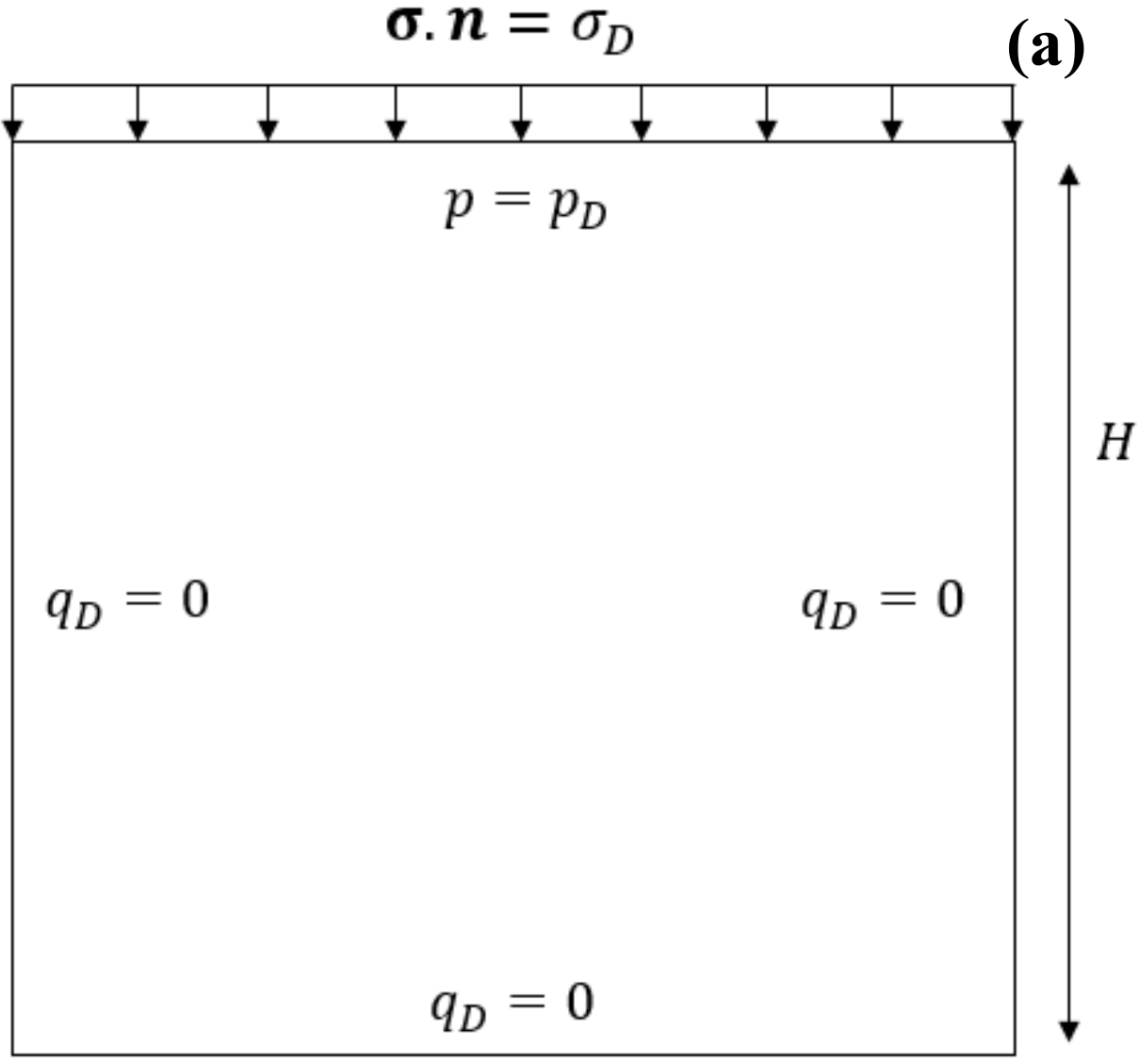}
    \includegraphics[width=9.0cm, height=7.0cm]{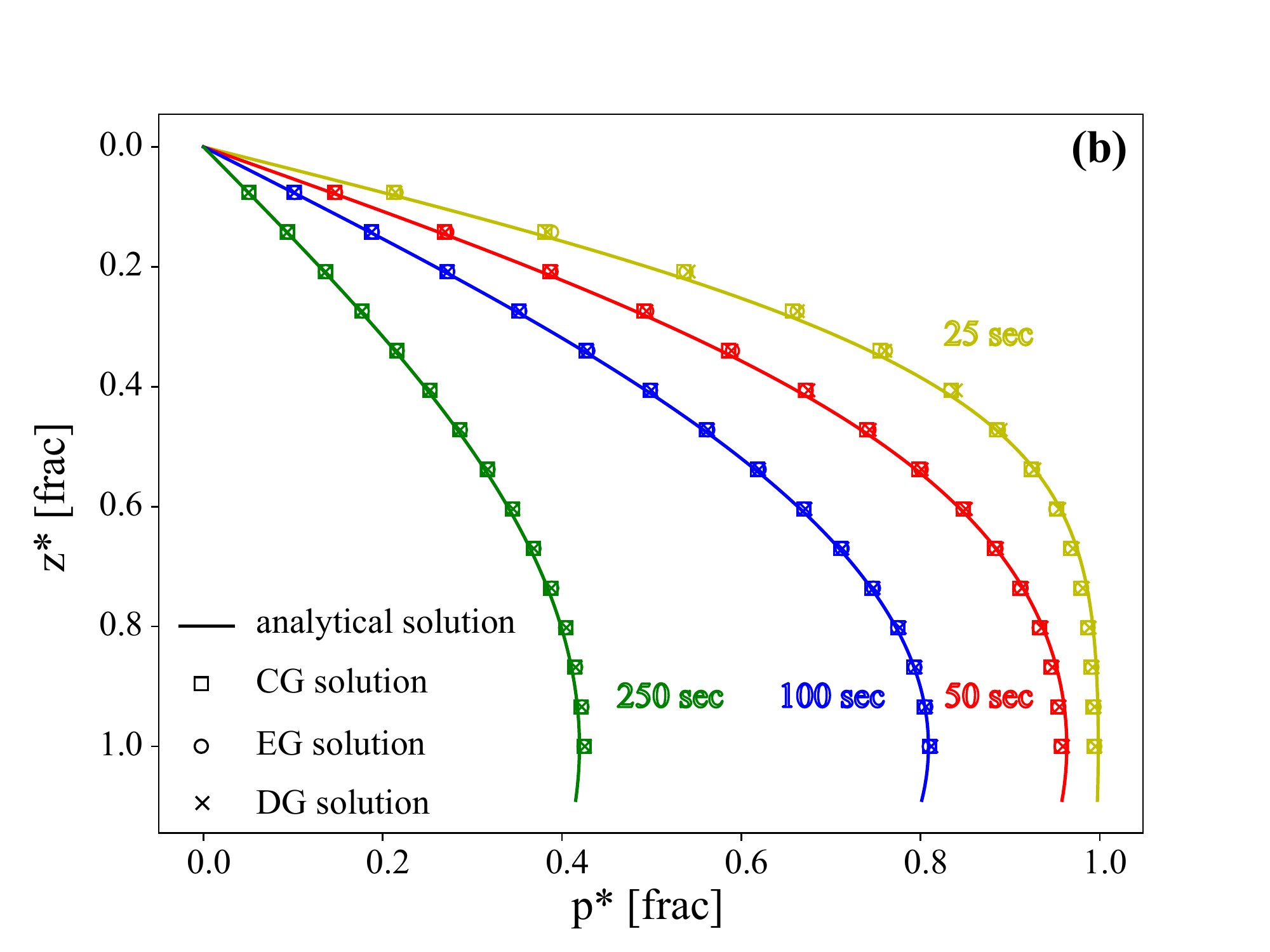}
   \caption{(\textbf{a}) geometry and boundary conditions and (\textbf{b}) the verification results of the CG, EG, and DG methods for Terzaghi\textquotesingle s 1D consolidation problem.}
   \label{fig:validation_case}
\end{figure}

The numerical discretizations, CG, EG, and DG methods, discussed in the methodology section are verified using Terzaghi\textquotesingle s 1D consolidation problem \cite{terzaghi1951theoretical}. Assuming the reservoir is homogeneous, isotropic, and saturated with a single-phase fluid, the non-dimensional parameters are defined as follows \cite{choo2018enriched}:

\begin{equation}\label{terzaghi_1d_1}
p^{*} :=\frac{p}{\sigma}, \: z^{*} :=\frac{z}{H}, \: t^{*} :=\frac{c_{v}}{H^{2}} t,
\end{equation}

\noindent
where $p$ is the fluid pressure, $\sigma$ is far-field stress or external load, $z$ is the distance from the drainage boundary, $H$ is the domain thickness, and $c_{v}$ is the coefficient of consolidation, which is calculated by:

\begin{equation}\label{terzaghi_1d_2}
c_{v}=3 K\left(\frac{1-v}{1+v}\right) \frac{k_{m,y}}{\mu}.
\end{equation}

\noindent
where $k_{m,y}$ is the matrix permeability in y-direction. Subsequently, the pressure solution of Terzaghi\textquotesingle s 1D consolidation is written as follows:

\begin{equation}\label{eq:terzaghi}
\begin{aligned} p^{*}\left(z^{*}, t^{*}\right)=& \sum_{m=0}^{\infty} \frac{2}{M} \sin \left(M z^{*}\right) e^{-M^{2} t^{*}}, \end{aligned}
\end{equation}

\noindent
where $M=\pi(2 m+1) / 2$. Geometry and boundary conditions utilized to verify the numerical models are presented in Figure \ref{fig:validation_case}a, and input parameters are $\bm{\sigma_{D}}=[0, 1]$ $\mathrm{kPa}$, $p_{D}=0$ $\mathrm{Pa}$, $\boldsymbol{k_{m}}=10^{-12} \bm{I}$ $\mathrm{m}^{2}$, $\mu=10^{-6}$ $\mathrm{kPa.s}$, $\rho=1000$ $\mathrm{kg} / \mathrm{m}^{3}$, $K=1000$ $\mathrm{kPa}$, mesh size($h$) $=0.05$ $m$, $K_{\mathrm{s}} \approx \infty$ $\mathrm{kPa}$, which leads to $\alpha \approx 1$, and $v=0.25$, $\Delta t^n$ $=$ 1.0 sec, $\lambda_l$ and $\mu_l$ are calculated by the following equations:

\begin{equation}\label{eq:lambda_l}
\lambda_{l}=\frac{3 K v}{1+v}, \text{ and } \mu_{l}=\frac{3 K(1-2 v)}{2(1+v)}.
\end{equation}

\noindent
Results of CG, EG, and DG methods are presented in Figure \ref{fig:validation_case}b illustrating reasonable matches between analytical and numerical solutions for all time steps, 25, 50, 100, and 250 sec. The solutions of CG, EG, and DG methods are approximately similar. More details regarding the convergence rate of the EG solution for Terzaghi\textquotesingle s equation can be found in Choo and Lee \cite{choo2018enriched}.  \par

\subsection{1D Consolidation of a Two-layered Material}
\begin{figure}[H]
   \centering
        \includegraphics[width=7.0cm, height=7.0cm]{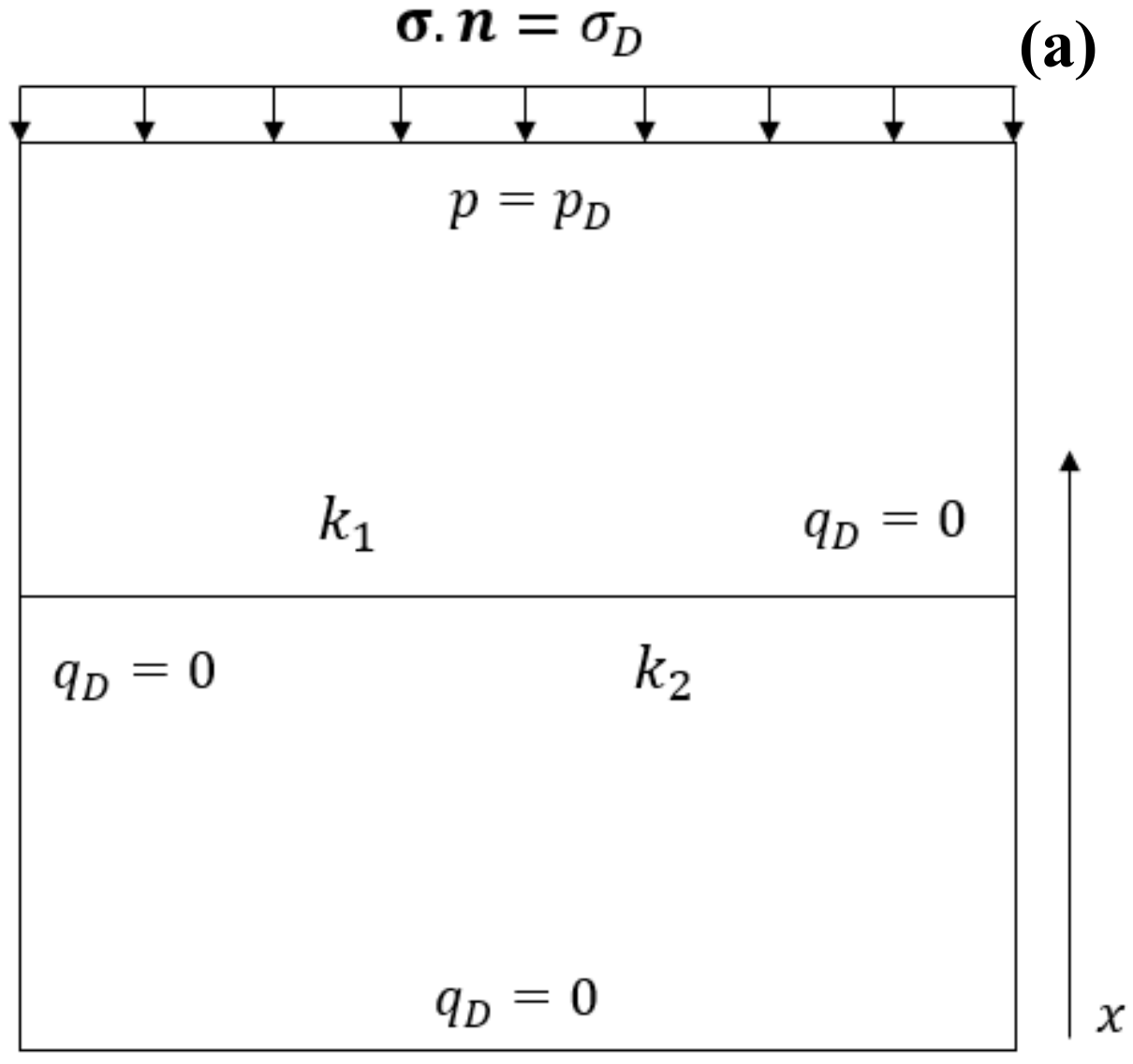}
        \includegraphics[width=9.0cm, height=7.0cm]{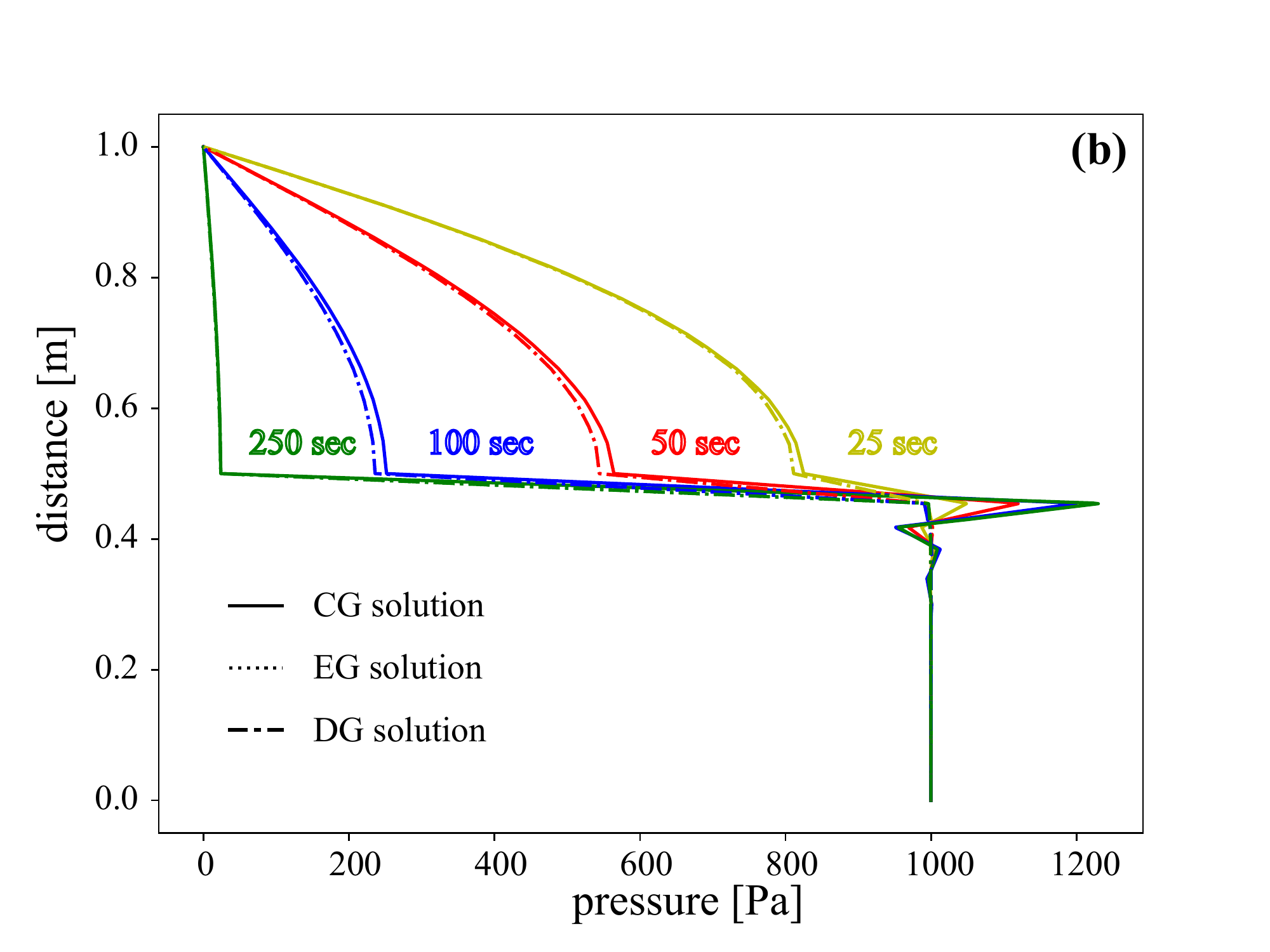}
        \includegraphics[width=9.0cm, height=7.0cm]{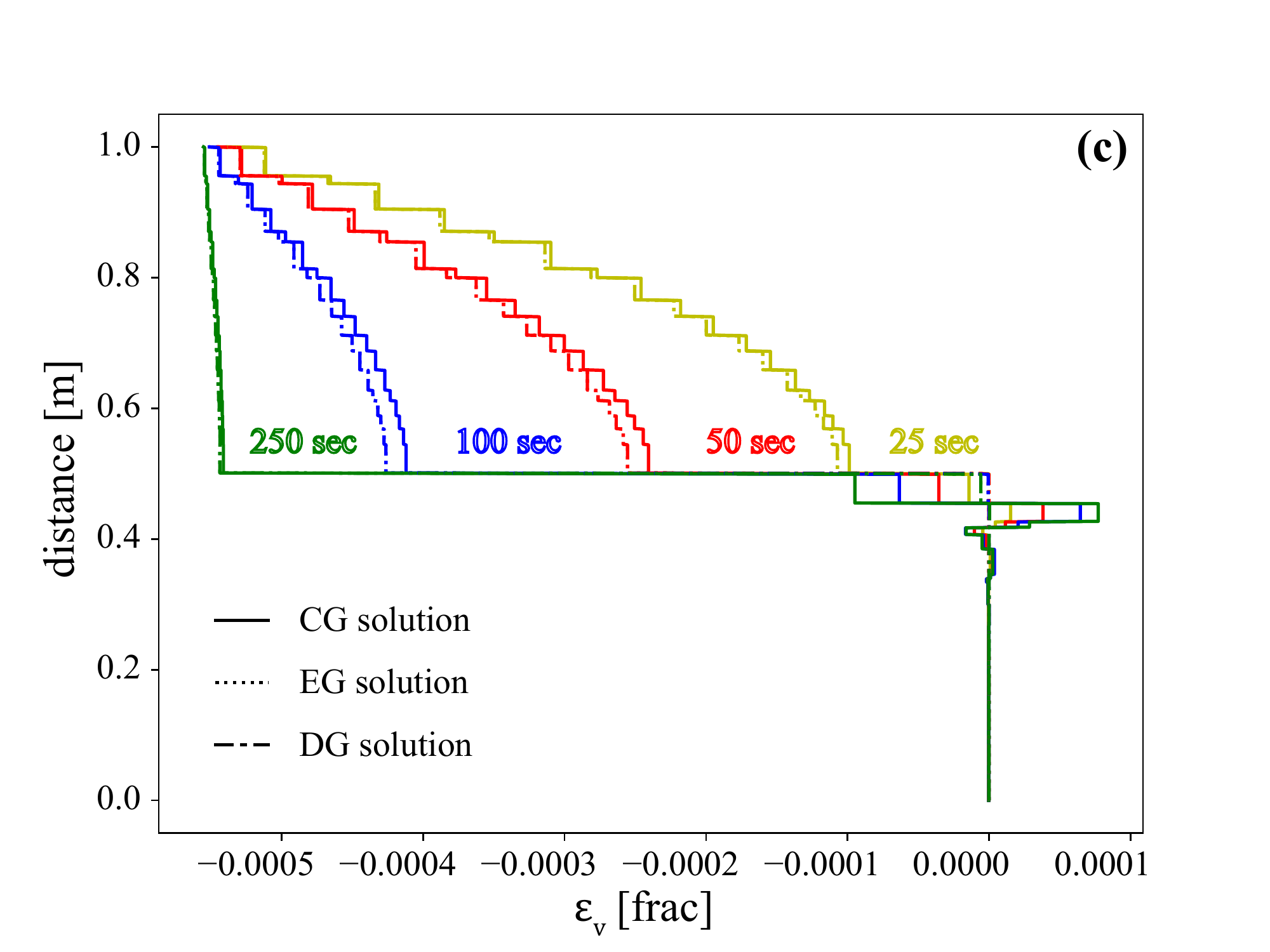}
   \caption{Two-layered material example:(\textbf{a}) geometry and boundary conditions, (\textbf{b}) pressure results, and (\textbf{c}) $\varepsilon_v$  results of the CG, EG, and DG methods. Note that the solutions of EG and DG methods are overlapping.}
   \label{fig:mat_two_case}
\end{figure}

This example is adapted from the previous studies, and it aims to evaluate the pressure, $p$, and volumetric strain, $\varepsilon_v$, results in two-layered material, which contain a large different permeability, among CG, EG, and DG methods \cite{choo2018enriched}. Figure \ref{fig:mat_two_case}a presents a geometry and boundary conditions utilized in this investigation. Although most of the input parameters, mesh size, and solving algorithm are similar to the validation case, in this example, $\bm{k_{m}}=$ $k_{1}\bm{I},$ if $1>x>0.5$ and $\bm{k_{m}}=k_{2}\bm{I},$ if $0.5>x>0,$ and $k_{1}=10^{-12}$ $\mathrm{m}^{2}$ and $k_{2}=10^{-16}$ $\mathrm{m}^{2}$. \par


Figure \ref{fig:mat_two_case}b shows the pressure solutions among the CG, EG, and DG methods. CG result illustrates unphysical (spurious) pressure oscillations at the interface, which complies with the previous studies \cite{choo2018enriched, Haga2012}. The EG and DG methods, on the other hand, provide a smooth pressure solution for all time steps. These results may reflect the fact that the CG method, in its classical form, does not conserve mass locally. \par

In addition to the pressure oscillation, the CG method also delivers the oscillation in displacement, $\bm{u}$, which affects the $\varepsilon_v$ calculation as illustrated in Figure \ref{fig:mat_two_case}c. These oscillations may lead to incorrect fluid flux calculation that may revoke the local mass conservation property of the CG method. Besides, the non-physical oscillation in $\varepsilon_v$ may cause incorrect permeability alteration. As expected, there is no $\varepsilon_v$ oscillation produced by the EG and DG methods for all time steps.

\subsection{2D Flow in a Deformable Media with structured heterogeneity}

\begin{figure}[H]
   \centering
    \includegraphics[width=7.0cm, height=7.0cm]{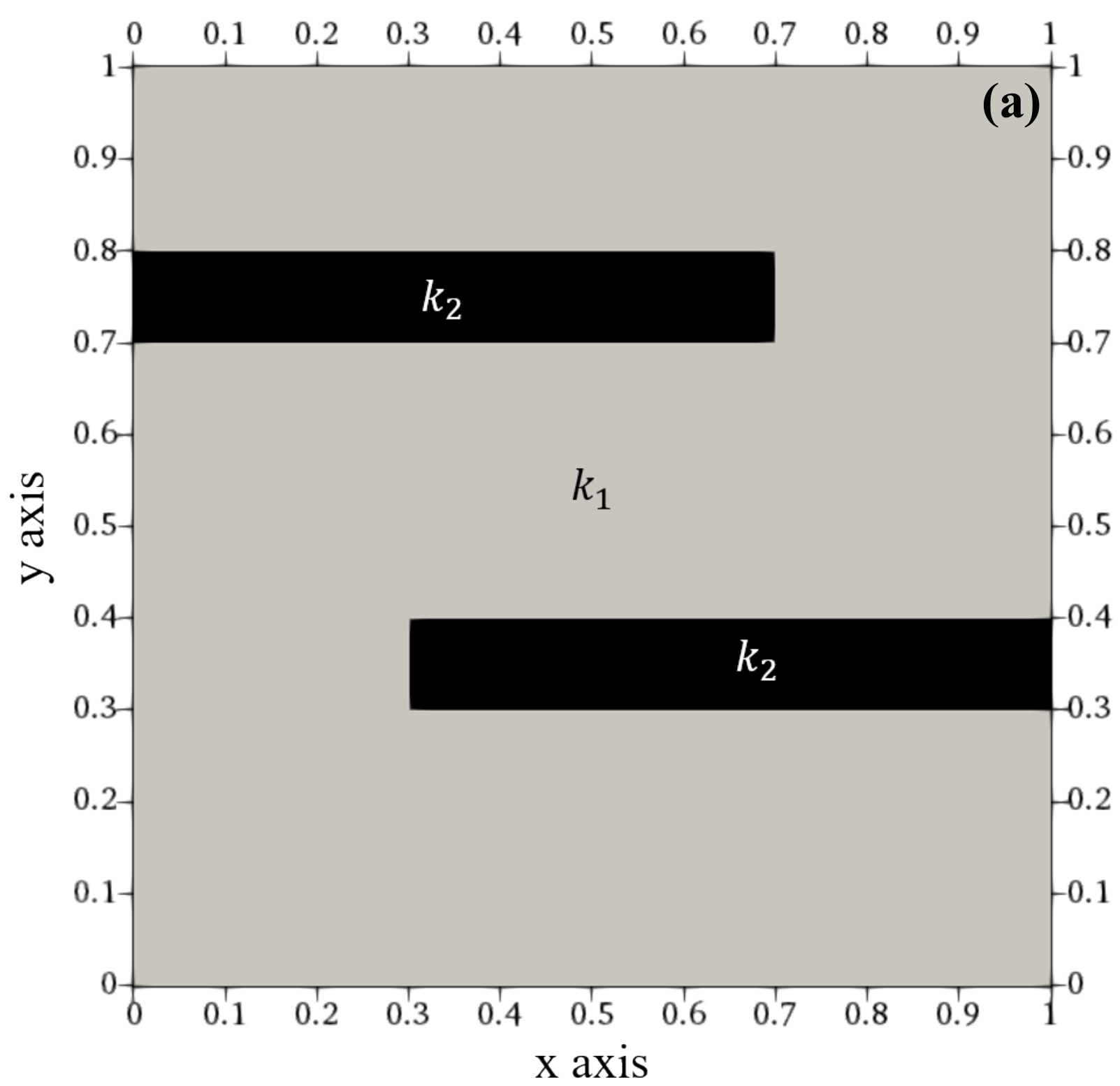}
    \includegraphics[width=7.0cm, height=7.0cm]{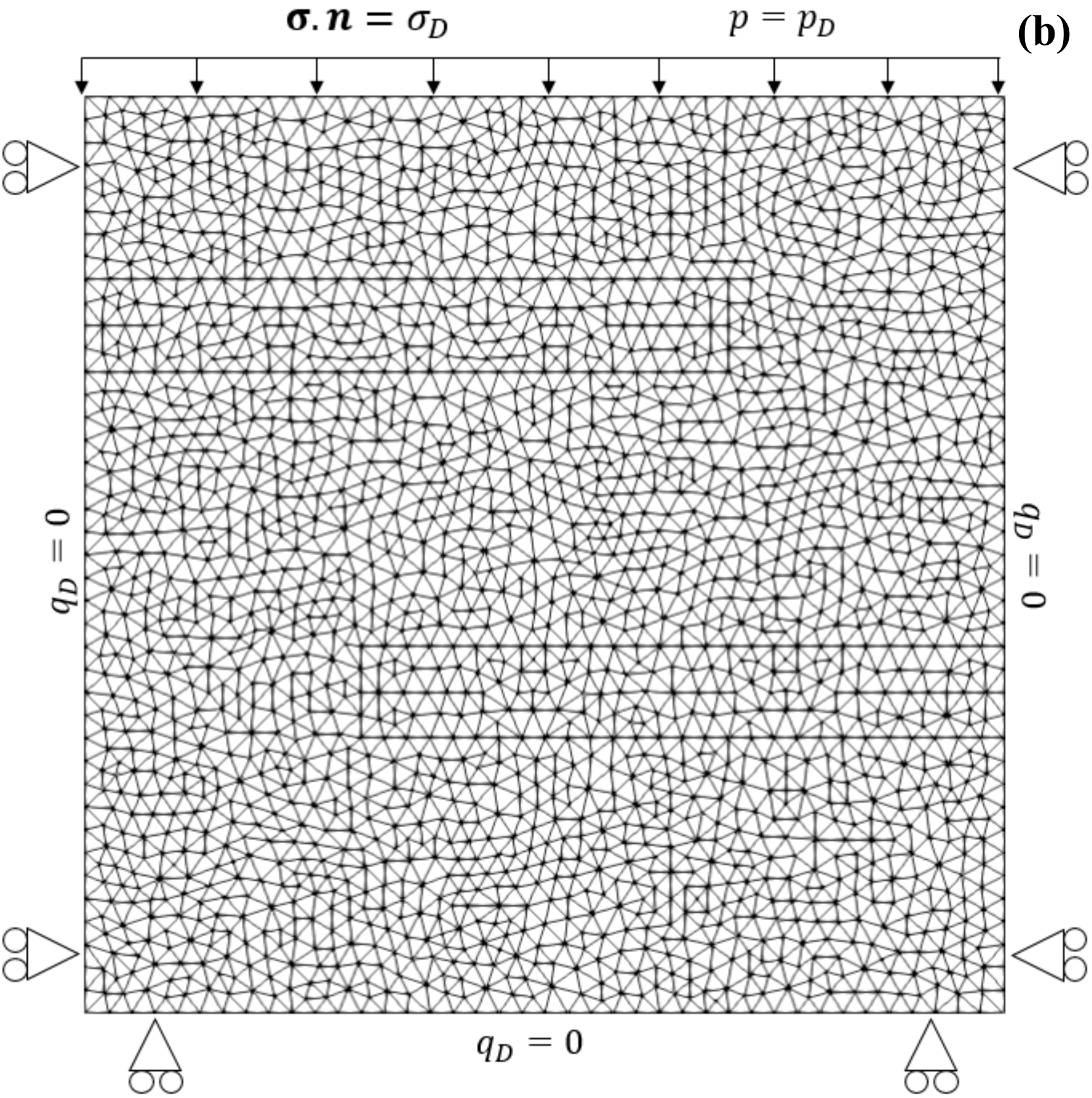}
   \caption{Structured heterogeneity example: (\textbf{a}) geometry and properties (\textbf{b}) mesh and boundary conditions}
   \label{fig:layered_case}
\end{figure}

The geometry, input properties, mesh, and boundary conditions used in this example are presented in Figure \ref{fig:layered_case}. This example aims to compare the local mass conservative property, flux approximation, $\Bar{\bm{\kappa}}$ and $\Bar{\varepsilon_v}$ alteration, number of iterations, and DOF among CG, EG, and DG methods. Note that $\Bar{\bm{\kappa}}$ and $\Bar{\varepsilon_v}$ are an arithmetic average of $\bm{\kappa}$ and $\varepsilon_v$, respectively. The model input parameters are $\bm{\sigma_{D}}=[0, 20]$ $\mathrm{MPa}$, $p_{D}=1$ $\mathrm{MPa}$, $p_{0}=10$ $\mathrm{MPa}$, $\alpha=0.79$, $c_{f}=1$ $\times 10^{-10}$ $\mathrm{Pa}$, $\rho=1000$ $\mathrm{kg} / \mathrm{m}^{3}$, $\mu=10^{-6}$ $\mathrm{kPa.s}$, $h=$ $0.025 \mathrm{m}$, and $v=0.2$. $K_{S}$, $\Delta t^n$ $=$ 1.0 sec, $\lambda_{l}$, and $\mu_{l}$ are calculated by \eqref{eq:lambda_l}. $\bm{k_m}$ is assigned to each subdomain as shown in Figure \ref{fig:layered_case}a, and $\bm{k_{1}}=10^{-12}\bm{I}$ $\mathrm{m}^{2}$ and $\bm{k_{2}}=10^{-16}\bm{I}$ $\mathrm{m}^{2}$. Here, $K$ is varied for three values: $K= 8$, $2$, and $1$ GPa for case 1, 2, and 3, respectively. The input values of case 1, $K= 8$ GPa, is used according to Jaeger et al. \cite{jaeger2009fundamentals}. This setting, three different $K$ values, is proposed to evaluate the interaction between discretization method and saturated rock stiffness ($K$).

The local mass conservation of each cell at each time step, $\mathrm{r^n_{ mass }}$, is calculated as follows:

\begin{equation}\label{eq:mass_loss}
\mathrm{r^n}_{\mathrm{mass}}:=\int_{T} \rho\left(\phi c_{f}+\frac{\alpha-\phi}{K_{s}}\right) \frac{p^{n}-p^{n-1}}{\Delta t^n}+\rho \alpha \nabla \cdot \frac{\boldsymbol{u}^{n}-\boldsymbol{u}^{n-1}}{\Delta t^n} d V+\sum_{e \in \mathcal{E}_{h}} \int_{e} \bm{v}^{n} \cdot\left.\boldsymbol{n}\right|_{\mathrm{e}} d S,
\end{equation}

\noindent
where the numerical flux, $\bm{v}^{n} \cdot\left.\boldsymbol{n}\right|_{\mathrm{e}}$, is described as follows \cite{sun2009locally, lee2018enriched}:

\begin{equation}\label{eq:flux_internal}
\bm{v}^{n} \cdot \left.\boldsymbol{n}\right|_{\mathrm{e}}=-\left\{\bm{\kappa^n}\left(\nabla p_{h}^{n}-\rho \bm{g}\right) \cdot \bm{n}\right\}+\frac{\beta}{h_{e}} \bm{\kappa^n}\llbracket p_{h}^{n}\rrbracket \quad \forall e \in \mathcal{E}_{h}^{0},
\end{equation}

\begin{equation}\label{eq:flux_neuman}
\bm{v}^{n} \cdot\left.\bm{n}\right|_{\mathrm{e}}=q_{D} \quad \forall e \in \mathcal{E}_{h}^{N},
\end{equation}

\begin{equation}\label{eq:rf_dir}
\bm{v}^{n} \cdot\left.\boldsymbol{n}\right|_{\mathrm{e}}=-\bm{\kappa^n}\left(\nabla p_{h}^{n}-\rho \bm{g}\right) \cdot \boldsymbol{n}+\frac{\beta}{h_{e}} \bm{\kappa^n}\left(p_{h}^{n}-p_{D}\right) \quad \forall e \in \mathcal{E}_{h}^{D},
\end{equation}

\begin{figure}[H]
   \centering
        \includegraphics[width=8.0cm, height=7.0cm]{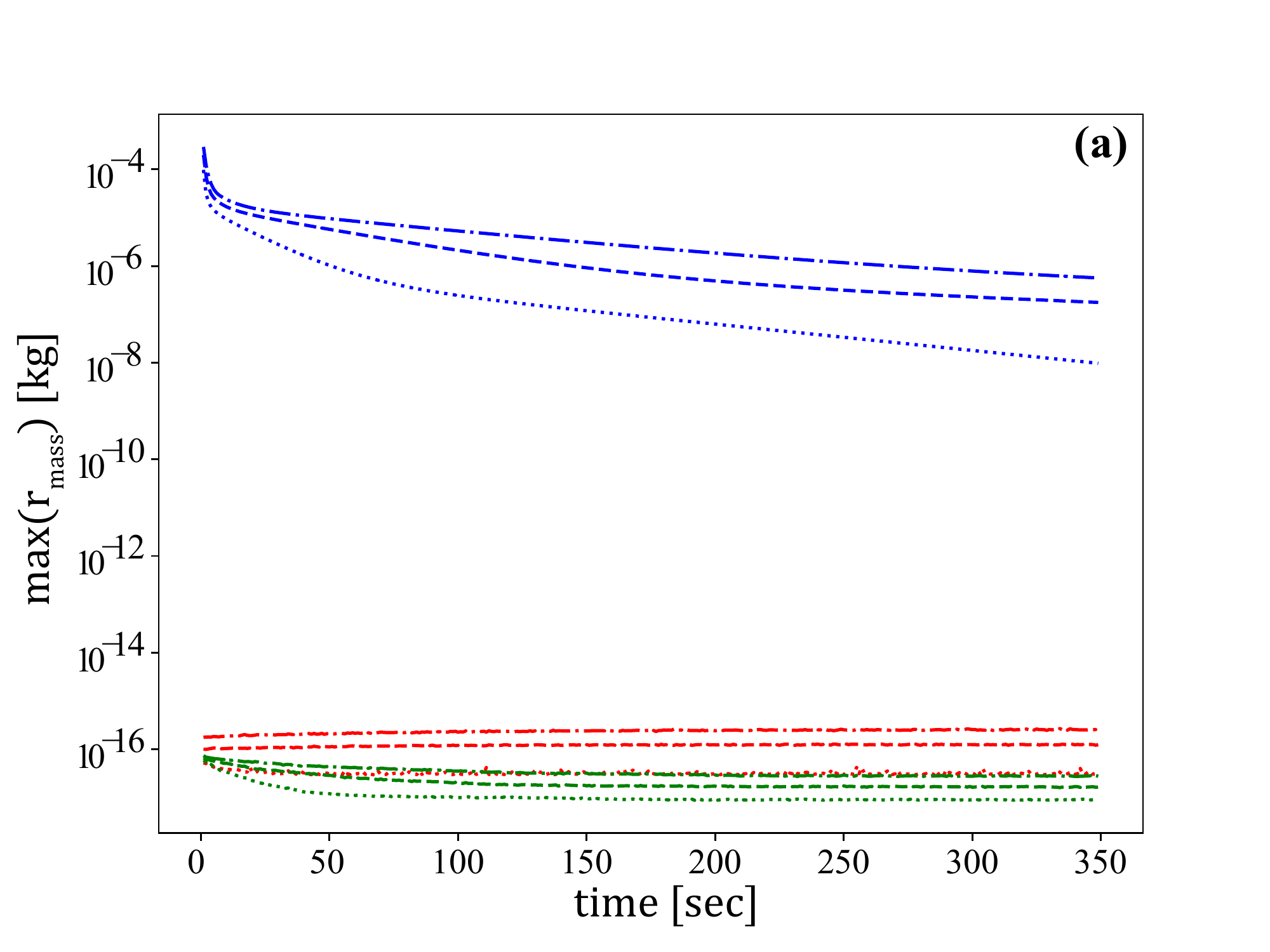}
        \includegraphics[width=8.0cm, height=7.0cm]{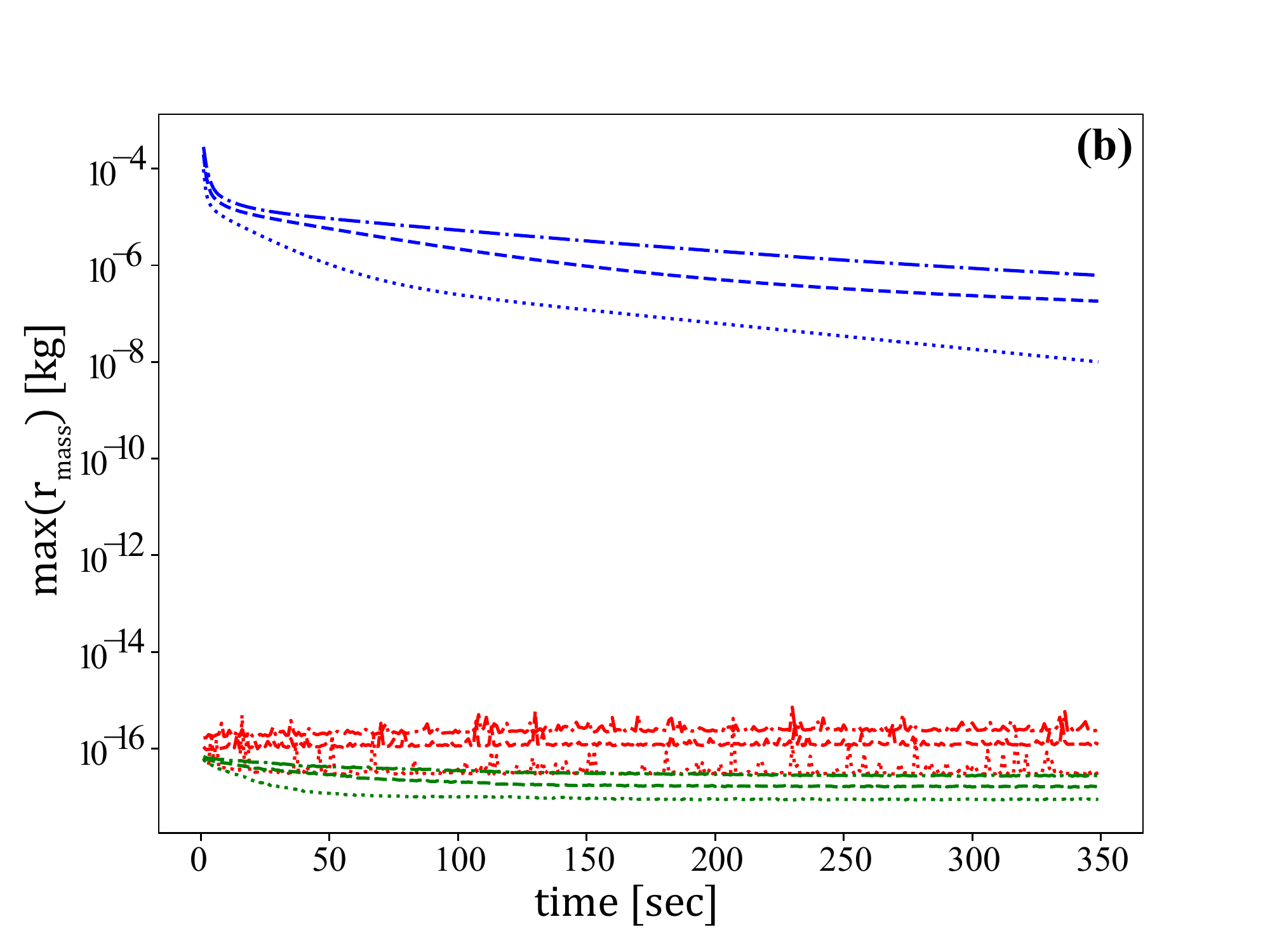}
        \includegraphics[width=8.0cm, height=0.9cm]{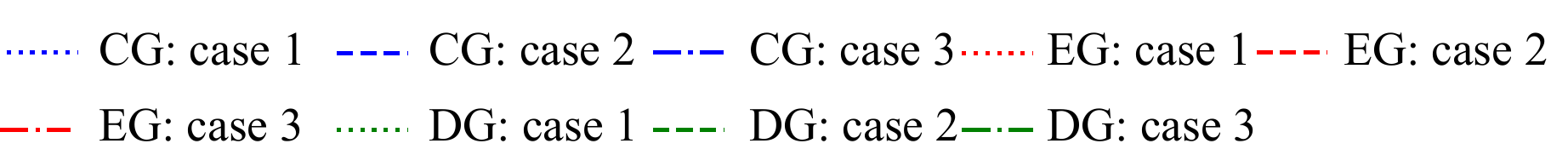}
   \caption{Structured heterogeneity example: the comparison of $\mathrm{max(r_{mass})}$ values among the CG (blue line), EG (red line), and DG (green line) methods using (\textbf{a}) pressure-independent $\bm{k_m}$ and (\textbf{b}) pressure-dependent $\bm{k_m}$ models for case 1, $K = 8$ GPa, case 2, $K = 2$ GPa, and case 3, $K = 1$ GPa}
   \label{fig:mass_loss_layered}
\end{figure}

\noindent
The local mass conservation results of structured heterogeneity example between pressure-independent $\bm{k_m}$ and pressure-dependent $\bm{k_m}$ models are presented in Figures \ref{fig:mass_loss_layered}a-b, respectively. The local mass conservation property is well preserved for the EG and DG methods, i.e. the maximum value of $\mathrm{r_{mass}}$, $\mathrm{max(r_{mass})}$, is close to zero. The CG results, on the contrary, have much higher $\mathrm{max(r_{mass})}$ value. The lack of local mass conservation of the CG method is the effect induced by the spurious $p$ and $\varepsilon_v$ oscillations that lead to the incorrect flux approximation. Moreover, cases that contain a lower $K$ value have higher $\mathrm{max(r_{mass})}$ value because the change in $\bm{u}$, and subsequently, $p$ are more drastic in the softer material.

\begin{figure}[H]
   \centering
        \includegraphics[width=8.0cm, height=7.0cm]{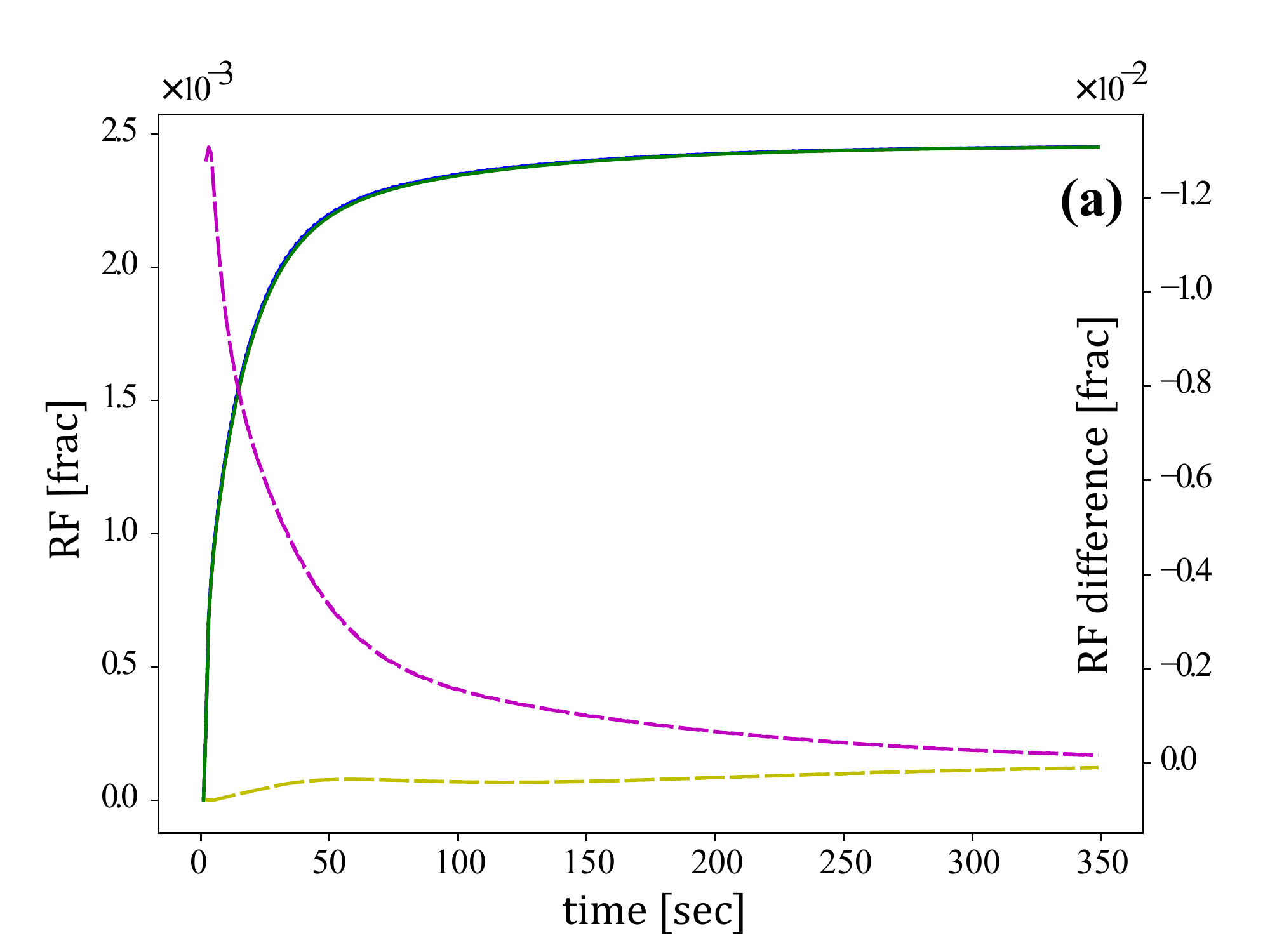}
        \includegraphics[width=8.0cm, height=7.0cm]{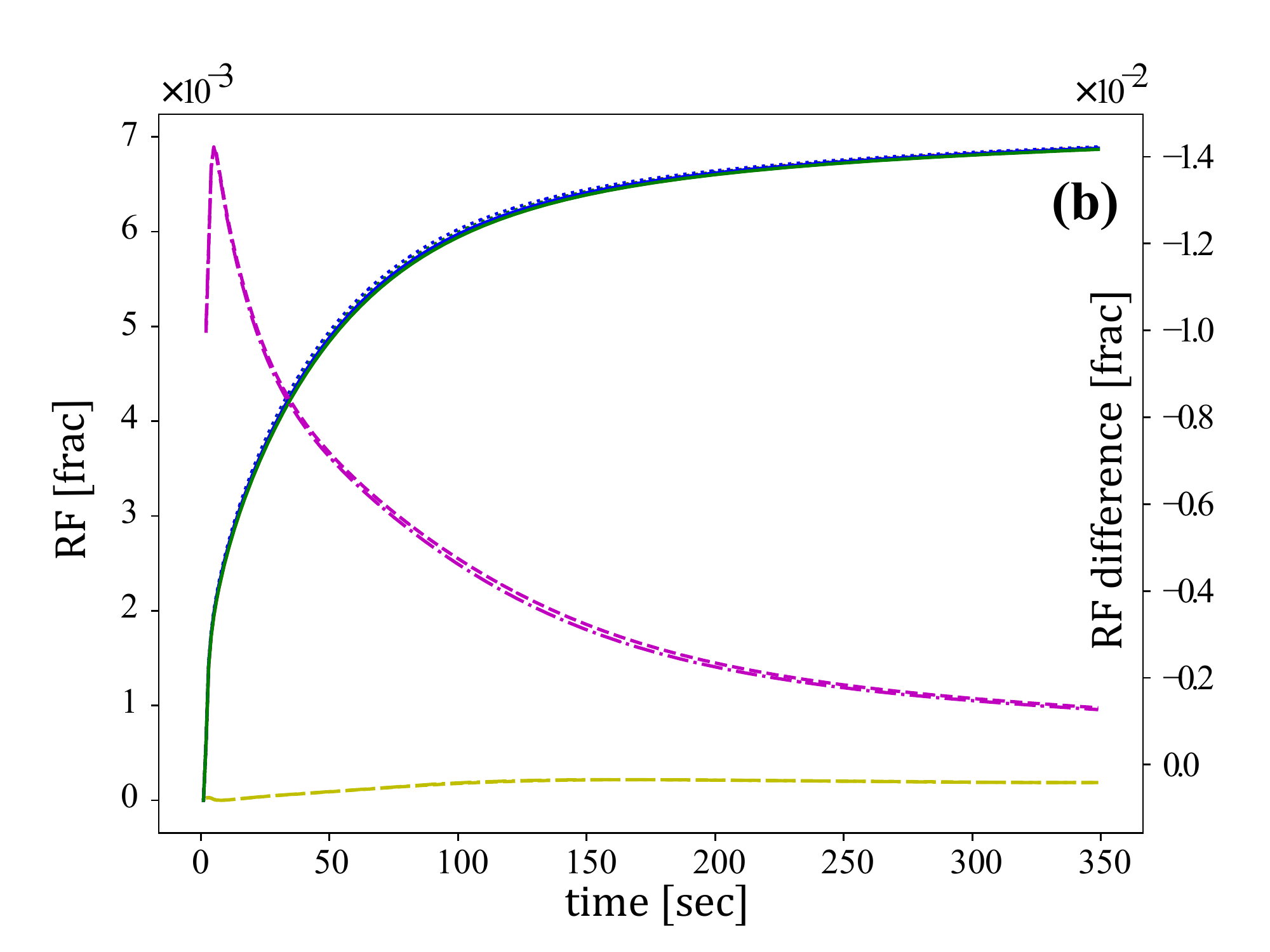}
        \includegraphics[width=8.0cm, height=7.0cm]{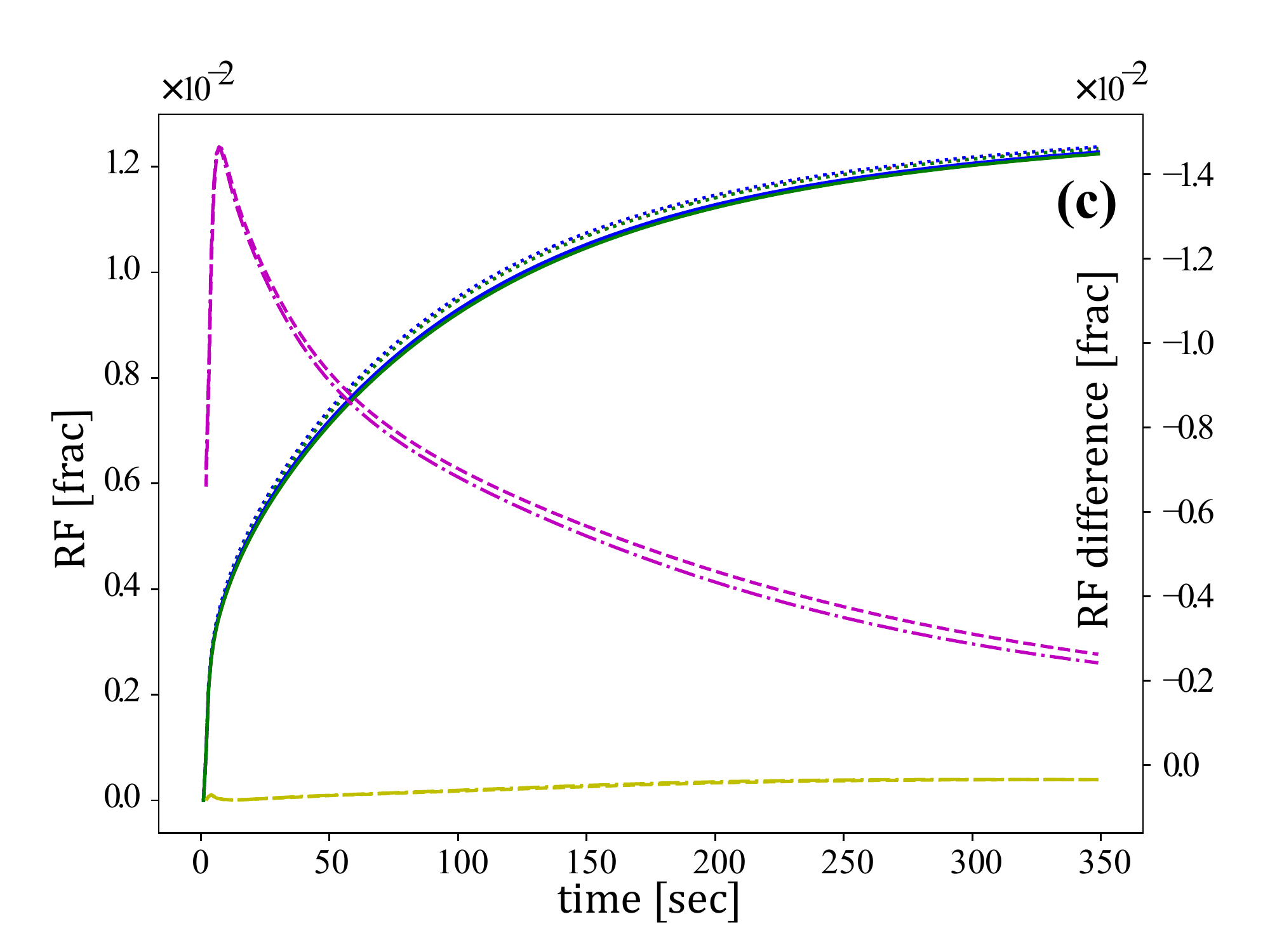}
        \includegraphics[width=8.0cm, height=5.0cm]{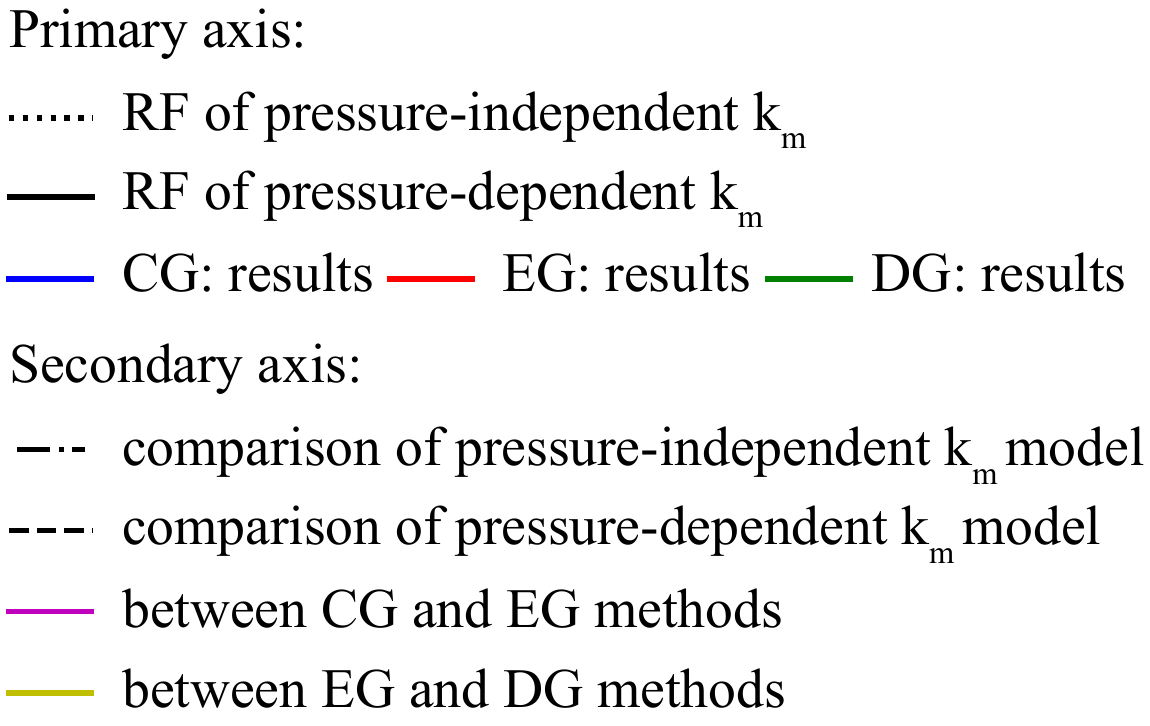}
   \caption{Structured heterogeneity example: the comparison of RF (primary axis) and the difference of RF (second axis) among the CG (blue line), EG (red line), and DG (green line) methods using pressure-dependent (solid) and pressure-independent (dotted) $\bm{k_m}$ models for (\textbf{a}) case 1, $K = 8$ GPa, (\textbf{b}) case 2, $K = 2$ GPa, and (\textbf{c}) case 3, $K = 1$ GPa. Note that CG, EG, and DG results overlay on top of each other.}
   \label{fig:rf_layered}
\end{figure}

Then, we investigate the error in the flux approximation using a recovery factor ($\mathrm{RF}$), which is defined at each time step as follows:

\begin{equation}\label{eq:rf}
\mathrm{RF}^n:=\frac{\sum_{{t=0}}^{t^{n}} \sum_{e \in \mathcal{E}_{h}} \int_{e} \bm{v}^{n} \cdot\left.\boldsymbol{n}\right|_{\mathrm{e}} dS}{\rho V_{0} \phi},
\end{equation}

\noindent
where $V_0$ is an initial reservoir volume. In this example, $\bm{v}^{n} \cdot\left.\boldsymbol{n}\right|_{\mathrm{e}}$ is calculated for all outlet surfaces, i.e. the top surface (Figure \ref{fig:layered_case}b). The results of case 1, case 2, and case 3 are presented in Figures \ref{fig:rf_layered}a-c, respectively. The primary axis shows RF values, while the secondary axis illustrates the difference between CG and EG solutions and the difference between DG and EG solutions, which are called RF difference hereafter. The following observations can be made from these figures:

\begin{enumerate}
    \item Even though the results between pressure-independent and -dependent $\bm{k_m}$ are different, and the difference grows when the value of $K$ is decreased (i.e., more deformable), the differences are still small (maximum difference between pressure-independent and -dependent $\bm{k_m}$ is approximately 3 \%.).
    \item The maximum difference of $\mathrm{RF}$ results between the CG and EG methods is approximately 1.4 $\%$. Besides, the maximum difference of $\mathrm{RF}$ results between the DG and EG methods is insignificant (the maximum difference is 0.08 \%).
    \item When the reservoir has a lower $K$ value, i.e. more deformable, $\mathrm{RF}$ becomes higher because of the compaction effect. Figures \ref{fig:rf_layered}a-c illustrate that the final $\mathrm{RF}$ is increased from $2.5 \times 10^{-3}$ to $1.2 \times 10^{-2}$ when $K$ is reduced from $8$ GPa to $1$ GPa. This observation holds even though the $\bm{k_m}$ reduction is larger in softer media. Therefore, the compaction effect may dominate over $\bm{k_m}$ reduction effect on reservoir productivity.
\end{enumerate}

\begin{figure}[H]
   \centering
        \includegraphics[width=8.0cm, height=7.0cm]{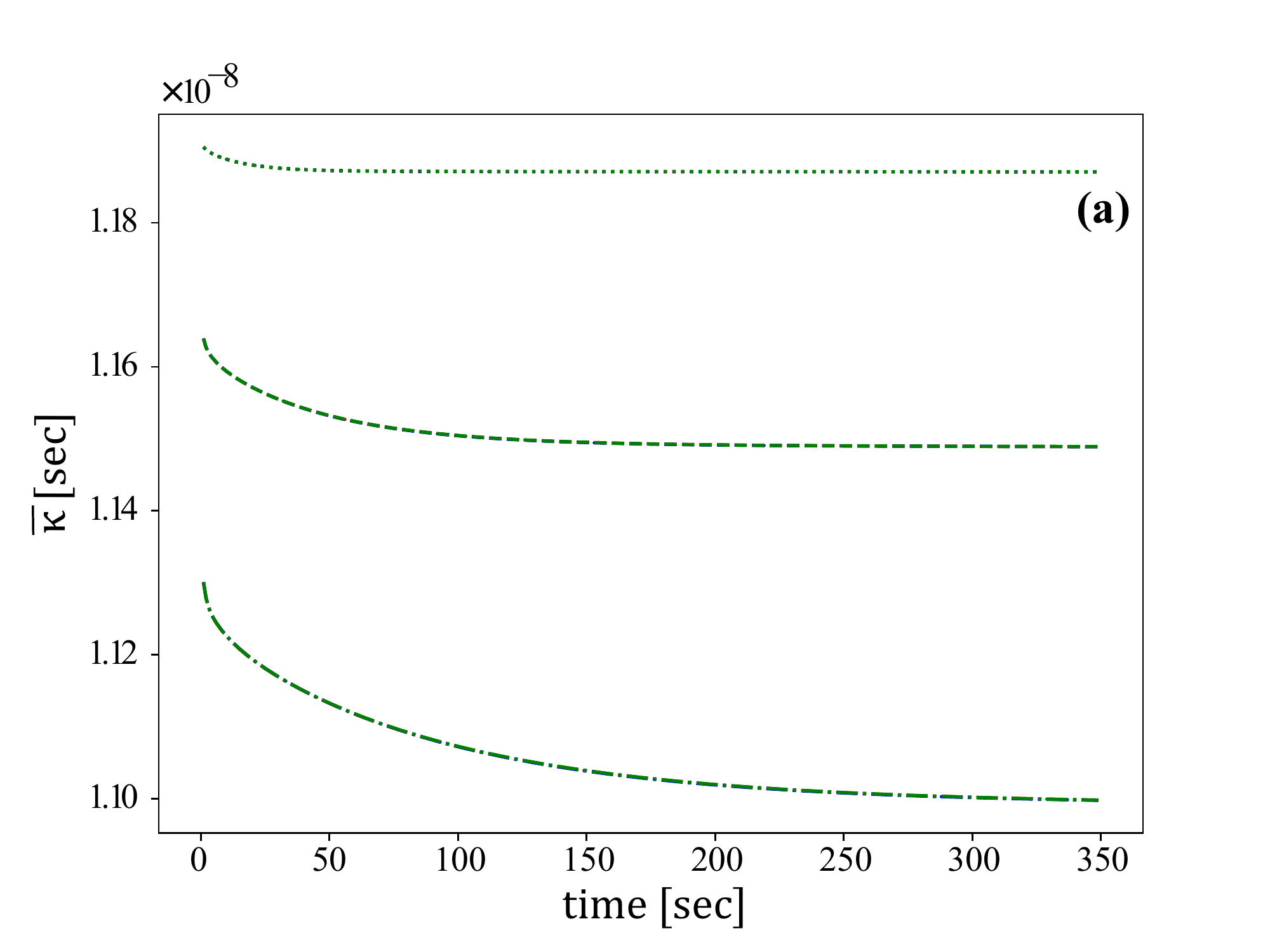}
        \includegraphics[width=8.0cm, height=7.0cm]{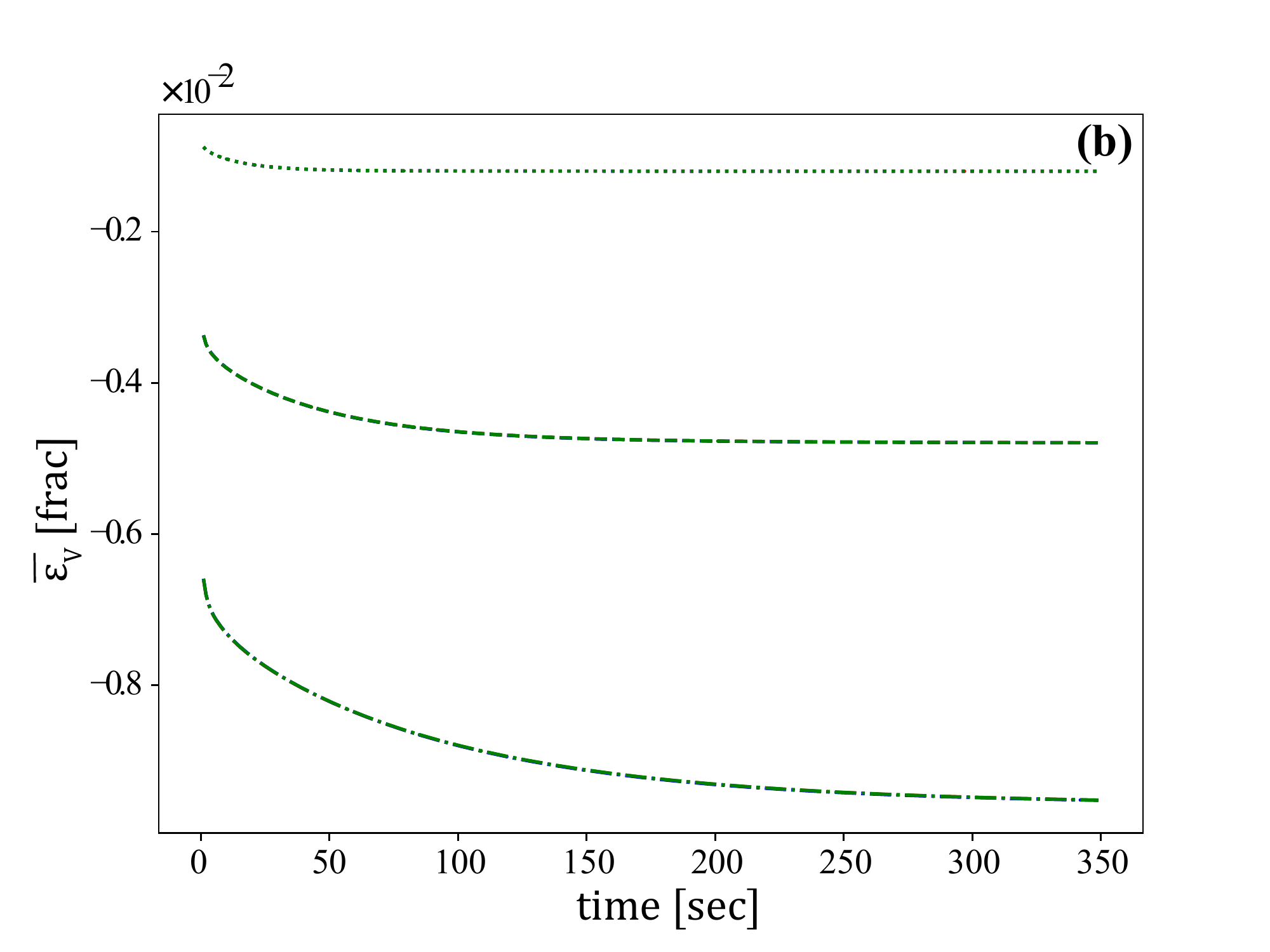}
        \includegraphics[width=8.0cm, height=7.0cm]{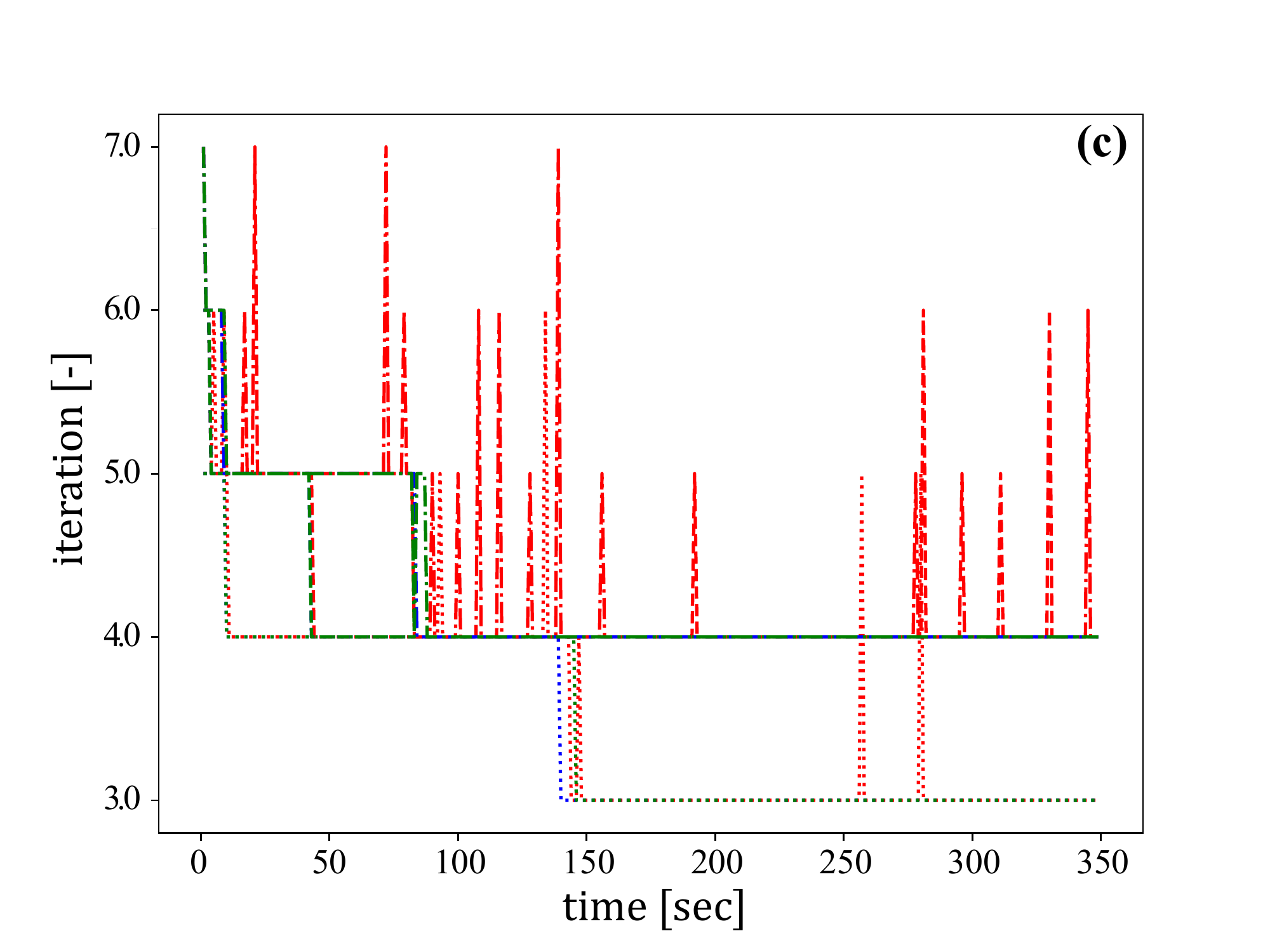}
        \includegraphics[width=8.0cm, height=0.9cm]{pictures/com_mass_no_deform_layered_legend.pdf}
   \caption{Structured heterogeneity example: the comparison of (\textbf{a}) $\mathrm{\bar{\bm{\kappa}}}$, (\textbf{b}) $\mathrm{\bar{\varepsilon_{v}}}$, and (\textbf{c}) number of iterations values among the CG (blue line), EG (red line), and DG (green line) methods using pressure-dependent $\bm{k_m}$ model for case 1, $K = 8$ GPa, case 2, $K = 2$ GPa, and case 3, $K = 1$ GPa. Note that CG, EG, and DG results are overlapped on top of each other in a and b.}
   \label{fig:com_layered}
\end{figure}

Results of $\Bar{\bm{\kappa}}$ and $\Bar{\varepsilon_v}$ illustrated in Figures \ref{fig:com_layered}a-b, respectively, supports the second statement. As observed from these figures, the $\Bar{\bm{\kappa}}$ and $\Bar{\varepsilon_v}$ among CG, EG, and DG methods are not significantly different. Therefore, the flux approximation, $\mathrm{RF}$, among these methods are almost the same.

Moreover, the results of the number of iterations (Algorithm \ref{al:perm_iteration} - Picard iteration) among all discretizations and cases are presented in Figures \ref{fig:com_layered}c. There is not much difference among CG, EG, and DG methods as the number of iterations in all cases range between three and seven. The softer media, however, needs more iteration than the tougher one, i.e. the number of iterations is larger in the material that has lower $K$ value than that of the higher $K$ value material. This is because the $\bm{\kappa}$ alteration is more drastic in the softer material than the tougher material; therefore, solutions take more iterations to converge.

The number of DOF among each method is shown in Table \ref{tab:struct_dof}. Note that all methods use the same CG vector function space for the displacement; hence, the DOF is the same for the displacement domain. For the pressure space, as expected, the CG method has the least DOF, i.e. approximately three times less than that of the EG method. The DG method, on the other hand, has the most DOF, i.e. about two times more than that of the EG method.

\begin{table}[H]
\begin{center}
\caption {Structured heterogeneity example: Degrees of freedom (DOF) comparison among CG, EG, and DG methods}\label{tab:struct_dof}
\begin{tabular}{|l|c|c|c|} \hline
\textbf{} & \textbf{CG} & \textbf{EG} & \textbf{DG}   \\\hline
Displacement     & 17,258  & 17,258     & 17,258     \\\hline
Pressure         & 2,198   & 6,432      & 12,702     \\\hline
Total      & 19,456  & 23,690     & 29,960     \\\hline
\end{tabular}
\end{center}
\end{table}

\subsection{2D Flow in a Deformable Media with Random Porosity and Permeability}

\begin{figure}[H]
   \centering
    \includegraphics[width=6.5cm, height=7.0cm]{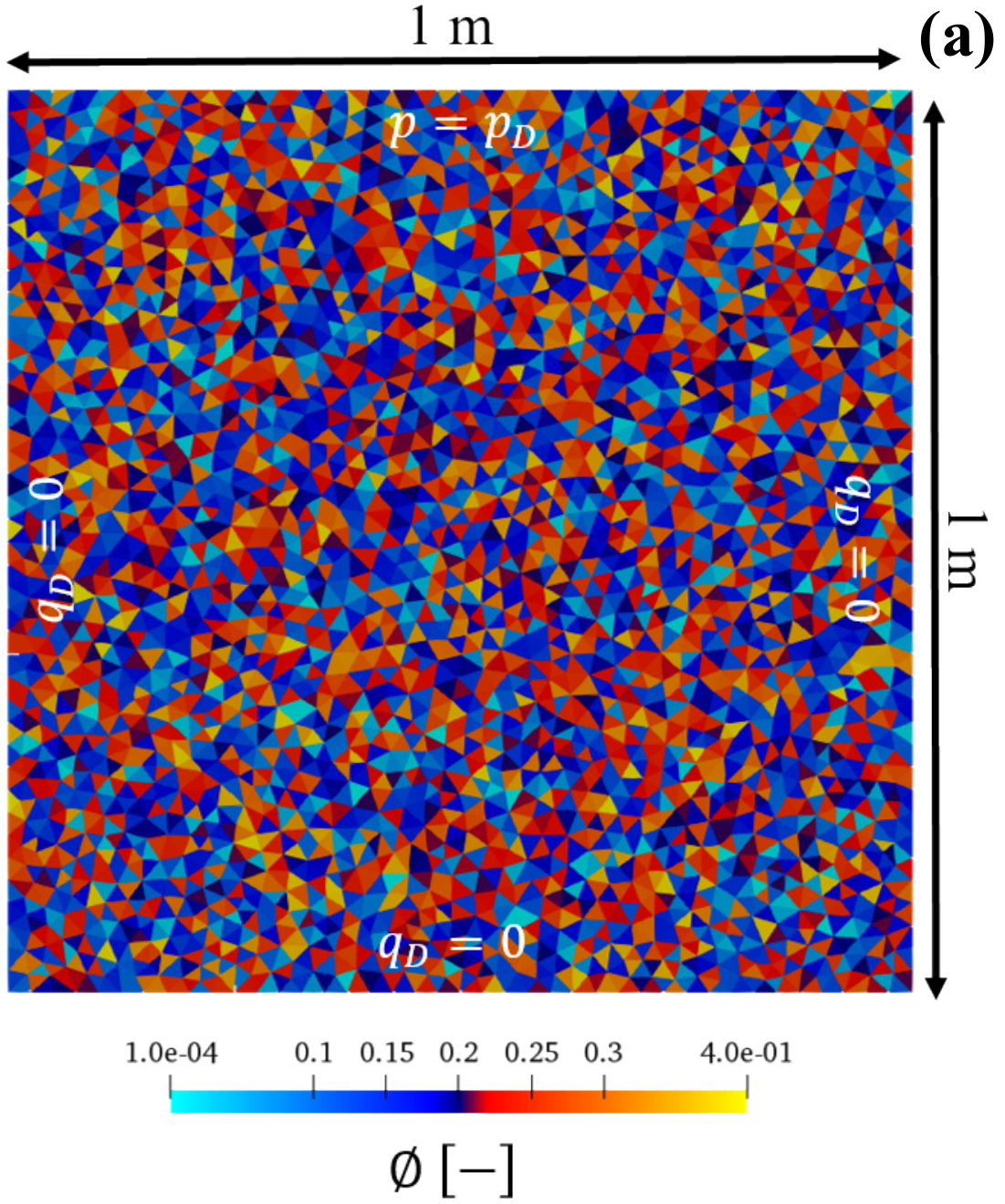}
    \includegraphics[width=7.5cm, height=7.0cm]{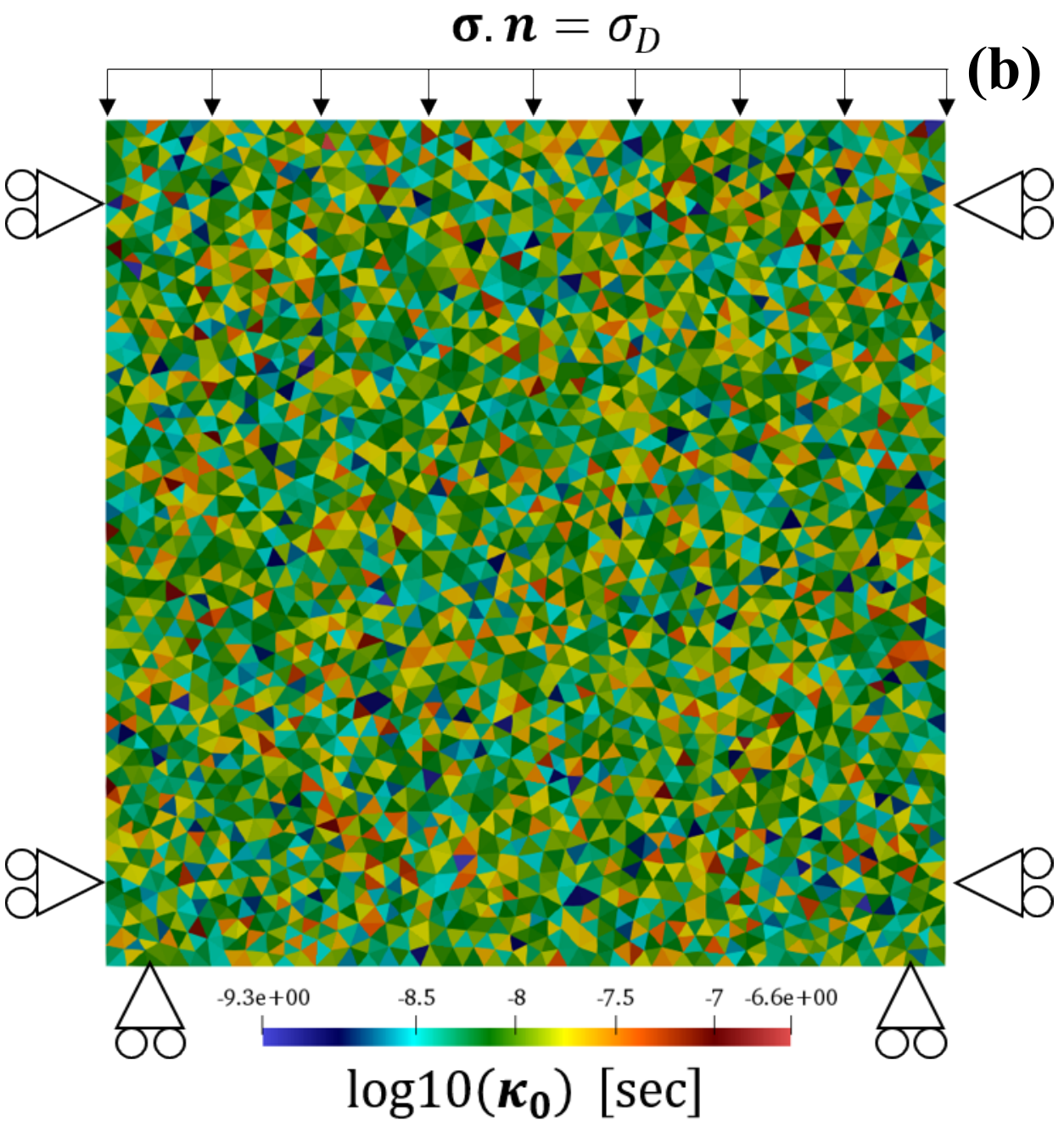}
   \caption{2D heterogeneous example: (\textbf{a}) $\phi$ field and mass boundary conditions and (\textbf{b}) $\mathrm{log10(\bm{\kappa_0})}$ field in x-direction and force boundary conditions}
   \label{fig:2d_case}
\end{figure}

The 2D heterogeneous example presented in Figure \ref{fig:2d_case} targets to evaluate the local mass conservative property ($\mathrm{max(r_{mass})}$), flux approximation ($\mathrm{RF}$), $\Bar{\bm{\kappa}}$ and $\Bar{\varepsilon_v}$ alteration, number of iterations, and DOF among CG, EG, and DG methods. Its geometry, input parameters, and boundary conditions are adapted from the previous study \cite{Kadeethum2019ARMA,kadeethum2020finite}. Most of the input parameters have the same values as the structured heterogeneity example. $K$ values are applied similarly to the structured heterogeneity case, i.e. $K= 8$, $2$, and $1$ GPa for case 1, 2, and 3, respectively. On the other hand, $\phi$ and $\bm{\kappa_0}$ fields are populated based on normal and log-normal distributions as illustrated in Figure \ref{fig:2d_case}a-b, respectively. The heterogeneous parameters are provided based on the following specifications, $\Bar{\phi}=0.2,$  $\operatorname{var}(\phi)=0.01,$ $\phi_{\min }=0.001,$ $\phi_{\max }=0.4,$  $\Bar{\bm{\kappa_0}}=1.2 \times 10^{8}\bm{I} \: \mathrm{s}, \operatorname{var}(\bm{\kappa_0})=1.4 \times 10^{-16} \bm{I}\: \mathrm{s}, \bm{\kappa_{0 \min}}=1.2 \times 10^{-13}\bm{I} \: \mathrm{s}$, and $\bm{\kappa_{0\max}}=1.2 \times 10^{-6}\bm{I} \mathrm{s}.$ $\overline{( .)}$ is an arithmetic average, var$( .)$ is the variance, $( .)_{\min }$ and $( .)_{\max }$ are the minimum and maximum values, respectively. $\Bar{\phi}$, $\phi_{\min }$, $\phi_{\max }$, $\Bar{\bm{\kappa_0}}$, $\bm{\kappa_{0 \min }}$, and $\bm{\kappa_{0 \max }}$ are selected based on the reported values of shales and sandstones from Jaeger et al. \cite{jaeger2009fundamentals}. Note that $\Bar{\bm{\kappa_0}}$, $\bm{\kappa_{0 \min }}$, and $\bm{\kappa_{0 \max }}$ reflect the $\Bar{\bm{k_{m0}}}$, ${\bm{k_{m0 \min}}}$, and ${\bm{k_{m0\max}}}$ values of $1.2 \times 10^{-14}\bm{I}$, $1.2 \times 10^{-19} \bm{I} $, $1.2 \times 10^{-12} \bm{I}$, respectively. \rev{In the nature, the $\phi$ and $\bm{\kappa_0}$ fields are normally correlated \cite{jensen2000statistics}. In this study, however, these two heterogeneous fields are populated independently, i.e., there is no correlation between $\phi$ and $\bm{\kappa_0}$. Since the completely random fields are the most challenge setting, i.e., the sharp discontinuity could appear in many places, and the permeability alteration resulted from the solid deformation \cite{abou2013petroleum, Du2007} could result in a negative value. Therefore, if the proposed approximation passes this case, it would be able to handle a correlated one.}

\begin{figure}[H]
   \centering
        \includegraphics[width=8.0cm, height=7.0cm]{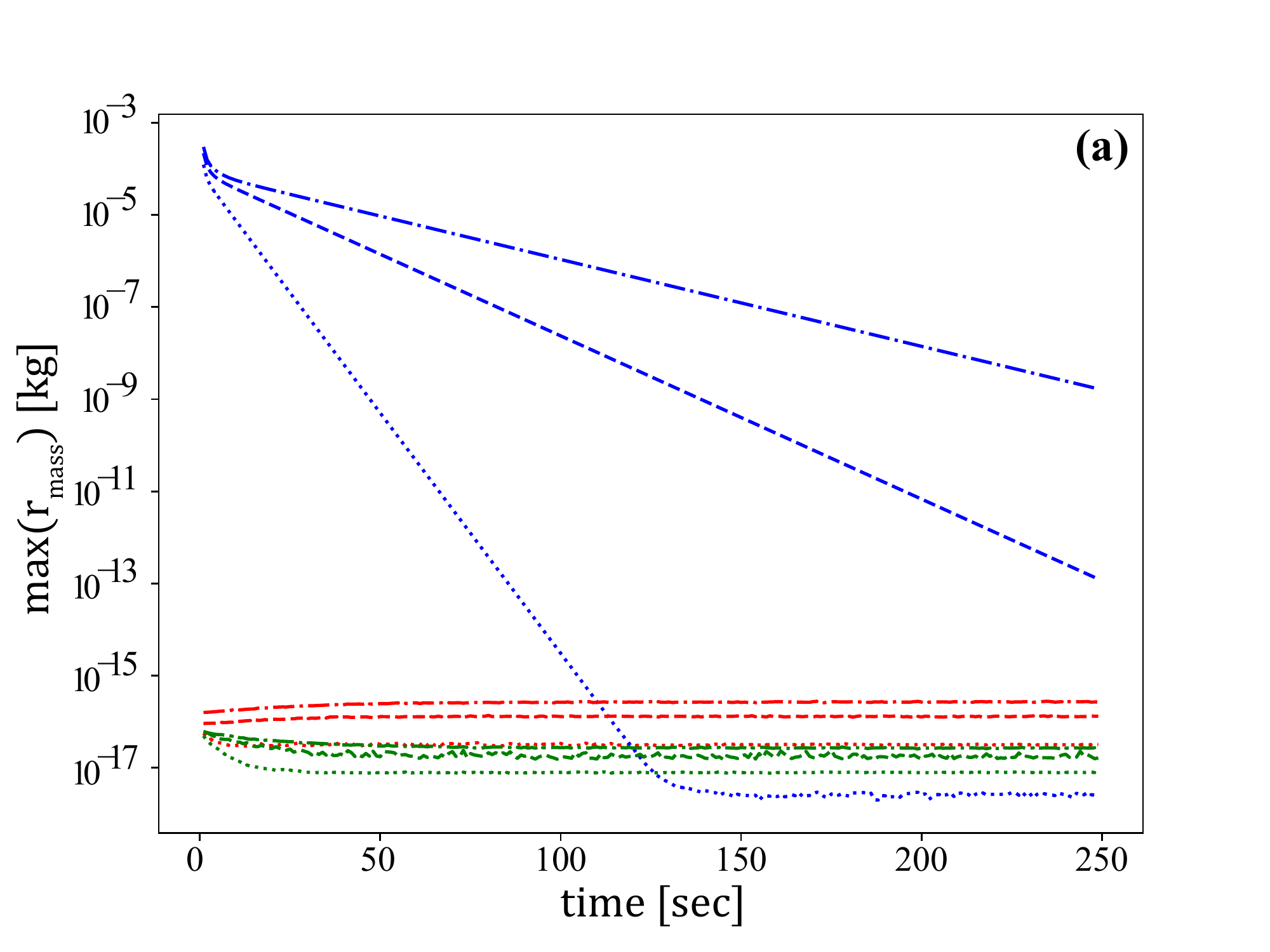}
        \includegraphics[width=8.0cm, height=7.0cm]{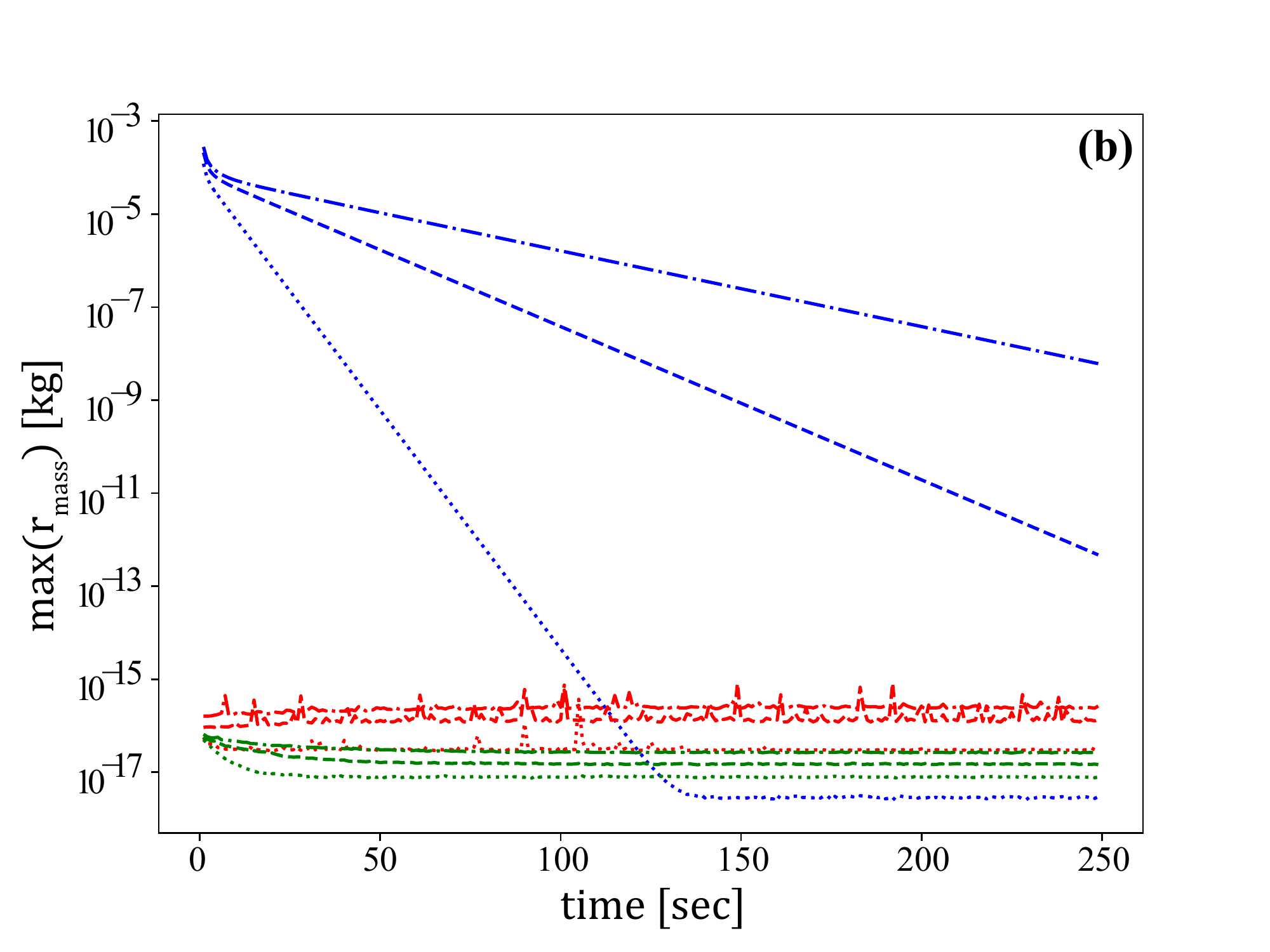}
        \includegraphics[width=9.0cm, height=0.9cm]{pictures/com_mass_no_deform_layered_legend.pdf}
   \caption{2D heterogeneous example: the comparison of $\mathrm{max(r_{mass})}$ values among the CG (blue line), EG (red line), and DG (green line) methods using (\textbf{a}) pressure-independent $\bm{k_m}$ and (\textbf{b}) pressure-dependent $\bm{k_m}$ models for case 1, $K = 8$ GPa, case 2, $K = 2$ GPa, and case 3, $K = 1$ GPa}
   \label{fig:mass_loss_2d}
\end{figure}

Figures \ref{fig:mass_loss_2d}a-b present $\mathrm{max(r_{mass})}$ results for pressure-independent and -dependent $\bm{k_m}$ models, respectively. The CG method illustrates a lack of local mass conservation property, and $\mathrm{max(r_{ mass })}$ value is significantly high at the beginning ($1 \times 10^{-3}$ kg). On the contrary, the EG and DG methods illustrate preservation of local mass, and their $\mathrm{max(r_{ mass })}$ values are almost equivalent to zero. Moreover, the case with a lower $K$ value have higher $\mathrm{max(r_{mass})}$ value because the change in $\bm{u}$ is more severe in the softer material.

\begin{figure}[H]
   \centering
        \includegraphics[width=8.0cm, height=7.0cm]{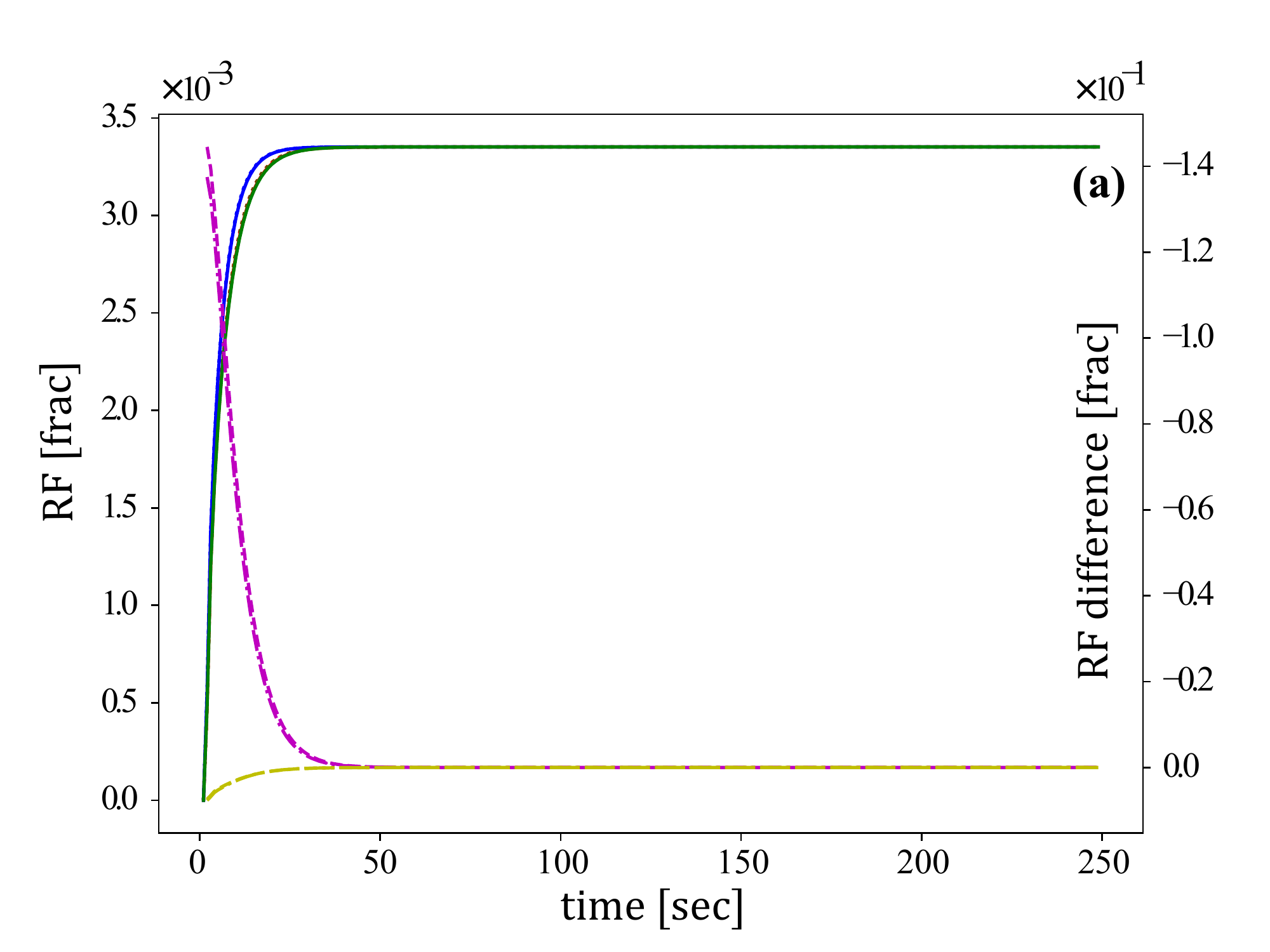}
        \includegraphics[width=8.0cm, height=7.0cm]{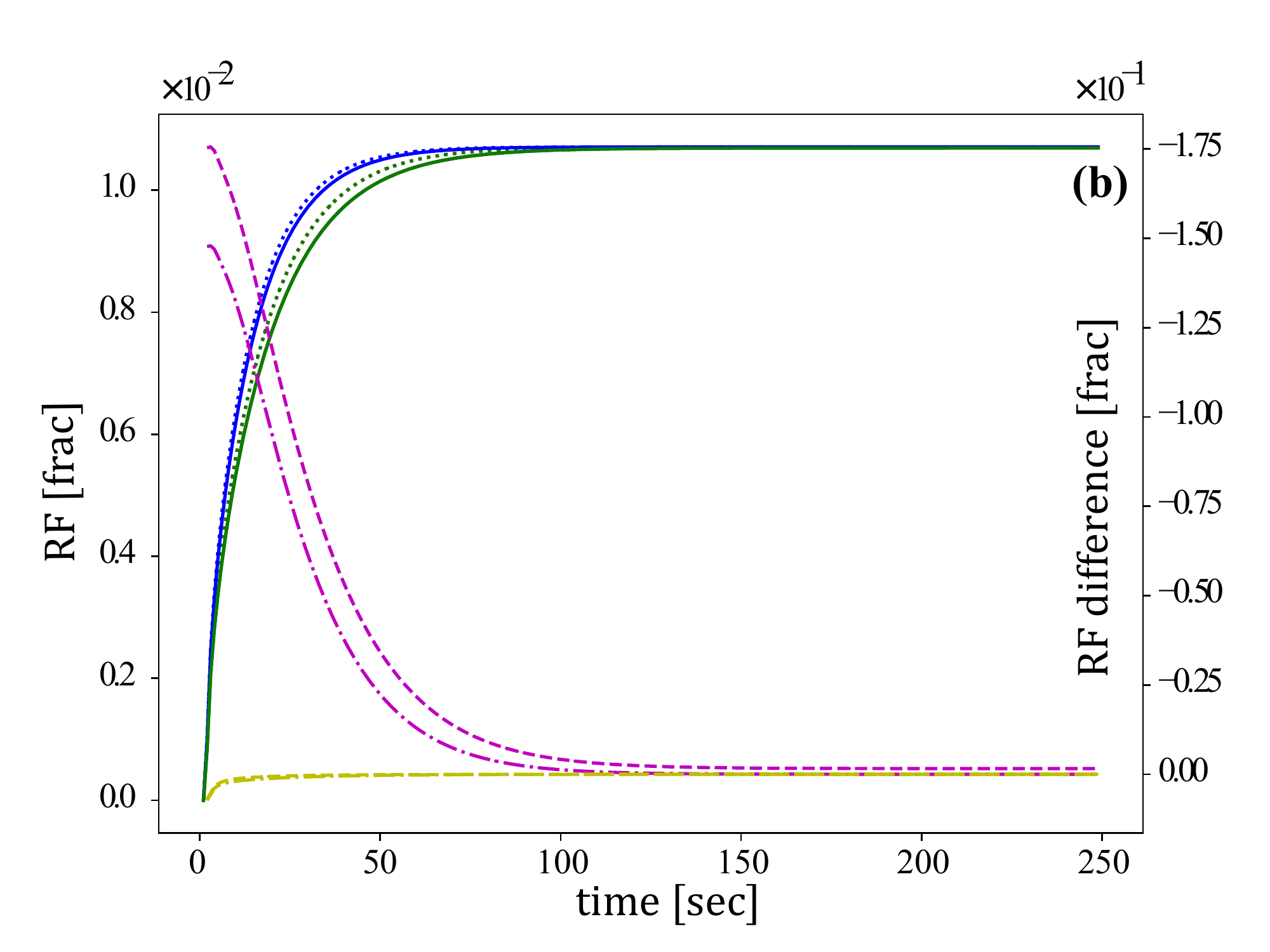}
        \includegraphics[width=8.0cm, height=7.0cm]{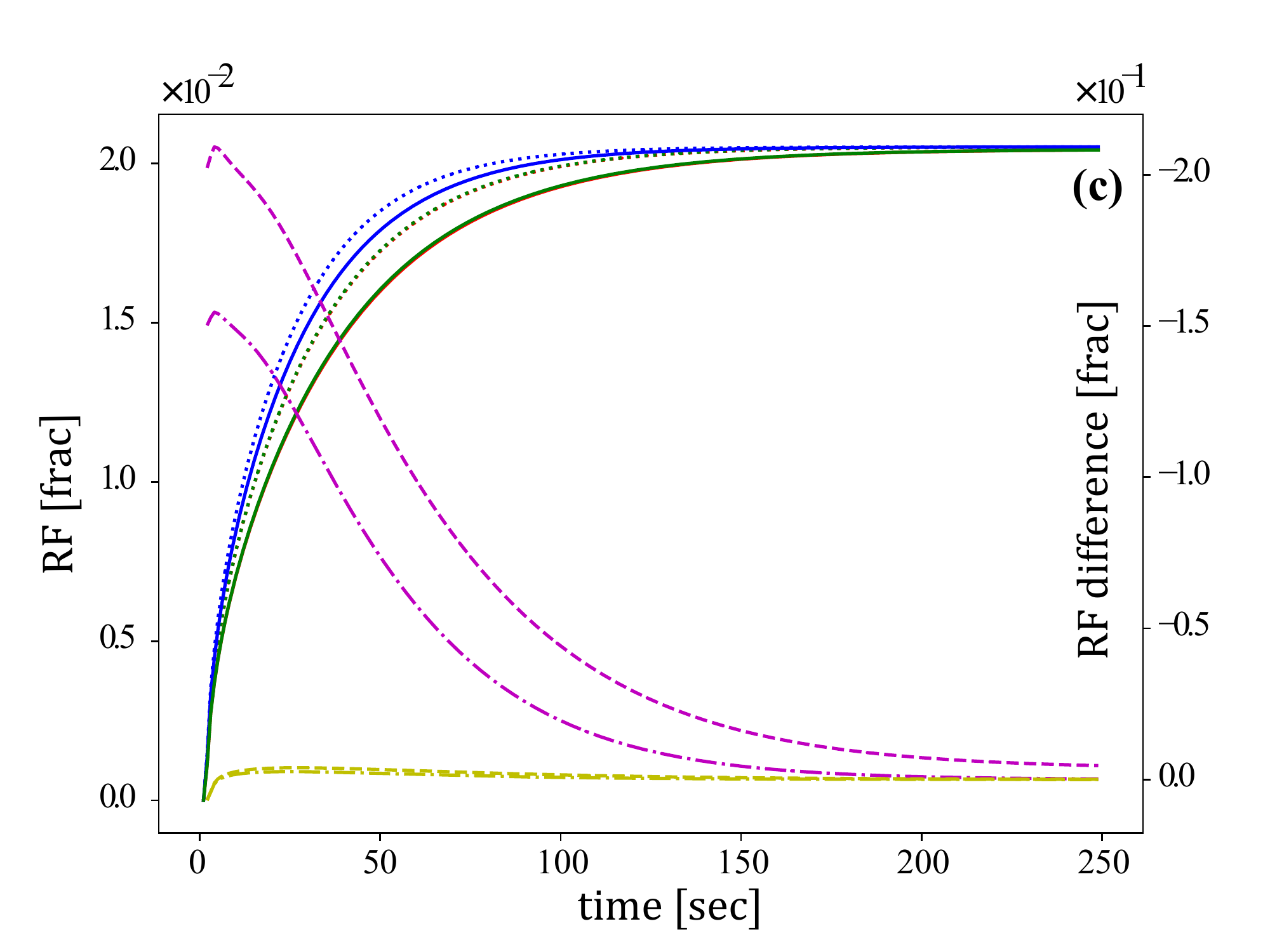}
        \includegraphics[width=8.0cm, height=5.0cm]{pictures/rf_layered_legend.pdf}
   \caption{2D heterogeneous example: the comparison of RF (primary axis) and the difference of RF (second axis) among the CG (blue line), EG (red line), and DG (green line) methods using pressure-dependent (solid) and pressure-independent (dotted) $\bm{k_m}$ models for (\textbf{a}) case 1, $K = 8$ GPa, (\textbf{b}) case 2, $K = 2$ GPa, and (\textbf{c}) case 3, $K = 1$ GPa. Note that EG and DG results overlay on each other.}
   \label{fig:rf_2d}
\end{figure}

The $\mathrm{RF}$ results of case 1, $K = 8$ GPa, case 2, $K = 2$ GPa, and case 3, $K = 1$ GPa are presented in Figures \ref{fig:rf_2d}a-c, respectively. The following observations can be drawn from these figures:

\begin{enumerate}
    \item The pressure-dependent $\bm{k_m}$ model delivers fluid slower than the pressure-independent $\bm{k_m}$ model. Since the fluid is withdrawn from the reservoir, $\bm{k_m}$ is reduced through time. Therefore, the pressure-dependent $\bm{k_m}$ model has less ability to produce reservoir fluid as the effective stress increases. Moreover, the difference between the two models becomes larger, when the reservoir is more deformable, i.e. $K$ is lower.
    \item The difference of $\mathrm{RF}$ results between the CG and EG methods are from 14 $\%$ to 20 $\%$. The maximum difference of $\mathrm{RF}$ results between the DG and EG methods, on the other hand, are much lower. The maximum difference is approximately 0.5 $\%$ when $K = 1$ GPa.
    \item As also observed from the structured heterogeneity example, the media that has a lower $K$ value, i.e. more deformable, has a higher $\mathrm{RF}$. Figures \ref{fig:rf_2d}a-c present that the final $\mathrm{RF}$ is increased from $3.5 \times 10^{-3}$ to $2.0 \times 10^{-1}$ when $K$ is reduced from $8$ GPa to $1$ GPa.
\end{enumerate}

\begin{figure}[H]
   \centering
        \includegraphics[width=8.0cm, height=7.0cm]{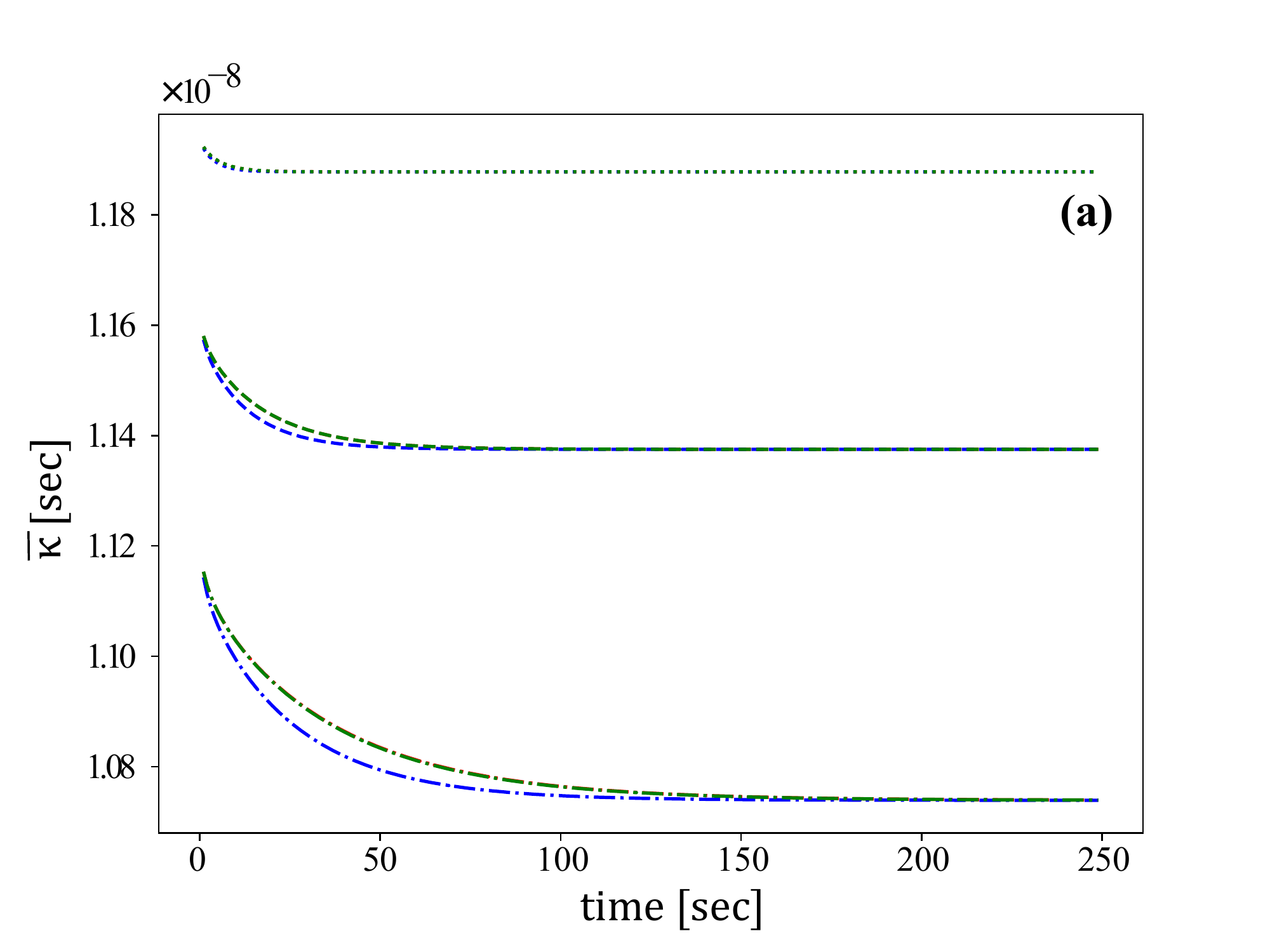}
        \includegraphics[width=8.0cm, height=7.0cm]{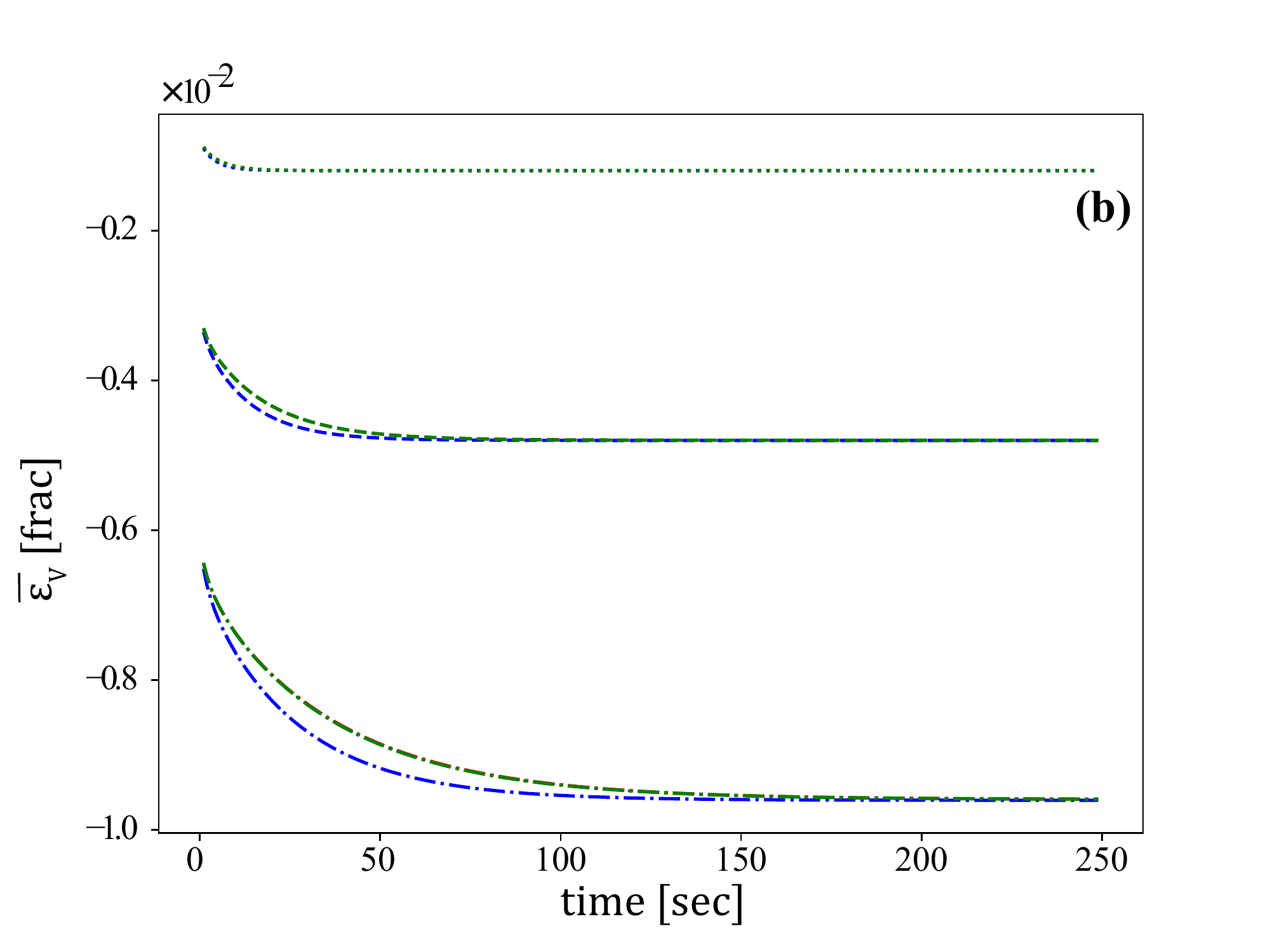}
        \includegraphics[width=8.0cm, height=7.0cm]{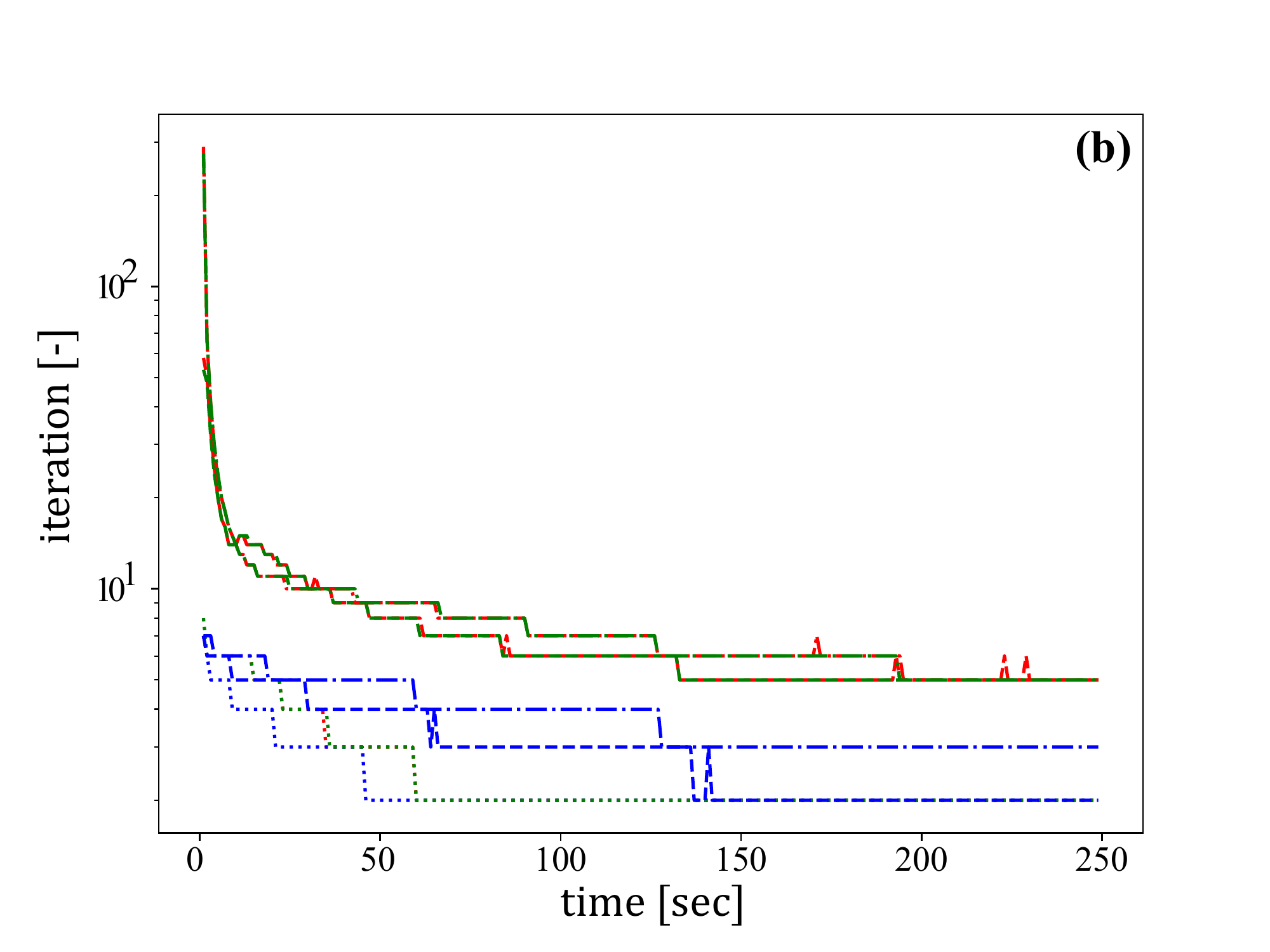}
        \includegraphics[width=8.0cm, height=0.9cm]{pictures/com_mass_no_deform_layered_legend.pdf}
   \caption{2D heterogeneous example: the comparison of (\textbf{a}) $\mathrm{\bar{\bm{\kappa}}}$, (\textbf{b}) $\mathrm{\bar{\varepsilon_{v}}}$, and (\textbf{c}) number of iterations values among the CG (blue line), EG (red line), and DG (green line) methods using pressure-dependent $\bm{k_m}$ model for case 1, $K = 8$ GPa, case 2, $K = 2$ GPa, and case 3, $K = 1$ GPa. Note that EG and DG results are on top of each other for (\textbf{a}) and (\textbf{b}).}
   \label{fig:com_2d}
\end{figure}

The disagreement of $\mathrm{RF}$ between the CG and EG/DG methods can be explained using the results of $\Bar{\bm{\kappa}}$ and $\Bar{\varepsilon_v}$ presented in Figures \ref{fig:com_2d}a-b, respectively. The $\Bar{\bm{\kappa}}$ and $\Bar{\varepsilon_v}$ values of the CG method are significantly different from the other two methods at the beginning, but they converge to the same values at the later time. On the contrary, the $\Bar{\bm{\kappa}}$ and $\Bar{\varepsilon_v}$ results of the EG and DG methods are approximately similar.

Besides, we compare the number of iterations of Algorithm \ref{al:perm_iteration}. We note that the number of iterations results of the EG and DG methods are almost the same, but they are much higher than that of the CG method (see Figure \ref{fig:com_2d}c). This situation is even more pronounced for the softer material. Comparing with the structured heterogeneity example, the number of iterations of this example is also much higher.

Next, the number of DOF among each method is presented in Table \ref{tab:2d_dof}. As also observed from the structured heterogeneity example, the DOF of EG method is approximately three times higher than that of the CG method, but it is almost two times lower than that of the DG method. Note that CG, EG, and DG methods have the same DOF for displacement space because it is solved on the same CG vector function space with a quadratic approximation.

\begin{table}[H]
\begin{center}
\caption {Degrees of freedom (DOF) comparison among CG, EG, and DG methods for 2D heterogeneous example}\label{tab:2d_dof}
\begin{tabular}{|l|c|c|c|} \hline
\textbf{} & \textbf{CG} & \textbf{EG} & \textbf{DG}   \\\hline
Displacement     & 17,370  & 17,370     & 17,370     \\\hline
Pressure         & 2,212   & 6,474      & 12,786     \\\hline
Total      & 19,582  & 23,844     & 30,156     \\\hline
\end{tabular}
\end{center}
\end{table}

\subsection{3D Flow in a Deformable Media with Random Porosity and Permeability}

\begin{figure}[H]
   \centering
   \includegraphics[width=8.0cm, height=7.0cm]{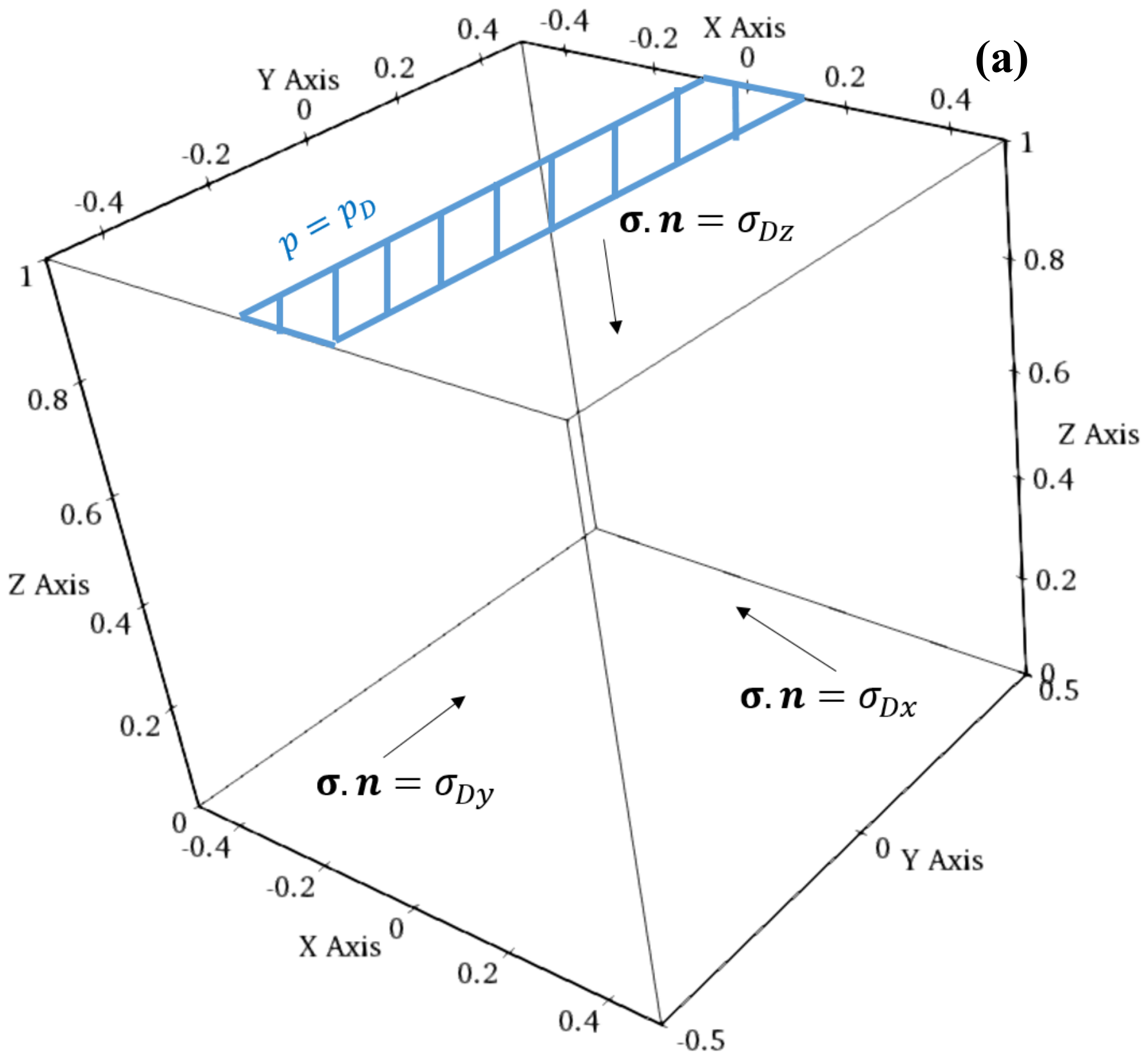}
    \includegraphics[width=7.0cm, height=7.0cm]{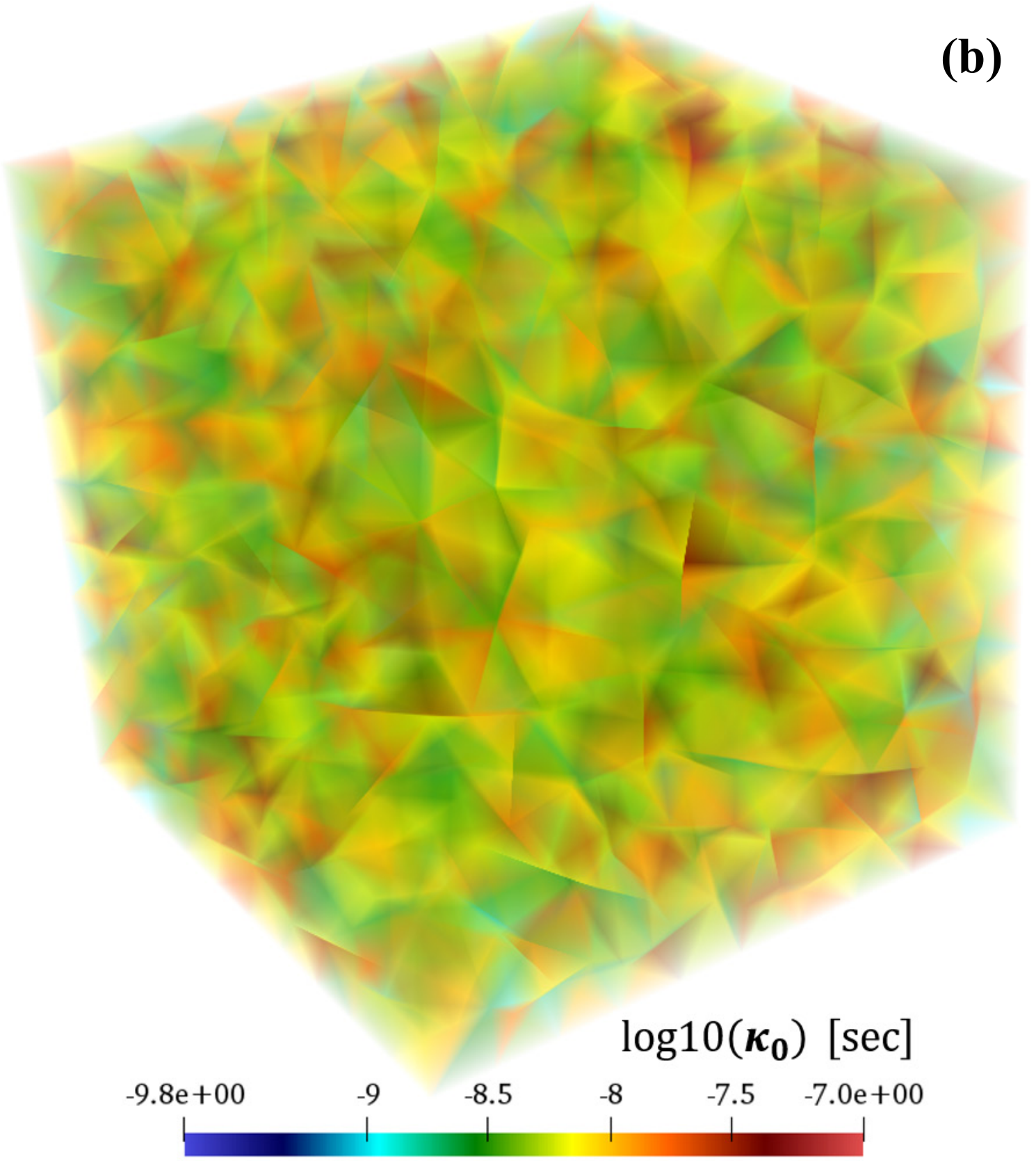}
    \includegraphics[width=8.0cm, height=7.0cm]{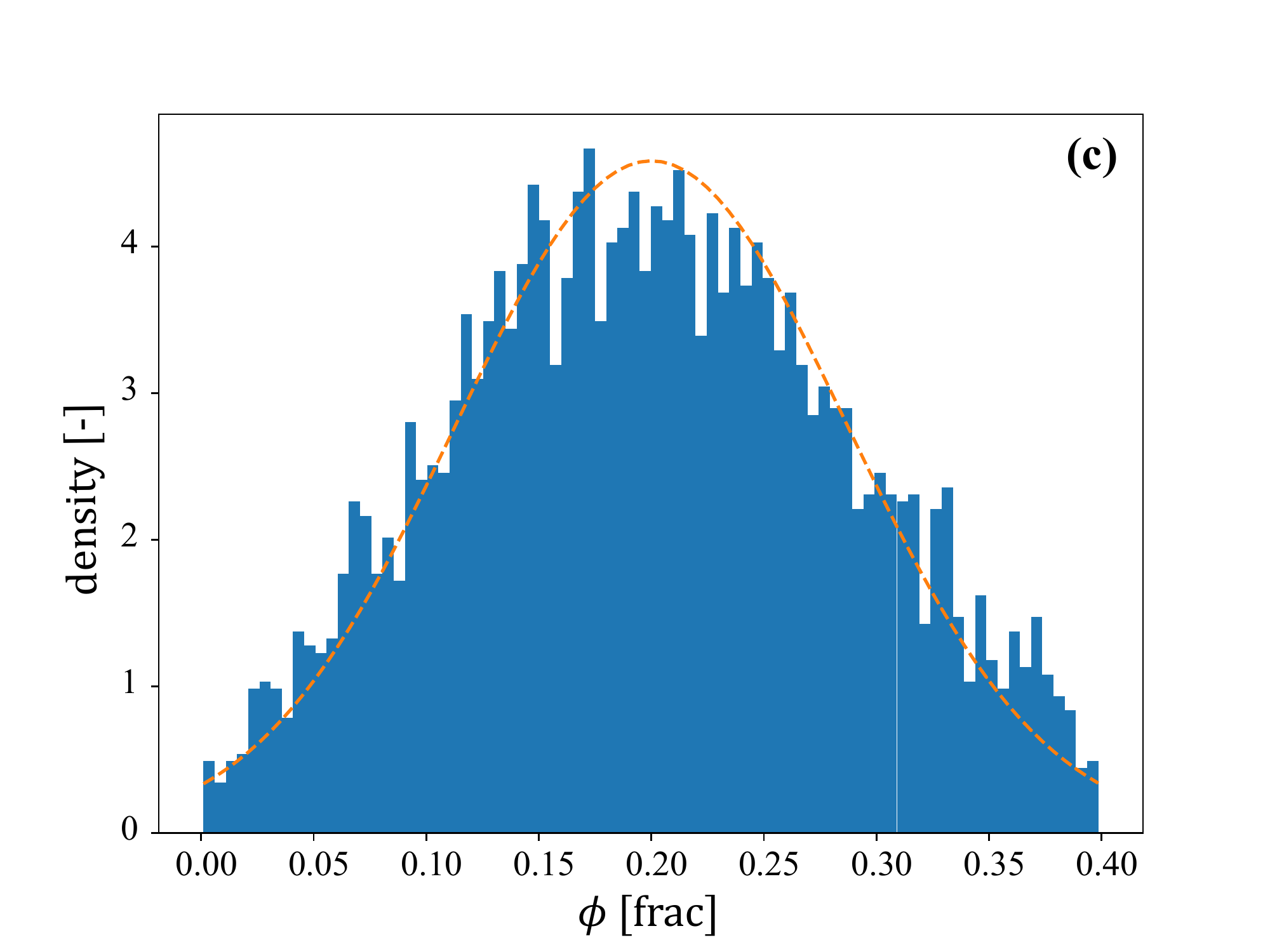}
    \includegraphics[width=8.0cm, height=7.0cm]{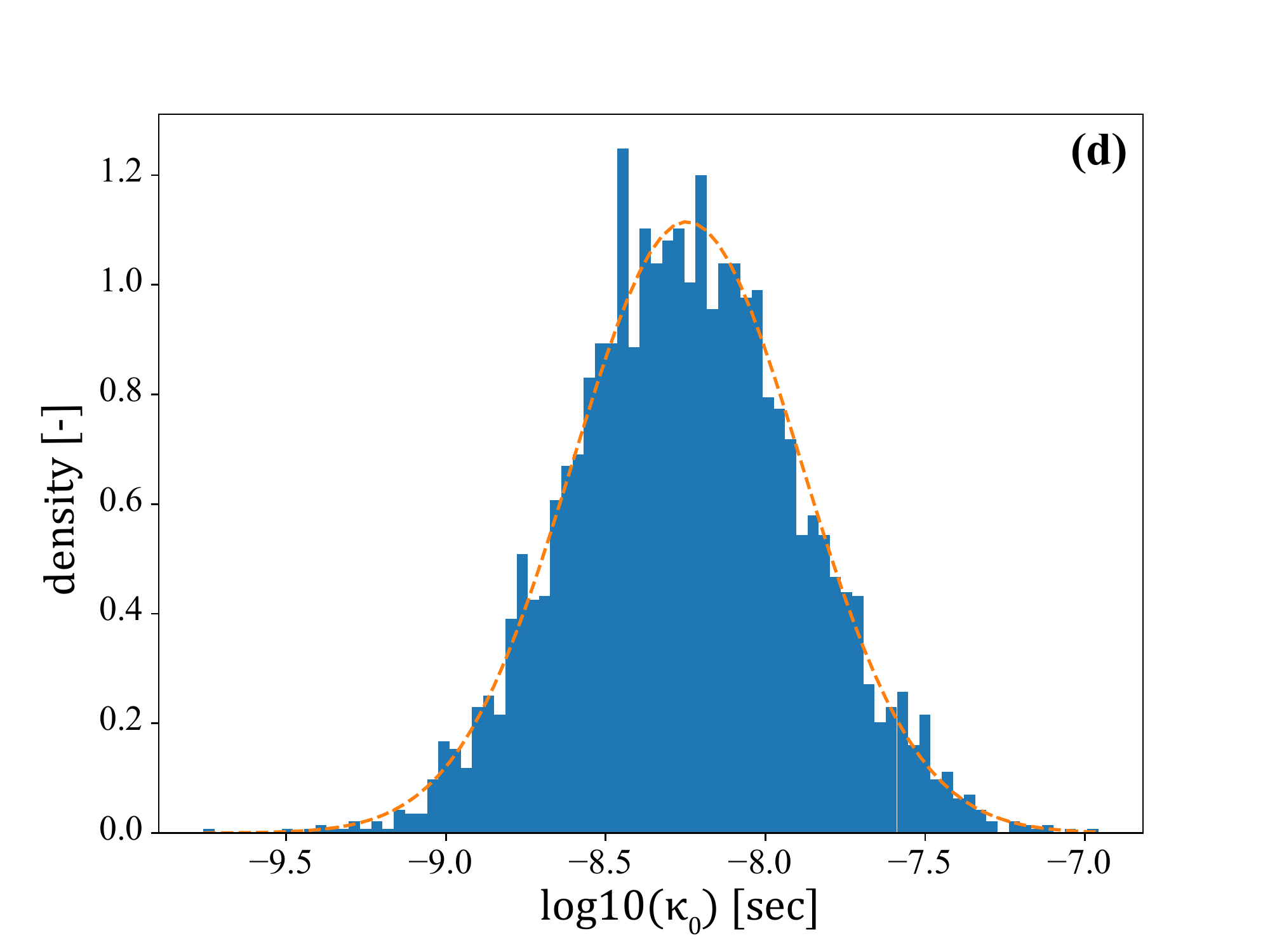}
   \caption{3D heterogeneous example: (\textbf{a}) geometry and boundary conditions (note that 1. besides the surface that is imposed by the constant pressure boundary, other surfaces are imposed by zero-flux boundary and 2. surfaces that are opposite of the stress boundaries are imposed with zero normal displacement), and (\textbf{b}) illustration of $\mathrm{log10(\bm{\kappa_0})}$ field (\textbf{c}) $\phi$ histogram, and (\textbf{d}) $\mathrm{log10(\bm{\kappa_0})}$ histogram}
   \label{fig:3d_case}
\end{figure}

This example aims to extend the investigation from the 2-Dimensional domain, which is presented in the previous section, to a 3-Dimensional domain. The geometry and boundary conditions are presented in Figure \ref{fig:3d_case}a. The input parameters that are applied to the structured heterogeneity and 2D heterogeneous examples are also used in this example. $\bm{\sigma_{D}}$ for this case is applied as follows: $\bm{\sigma_{Dx}}=[15, 0, 0]$, $\bm{\sigma_{Dy}}=[0, 15, 0]$, and $\bm{\sigma_{Dz}}=[0, 0, 20]$ $\mathrm{MPa}$. $\phi$ and $\bm{\kappa_0}$ fields are populated with the same specifications as 2D heterogeneous example as illustrated in \ref{fig:3d_case}c-d, respectively. The illustration of $\bm{\kappa_0}$ field is presented in Figure \ref{fig:3d_case}b.

\begin{figure}[H]
   \centering
        \includegraphics[width=8.0cm, height=7.0cm]{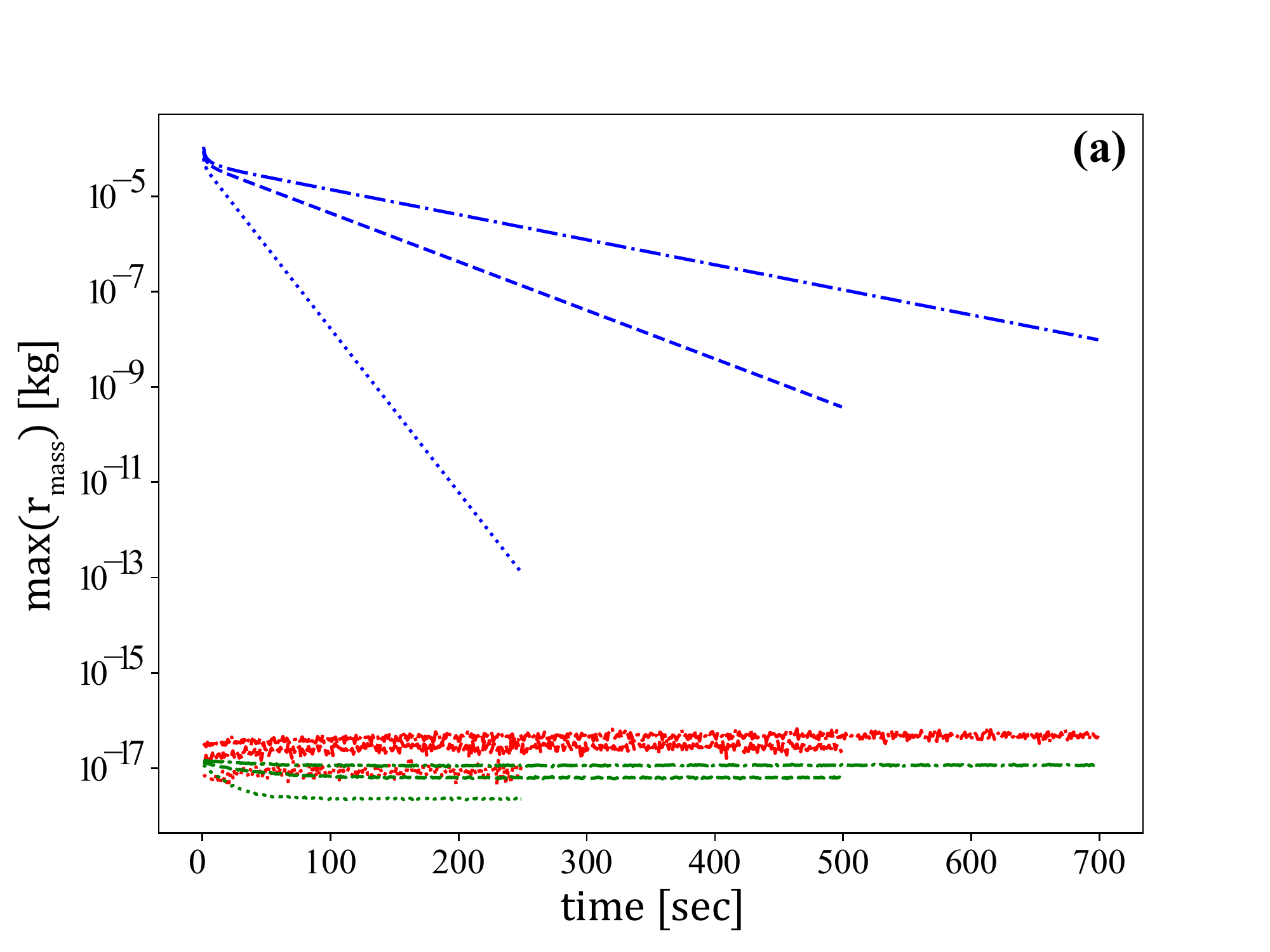}
        \includegraphics[width=8.0cm, height=7.0cm]{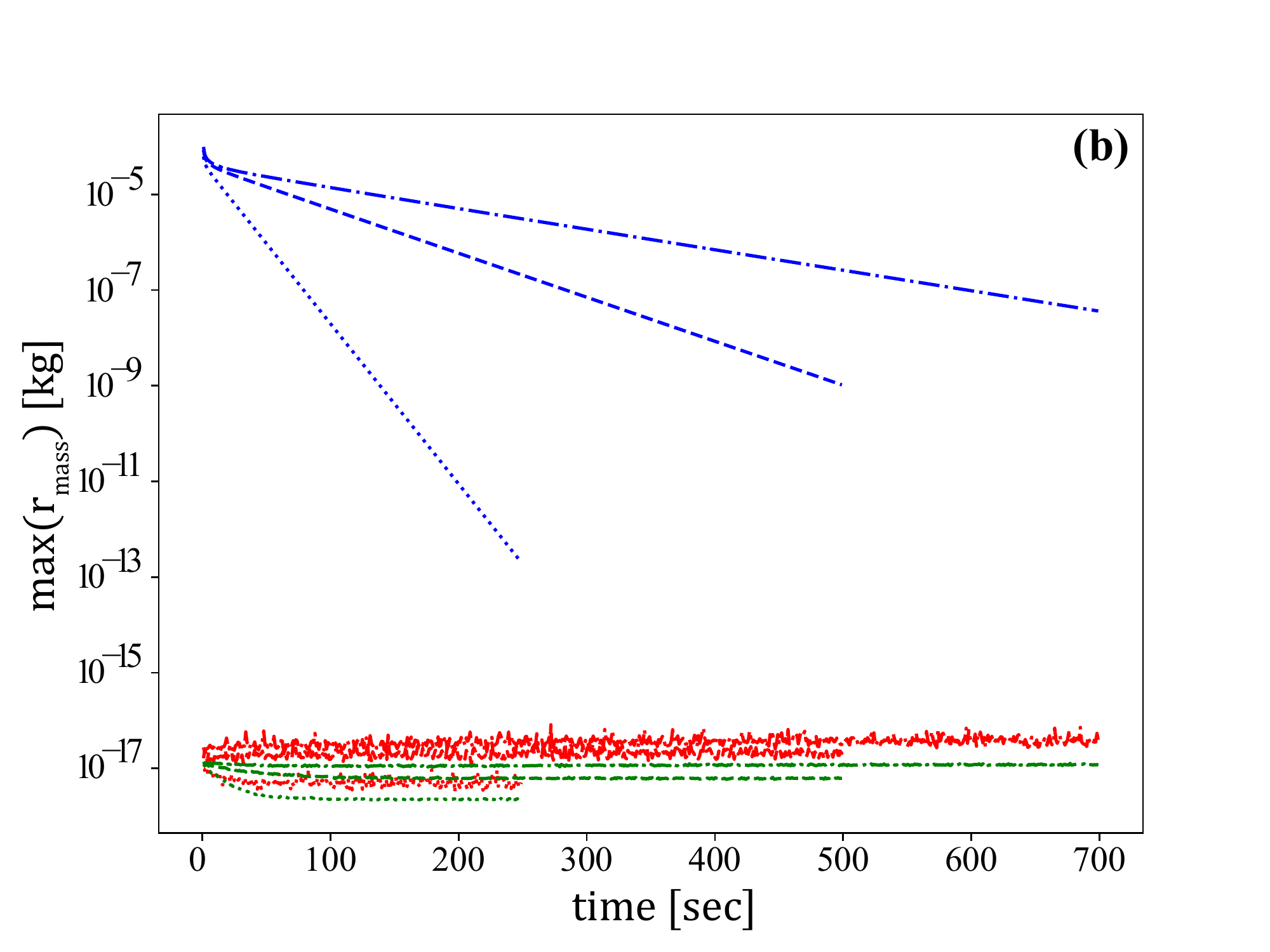}
        \includegraphics[width=8.0cm, height=0.9cm]{pictures/com_mass_no_deform_layered_legend.pdf}
   \caption{3D heterogeneous example: the comparison of $\mathrm{max(r_{mass})}$ values among the CG (blue line), EG (red line), and DG (green line) methods using (\textbf{a}) pressure-independent $\bm{k_m}$ and (\textbf{b}) pressure-dependent $\bm{k_m}$ models for case 1, $K = 8$ GPa, case 2, $K = 2$ GPa, and case 3, $K = 1$ GPa}
   \label{fig:mass_loss_3d}
\end{figure}

Figures \ref{fig:mass_loss_3d}a-b present $\mathrm{max(r_{mass})}$ results for pressure-independent and -dependent $\bm{k_m}$ models, respectively. While the local mass conservation property of the EG and DG methods is well preserved, it is revoked when the CG method is used, i.e. $\mathrm{max(r_{ mass })}$ value is as large as $1 \times 10^{-4}$ kg at the beginning. As expected, the cases with a lower $K$ value have higher $\mathrm{max(r_{mass})}$ value.

\begin{figure}[H]
   \centering
        \includegraphics[width=8.0cm, height=7.0cm]{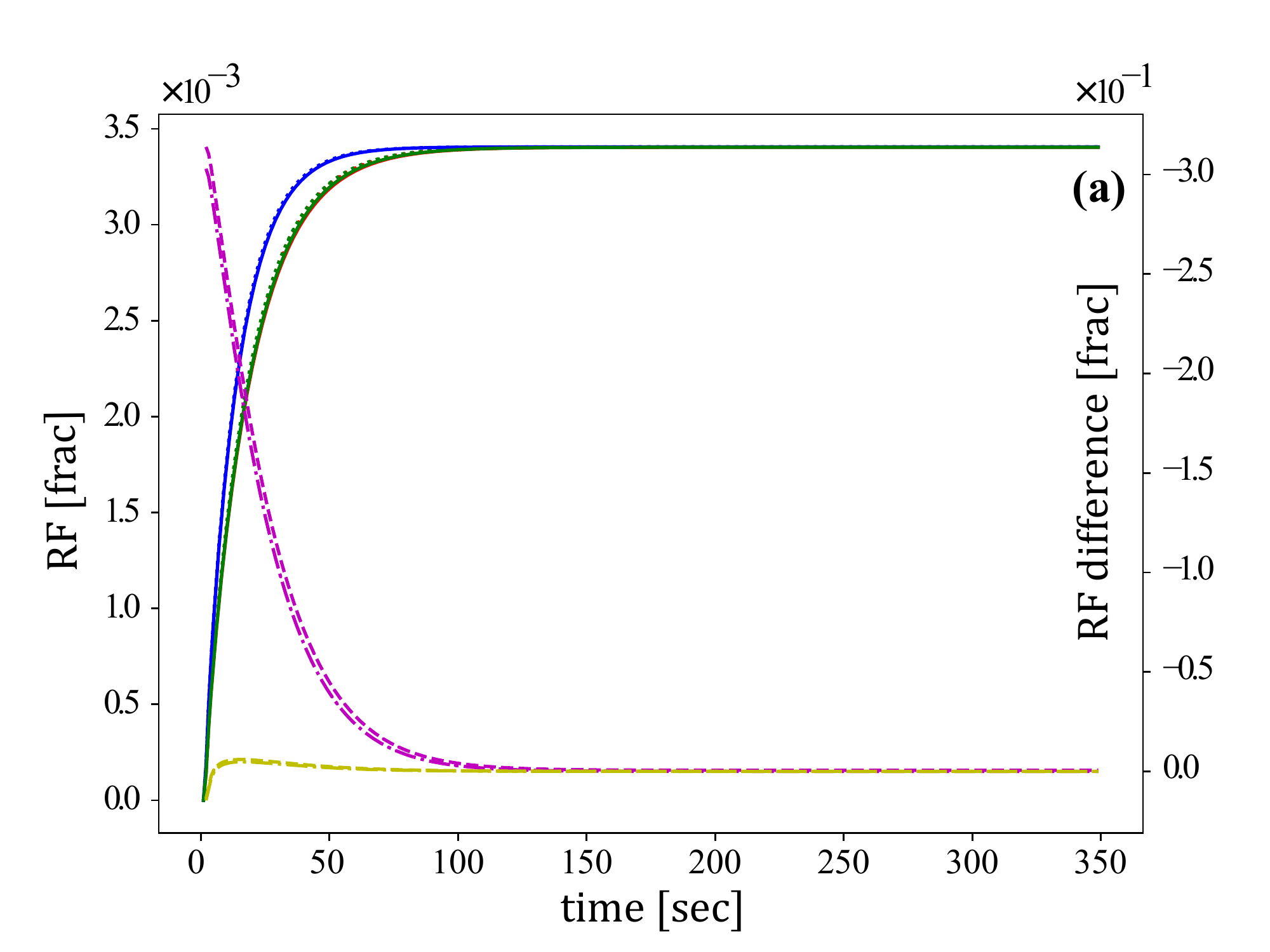}
        \includegraphics[width=8.0cm, height=7.0cm]{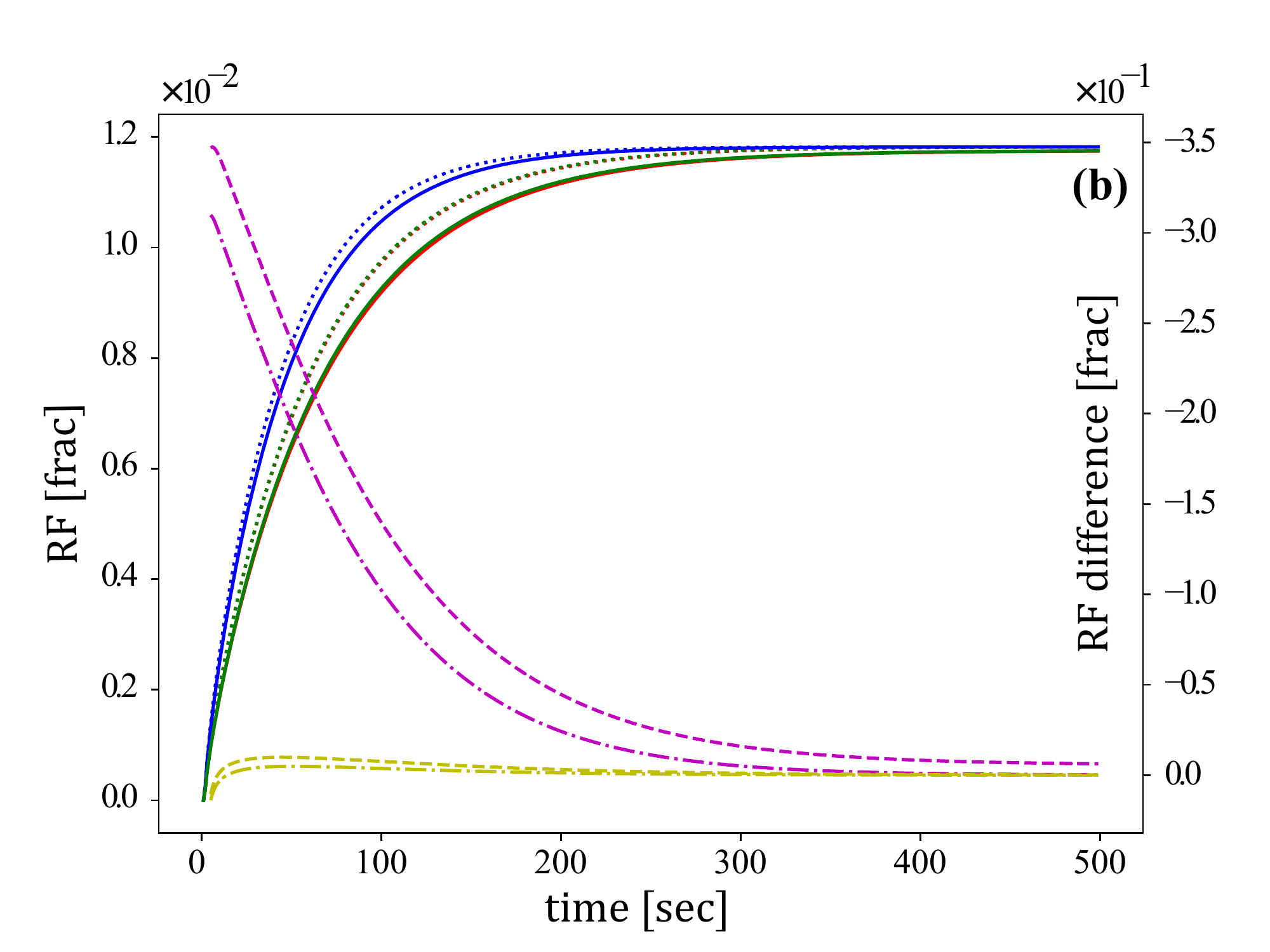}
        \includegraphics[width=8.0cm, height=7.0cm]{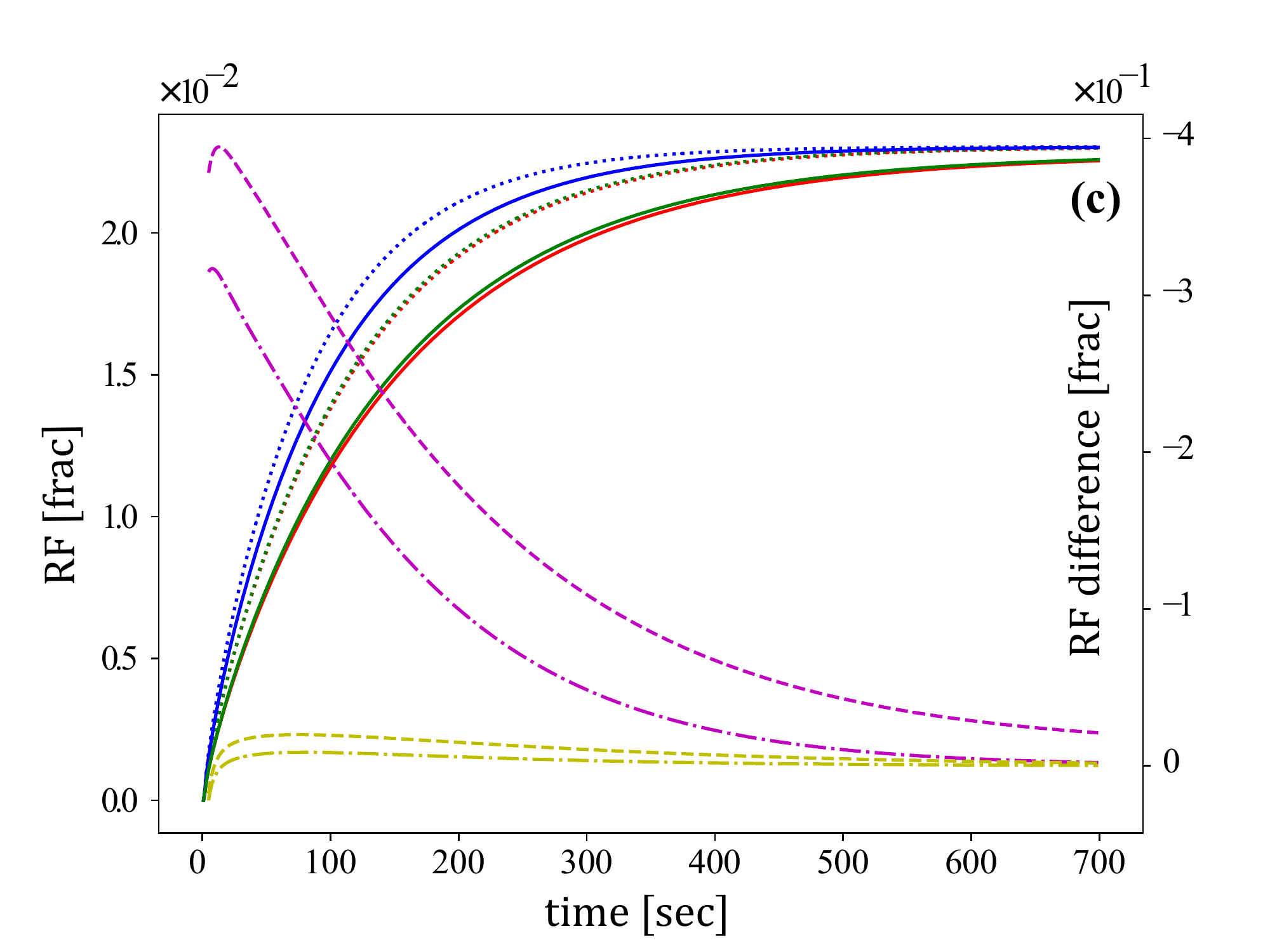}
        \includegraphics[width=8.0cm, height=5.0cm]{pictures/rf_layered_legend.pdf}
   \caption{3D heterogeneous example: the comparison of RF (primary axis) and the difference of RF (second axis) among the CG (blue line), EG (red line), and DG (green line) methods using pressure-dependent (solid) and pressure-independent (dotted) $\bm{k_m}$ models for (\textbf{a}) case 1, $K = 8$ GPa, (\textbf{b}) case 2, $K = 2$ GPa, and (\textbf{c}) case 3, $K = 1$ GPa}
   \label{fig:rf_3d}
\end{figure}

The $\mathrm{RF}$ results of case 1, $K = 8$ GPa, case 2, $K = 2$ GPa, and case 3, $K = 1$ GPa are presented in Figures \ref{fig:rf_3d}a-c, respectively. The observations that are drawn from the 2D heterogeneous example are still valid. The difference of $\mathrm{RF}$ results between the CG and EG methods, however, is now more pronounced, and it ranges from 30 $\%$ to 40 $\%$. On the other hand, the maximum difference between the DG and EG methods is still as small as approximately 3 $\%$.

\begin{figure}[H]
   \centering
        \includegraphics[width=8.0cm, height=7.0cm]{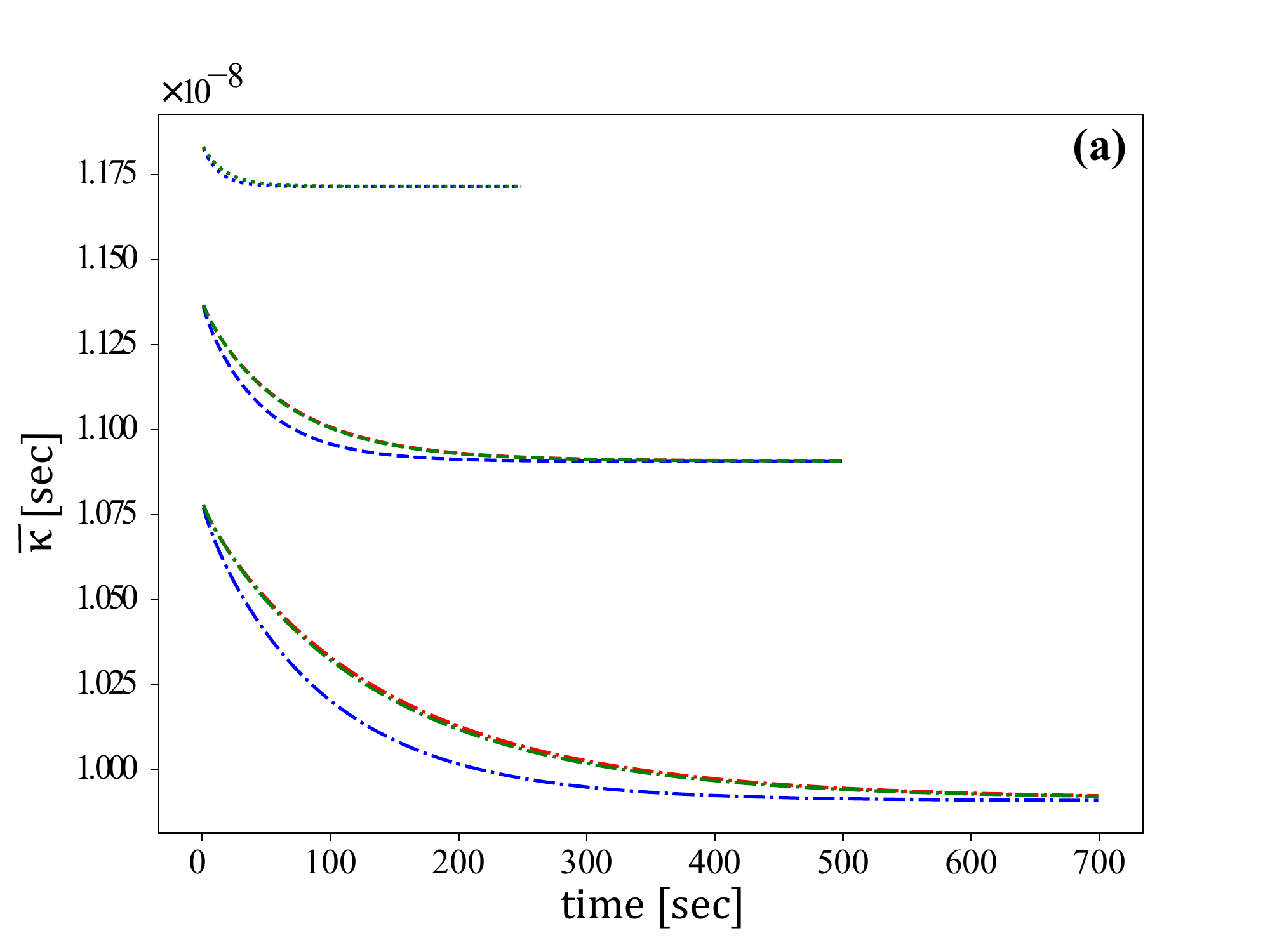}
        \includegraphics[width=8.0cm, height=7.0cm]{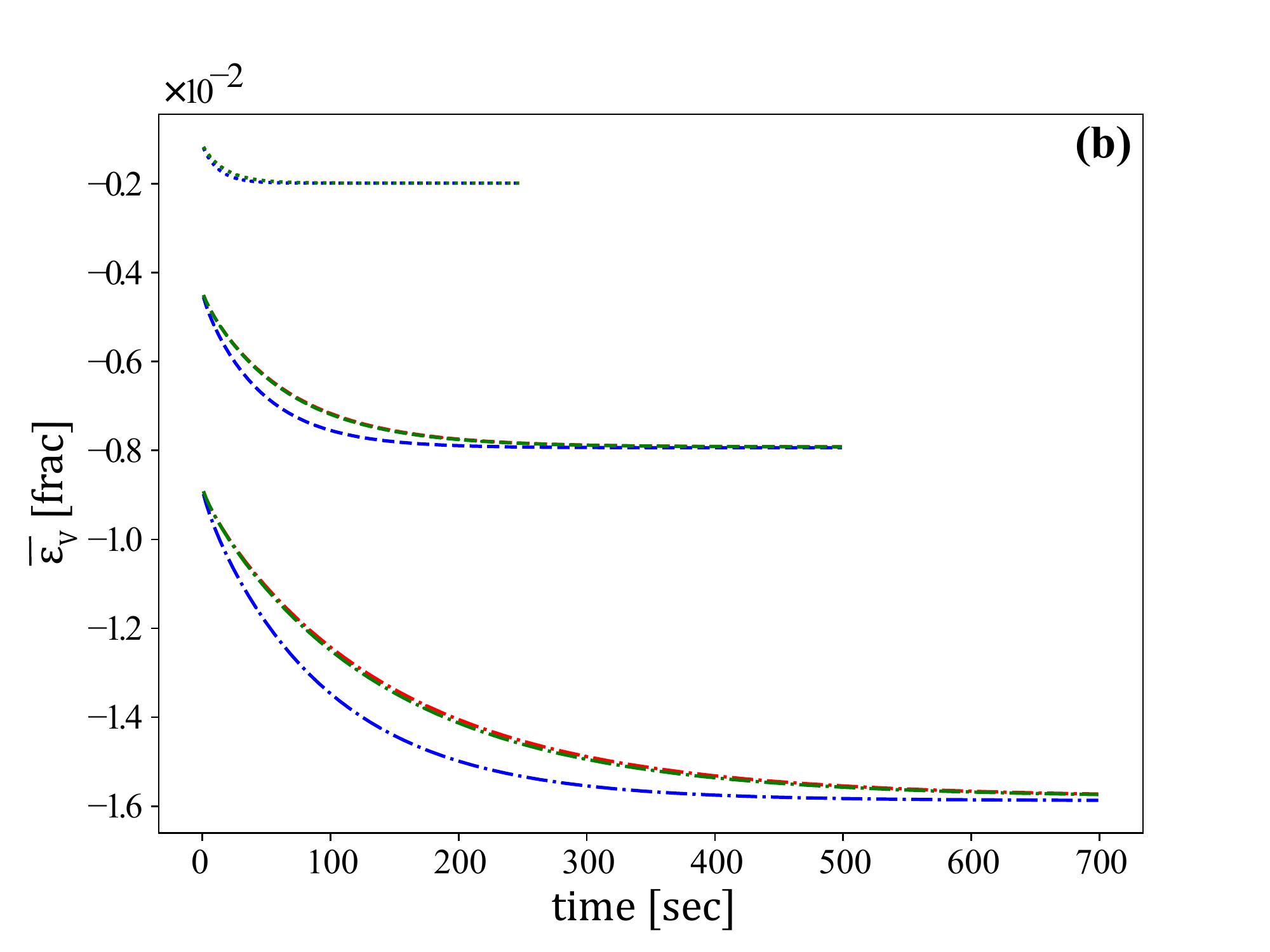}
        \includegraphics[width=8.0cm, height=7.0cm]{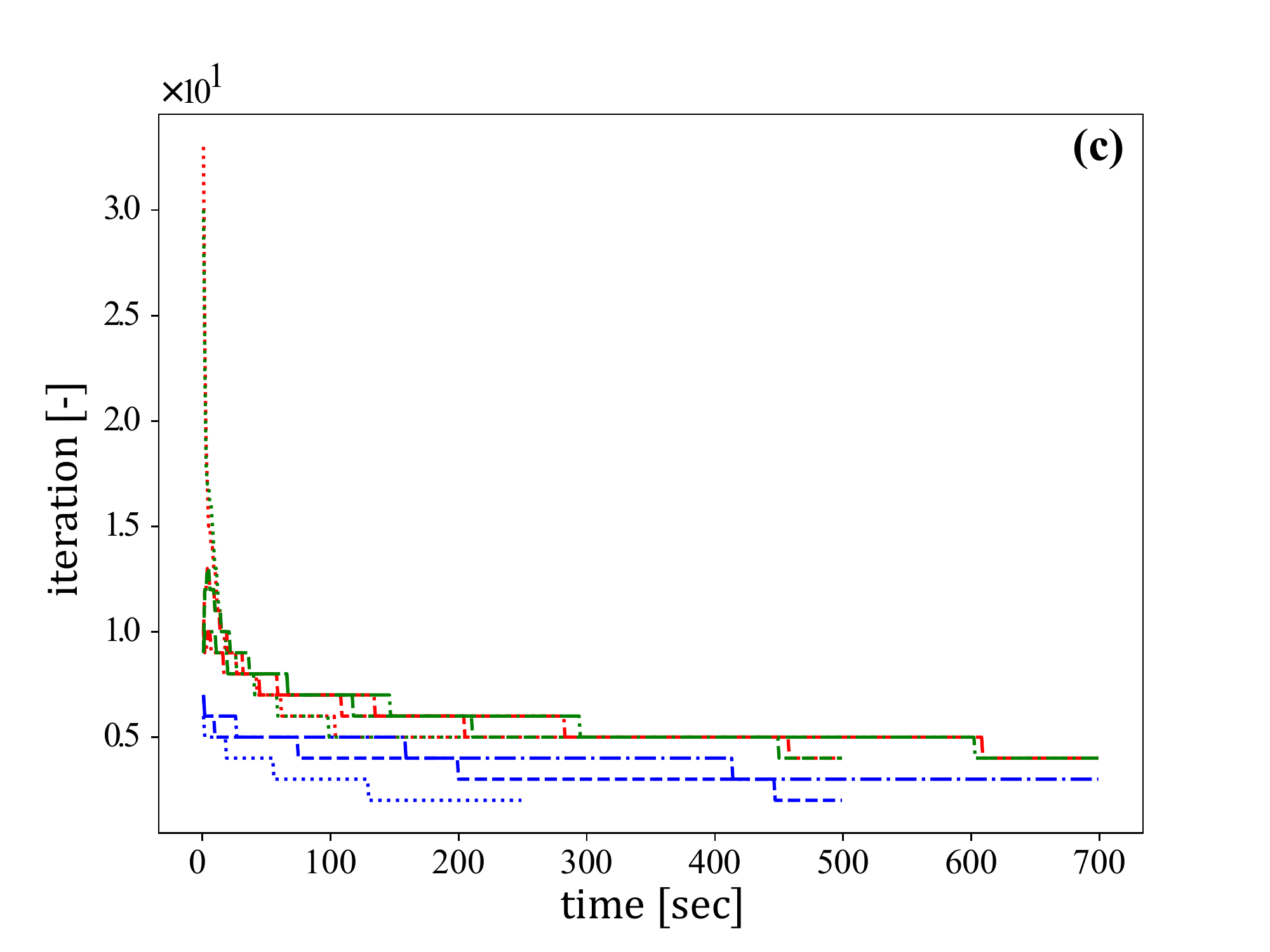}
        \includegraphics[width=8.0cm, height=0.9cm]{pictures/com_mass_no_deform_layered_legend.pdf}
   \caption{3D heterogeneous example: the comparison of (\textbf{a}) $\mathrm{\bar{\bm{\kappa}}}$, (\textbf{b}) $\mathrm{\bar{\varepsilon_{v}}}$, and (\textbf{c}) number of iterations values among the CG (blue line), EG (red line), and DG (green line) methods using pressure-dependent $\bm{k_m}$ model for case 1, $K = 8$ GPa, case 2, $K = 2$ GPa, and case 3, $K = 1$ GPa}
   \label{fig:com_3d}
\end{figure}

The results of $\Bar{\bm{\kappa}}$ and $\Bar{\varepsilon_v}$ are presented in Figures \ref{fig:com_3d}a-b, respectively. As also observed from 2D heterogeneous example, the $\Bar{\bm{\kappa}}$ and $\Bar{\varepsilon_v}$ values of the CG method are significantly different from the EG and DG methods, in the beginning, then they converge to the same values at the later time since the system reaches an equilibrium state. The results of the number of iterations comply with the 2D heterogeneous example, as shown in Figure \ref{fig:com_2d}c.

The number of DOF among each method is presented in Table \ref{tab:3d_dof}. Even though the CG, EG, and DG methods have the same DOF for the displacement space, the DOF of pressure space among these three methods are significantly different. The EG method is five times more expensive than the CG method, but it is three times less costly than the DG method. We note that the difference between the number of DOF is more significant in the 3D case than the 2D case.

\begin{table}[H]
\begin{center}
\caption {Degrees of freedom (DOF) comparison among CG, EG, and DG methods for 3D heterogeneous example}\label{tab:3d_dof}
\begin{tabular}{|l|c|c|c|} \hline
\textbf{} & \textbf{CG} & \textbf{EG} & \textbf{DG}   \\\hline
Displacement     & 20,901  & 20,901     & 20,901     \\\hline
Pressure         & 1,068   & 5,163      & 16,380     \\\hline
Total      & 21,969  & 26,064     & 37,281     \\\hline
\end{tabular}
\end{center}
\end{table}

\subsection{North Sea field example}

\begin{figure}[H]
   \centering
        \includegraphics[width=14.0cm, height=10.0cm]{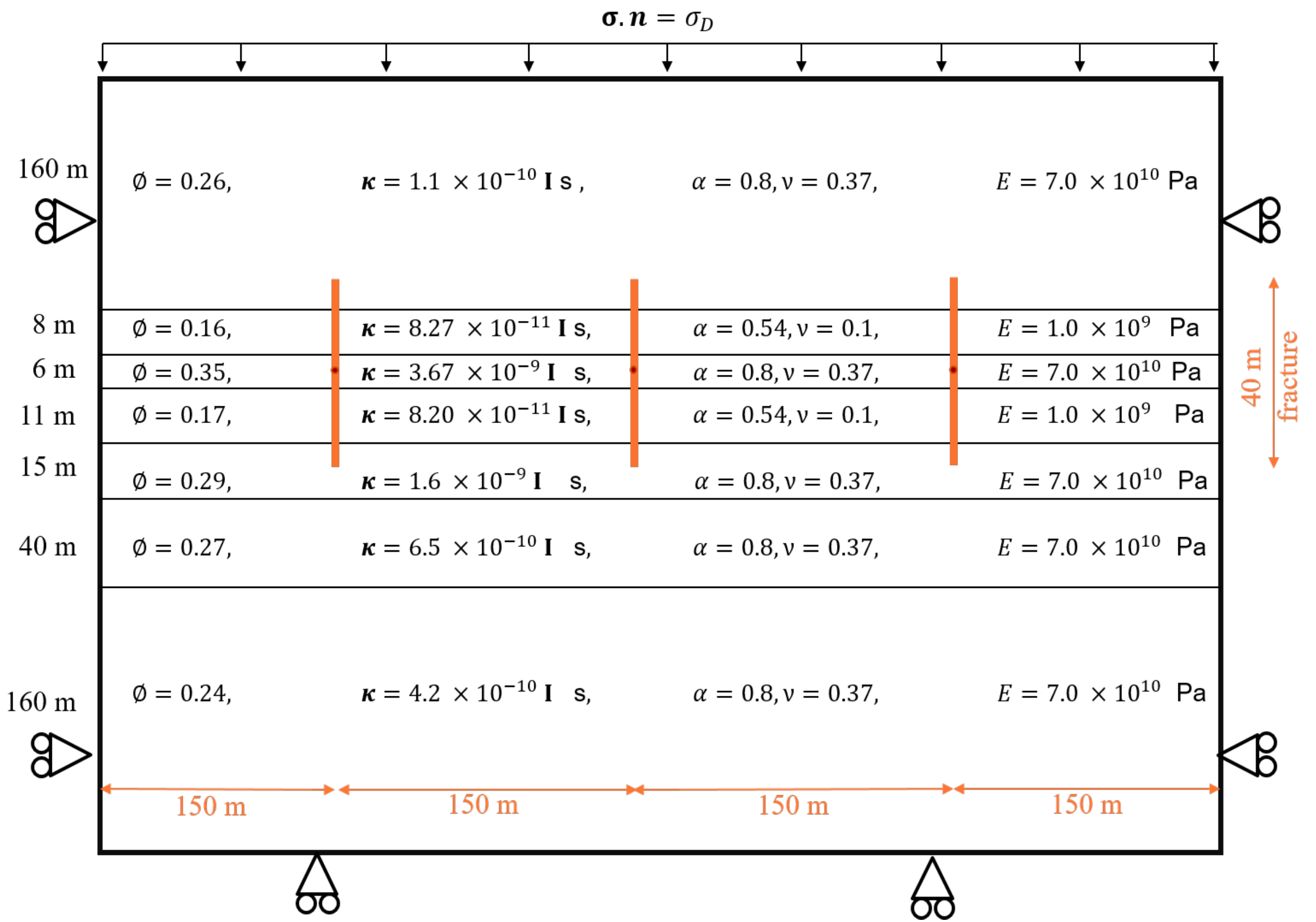}
   \caption{North Sea field example: the geometry and boundaries conditions. Note that the fractures are shown in orange, and the horizontal wells (red dots) are located in the middle of each fracture. Note that the drawing is not scaled to the actual dimension.}
   \label{fig:half_dan_geo}
\end{figure}

\rev{
The last example is based on the configuration of a layered hydrocarbon field in the North Sea \cite{calvert2014insights, calvert2018insights}. The reservoir is exploited by several long (2-5 kilometer) parallel horizontal wells. The wells are hydraulically fractured \citep[e.g.][]{salimzadeh2019effect} along the length of the wells. Figure \ref{fig:half_dan_geo} illustrates the model geometry, including the wells in the middle of each fracture, together with the material properties of each layer and boundary conditions. 
Note that no-flow boundary condition, i.e. $q_D = 0$, is applied to all external faces. In this example, we set $p_0 = 1.0 \times 10^{7}$ Pa, $\sigma_D = 30.0 \times 10^{6}$ Pa, $\rho = 897.0$ $\mathrm{kg/m^{3}}$, $c_f = 1.0 \times 10^{-9}$ Pa$^{-1}$, $\Delta t^n$ $=$ $2.628 \times 10^{6}$ sec,  $\tau$ $=$ $9.461 \times 10^{7}$ sec. We assume that the fracture permeability is much higher than the matrix permeability \cite{calvert2014insights, calvert2018insights}; hence, the constant pressure is applied to the fracture, i.e., $p_D = 1.0 \times 10^{5}$ Pa. We would like to emphasize that to mimic a field model, the robustness of the numerical method to different mesh sizes is one of the vital desire characteristics. 
}


\begin{figure}[H]
   \centering
        \includegraphics[width=8.0cm, height=6.0cm]{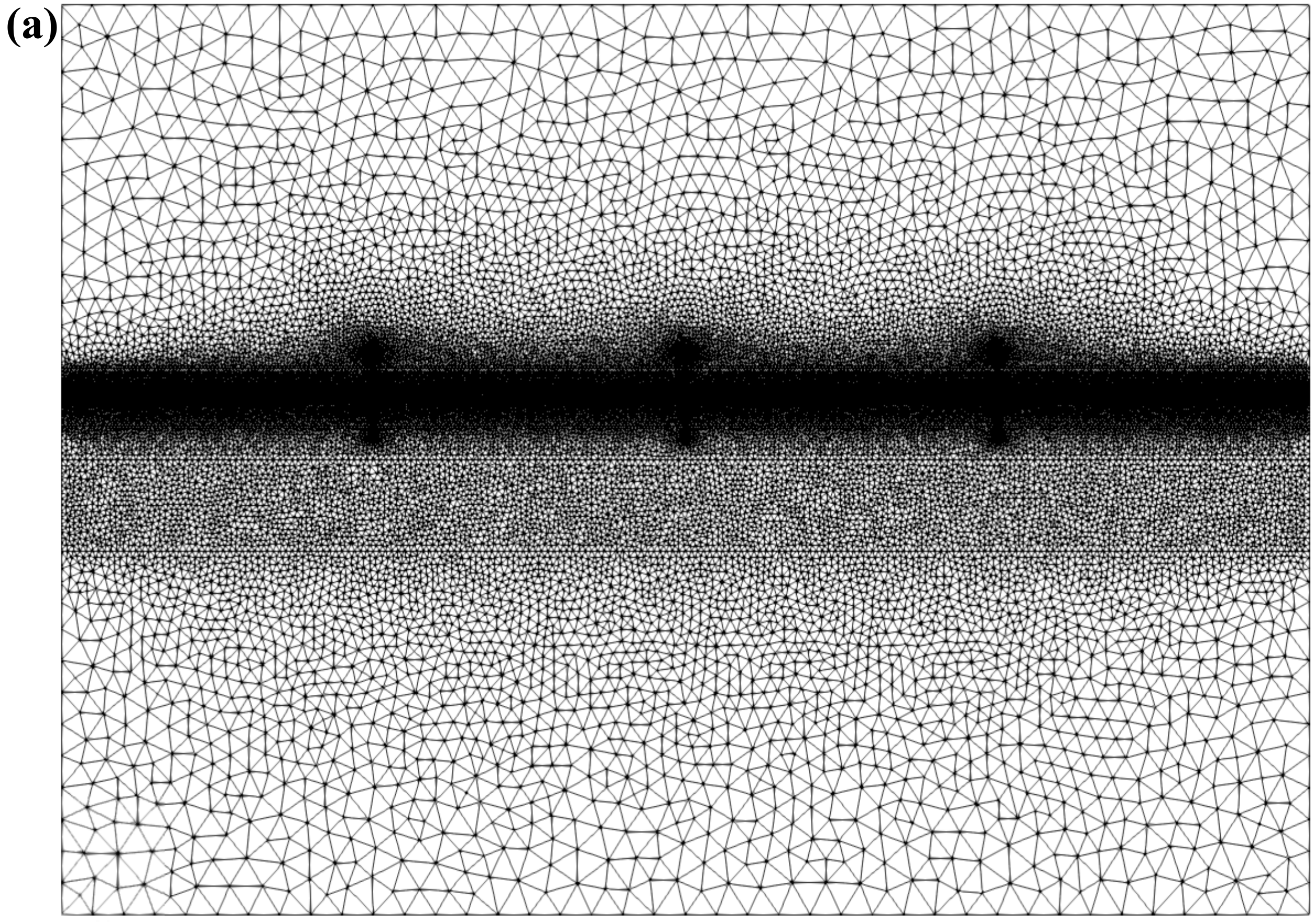}
        \includegraphics[width=8.0cm, height=6.0cm]{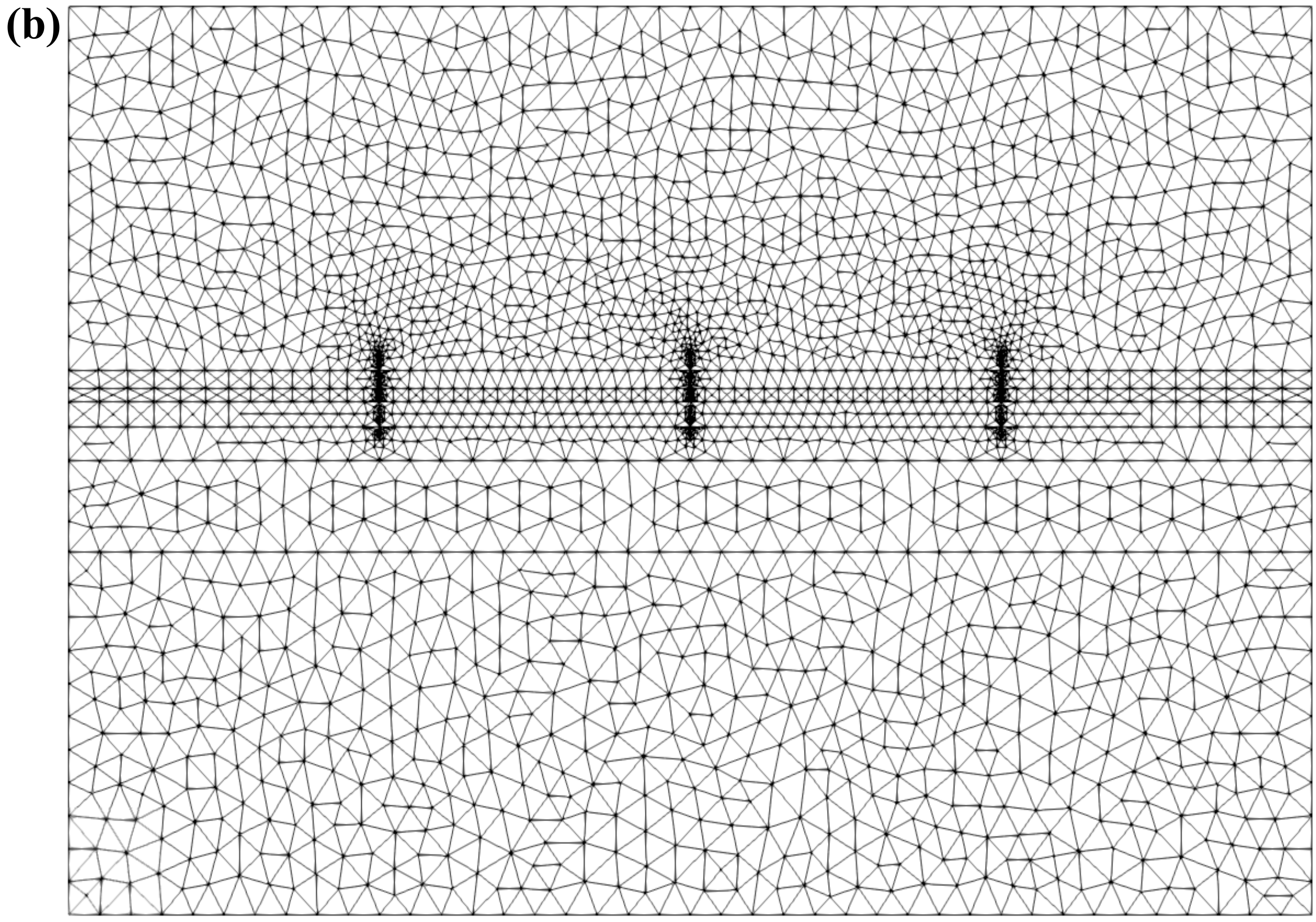}
        \includegraphics[width=8.0cm, height=7.0cm]{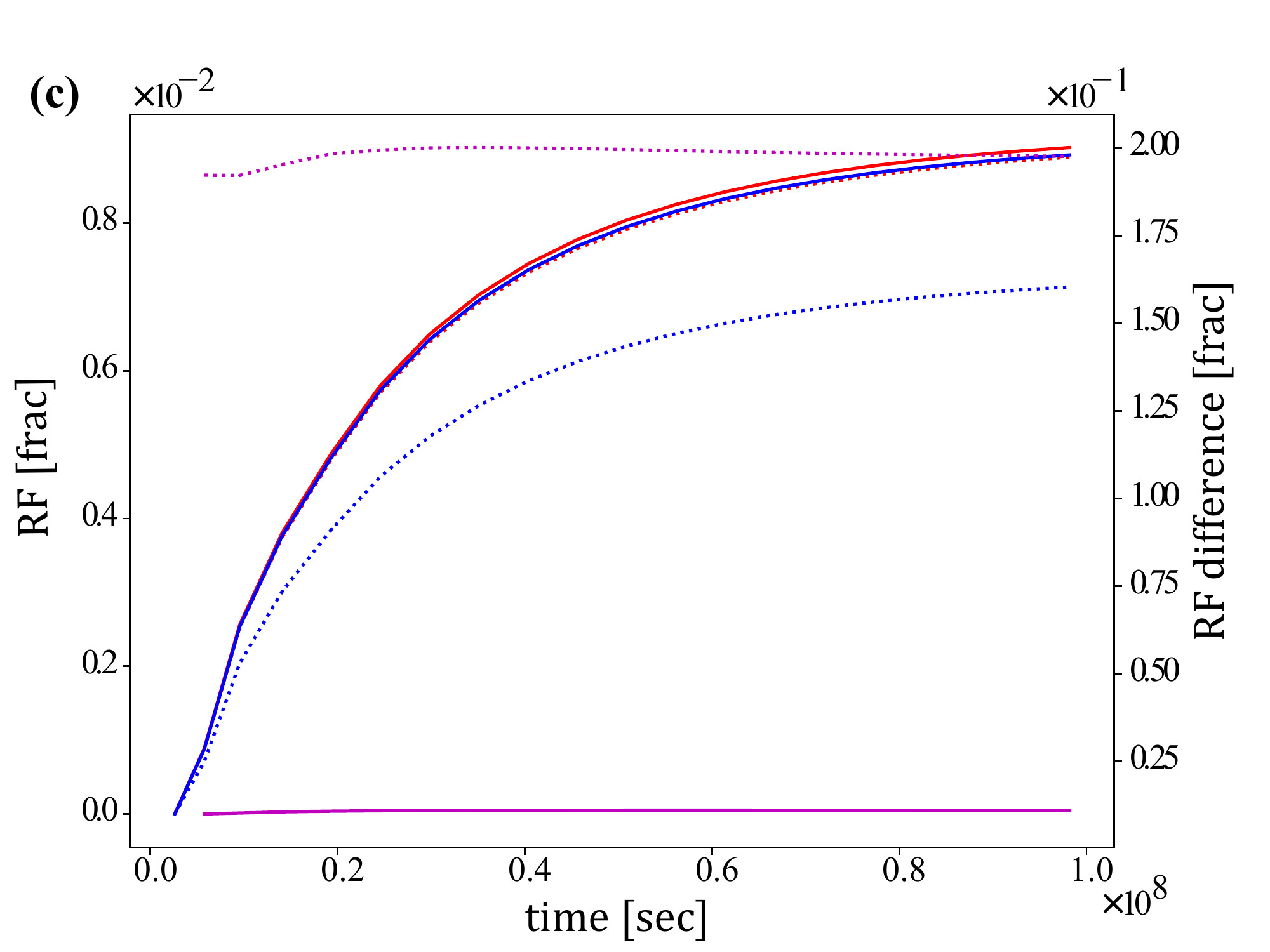}
        \includegraphics[width=8.0cm, height=5.0cm]{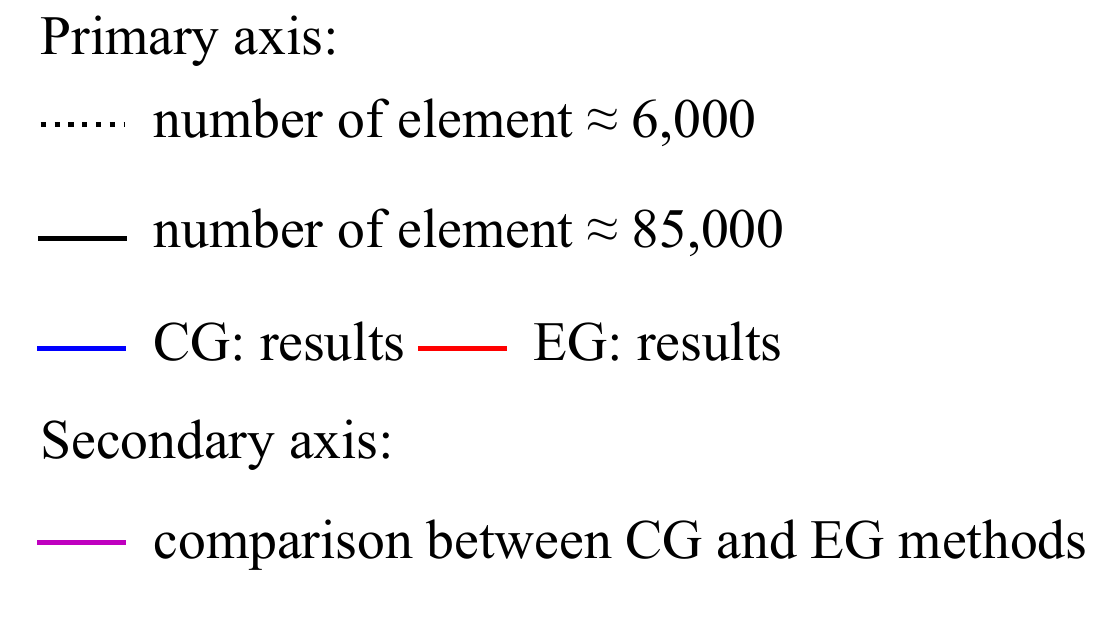}
   \caption{North Sea field example:  the comparison of (\textbf{a}) mesh with number of element $\approx 85,000$, (\textbf{b}) mesh with number of element $\approx 6,000$, and (\textbf{c}) RF (primary axis) and (\textbf{a}) the difference of RF (second axis) between the CG (blue line) and EG (red line) methods. Note that for (\textbf{c}) the results of EG for all cases and CG with number of elements $\approx$ 85,000 are almost overlapped.}
   \label{fig:half_dan_result}
\end{figure}

\begin{figure}[H]
   \centering
        \includegraphics[width=10.0cm, height=10.0cm]{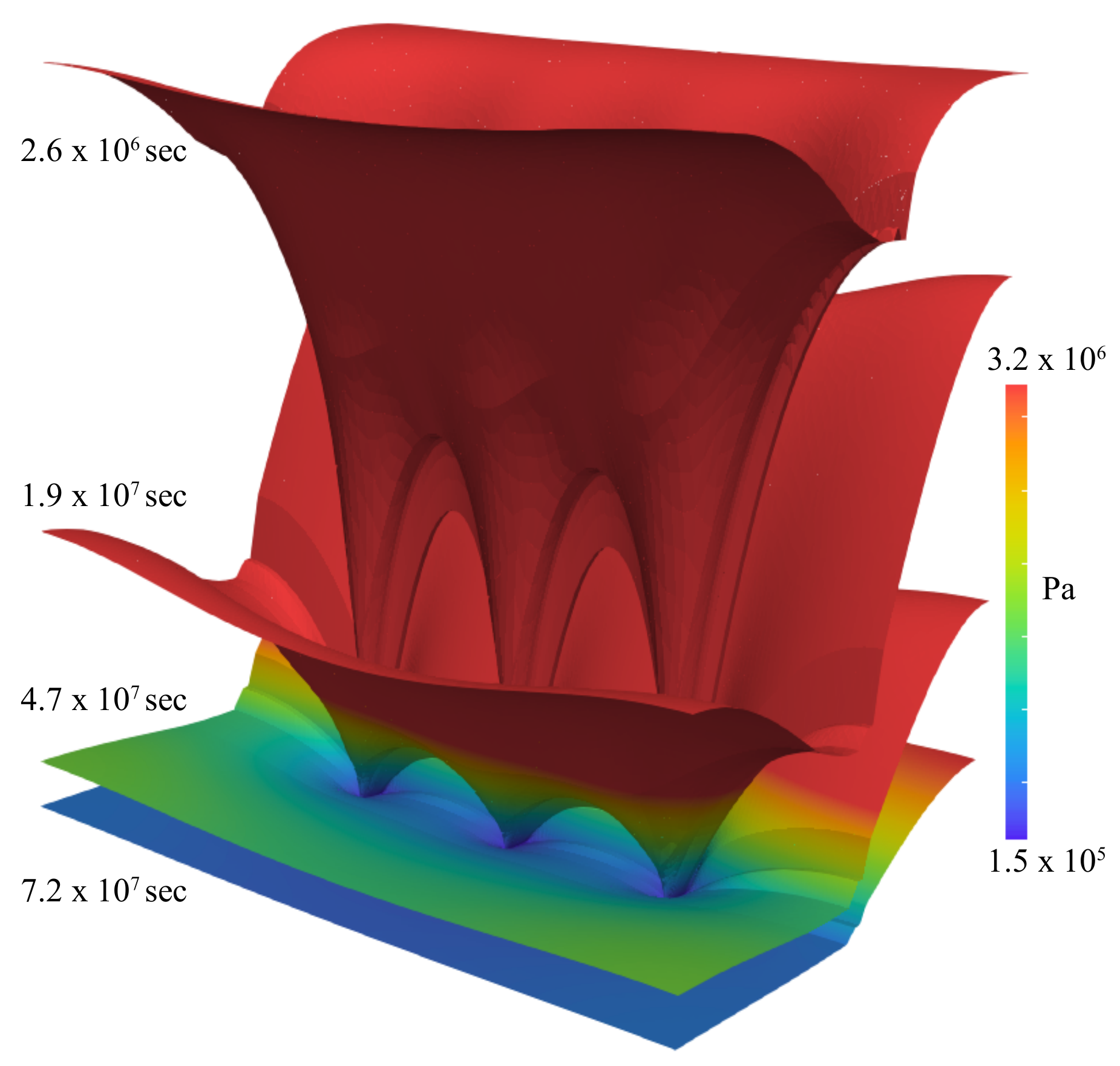}
   \caption{North Sea field example: the examples of pressure solution using mesh with number of element $\approx 6,000$ with different time.}
   \label{fig:half_dan_pressure}
\end{figure}

\rev{
From previous sections, one could see that the effect of permeability alteration resulted from the solid deformation is non-negligible. Therefore, it should be taken into account, and the EG and DG methods could provide comparable results. Hence, in this section, we focus on comparisons between the CG and EG methods using the pressure-dependent $\bm{k_m}$ model to illustrate that the results of EG method are less sensitive to the mesh size compared to the CG method. The CG and EG results of RF and two different meshes are presented in Figure \ref{fig:half_dan_result}. Besides, the examples of pressure solution of the coarse mesh (number of elements $\approx$ 6,000) are presented in Figure \ref{fig:half_dan_pressure}.
}

\rev{
The RF results (see Figure \ref{fig:half_dan_result}a) illustrate that the EG method provides consistent results for both fine and coarse mesh sizes (number of elements $\approx$ 6,000 and 85,000). The EG at the coarse mesh solution already is as precise as the CG on the fine mesh (Figure \ref{fig:half_dan_result}). While the CG results are mesh dependent and varied significantly among different mesh sizes, the results of the EG method are similar regardless of the mesh size. The EG method allows using a relatively coarse mesh to maintain accuracy. This is an essential feature of the EG method that makes it more suitable for large scale simulations. 
}




\rev{
The DOF comparison of the model used in this example is presented in Table \ref{tab:half_dan_dof}. The total DOF for the EG at the coarse mesh is about 11 times less than that for the CG at the fine mesh.
}

\begin{table}[H]
\begin{center}
\caption {Degrees of freedom (DOF) comparison between CG and EG methods using different mesh size for North Sea field example}\label{tab:half_dan_dof}
\begin{tabular}{|l|c|c|c|c|}
\hline
                   & \multicolumn{2}{c|}{\textbf{CG}} & \multicolumn{2}{c|}{\textbf{EG}} \\ \hline
number of elements $\approx$ & 6,000          & 85,000          & 6,000          & 85,000          \\ \hline
Displacement       &     24,486      &   331,062          &   24,486         &  331,062         \\ \hline
Pressure           &    3,096      &   41,438      &    9,148     &  124,094       \\ \hline
Total              &  27,582        &   372,500      &   33,634       & 455,156        \\ \hline
\end{tabular}
\end{center}
\end{table}

\section{Conclusion}\label{sec:conclusion}

This study presents the discretization, block structure, and coupling strategy for solving the single-phase fluid flow in the poroelastic media with permeability alteration using the EG method. Moreover, we provide comparisons of the positive and negative aspects of the CG (in its classical form), EG, and DG methods. Taking into consideration the limitations and assumptions made in this study, the following conclusions are drawn:

\begin{itemize}



\item The CG method suffers from the spurious oscillations in pressure and volumetric strain at the interfaces, especially, with significant permeability differences are observed. The EG and DG methods, on the contrary, deliver smooth solutions since the EG and DG methods can handle the discontinuity of pressure across the interfaces, which also results in the local mass conservation.

\item The CG, EG, and DG methods provide approximately the same flux approximation in the deformable media with the structured heterogeneity example. However, flow in the 2D deformable media with random porosity and permeability example illustrates that the flux approximation calculated by the CG method is different from the EG and DG method as high as 20 \%.
This difference becomes more substantial in the 3D domain (the maximum difference is 40 \%) since there are more material interfaces in the 3D than the 2D domain, which lead to the increase of pressure and $\varepsilon_v$ oscillation across interfaces.  The EG and DG methods deliver approximately the same flux approximation in all examples (the maximum difference is less than  3 \%).

\item If the lack of local mass conservation is not the primary concern, the CG method could be used to model the poroelastic problem for models with the little material discontinuity, e.g. the structured heterogeneity example as it requires the least computational resources. However, the CG method may not be suitable when the heterogeneity is distributed since it may overestimate the flux calculation.

\item The EG and DG methods require higher computational resources than the CG method. The number of Picard iteration of EG and DG methods is approximately similar, but it is larger than that of the CG method. Thus, providing or developing the optimal solver for both EG and DG cases are necessary. The number of iterations is increased if $K$ is lower, where the media is more deformable.
\item The DOF of EG is approximately three and five times larger than that of the CG method for 2D and 3D domains, respectively. However, it is almost two and three times fewer than that of the DG method for 2D and 3D domains, respectively.

\item The permeability reduction resulted in the solid deformation is significant, and it should be considered. Moreover, compaction effect may overcome $\bm{k_m}$ reduction effect on the reservoir deliverability, i.e. even though the softer reservoir has higher $\bm{k_m}$ reduction its ultimate RF is higher than that of the more hardened media.

\item \rev{Applying the EG method in the field setting, illustrates its capability to produce robust results, i.e., the mesh size effect is minimal. This experiment shows the possibility of extending this method to real-field application.}

\end{itemize}


\rev{Finally, in terms of future work, fast solvers of the EG method focusing on multiphysics problems should be investigated and developed. Moreover, an adaptive enrichment, i.e., the piecewise-constant functions only added to the elements where the sharp material discontinuities are observed, can be developed.  }

\section{Acknowledgements}

The research leading to these results has received funding from the Danish Hydrocarbon Research and Technology Centre under the Advanced Water Flooding program. The computations in this work have been performed with the multiphenics library \cite{Ballarin2019} library, which is an extension of FEniCS \cite{AlnaesBlechta2015a} for multiphysics problems. We acknowledge developers and contributors to both libraries. FB also thanks Horizon 2020 Program for Grant H2020 ERC CoG 2015 AROMA-CFD project 681447 that supported the development of multiphenics.






\bibliographystyle{elsarticle-num}

\bibliography{cites.bib}

\end{document}